\documentclass[12pt]{article}

\usepackage[utf8]{inputenc}
\usepackage[T1]{fontenc}    
\usepackage{textgreek}
\usepackage[english]{babel}  

\usepackage[a4paper,margin=1in]{geometry}
\usepackage{setspace}
\usepackage[parfill]{parskip}     

\usepackage{ragged2e}    
\usepackage{soul} 
\usepackage[normalem]{ulem}

\usepackage{xcolor}

\usepackage{amsmath, amsthm, amssymb}  
\usepackage{siunitx} 
\usepackage{booktabs}         

\usepackage[symbol]{footmisc}
\renewcommand{\thefootnote}{\fnsymbol{footnote}}

\usepackage{enumitem} 

\usepackage[super,numbers,sort&compress]{natbib}
\usepackage[colorlinks=true, linkcolor=black, citecolor=black, urlcolor=blue]{hyperref} 
\usepackage[version=4]{mhchem} 
\usepackage{bm}   
\usepackage{comment} 
\usepackage{seqsplit} 

\usepackage{xparse}
\usepackage{etoolbox} 
\usepackage{pgffor}

\NewDocumentCommand{\figref}{m}{%
  Fig.~\figrefaux{#1}%
}

\NewDocumentCommand{\sfigref}{m}{%
  Fig.~\figrefaux{#1}%
}

\newcommand{\figrefaux}[1]{%
  \def\templist{}%
  \foreach \x [count=\i from 1] in {#1} {%
    \ifnum\i=1
      \textnormal{\csuse{\x}}%
    \else
      ,\hspace{0.15em}{\csuse{\x}}%
    \fi
  }%
}

\usepackage{graphicx}   
\usepackage{epstopdf}       
\usepackage{float}          
\usepackage{placeins}
\usepackage[labelfont=bf,justification=centering]{caption} 
\usepackage{newfloat}
\DeclareFloatingEnvironment[
    fileext=los,
    name=Figure
]{suppfigure}
\newcounter{supfig}            
 
\newcounter{modelfigure}

\newcommand{\majorheading}[1]{%
  \vspace{1.5em}
  \begin{center}
    {\Large\bfseries #1}
  \end{center}
  \vspace{0.5em}
}

\usepackage{authblk}    

\begin{document}
%
\title{\Large{\textbf{Actin cross-linking organizes basal body patterning through anomalous diffusion transitions}}}
\author[1]{Raghavan Thiagarajan\textsuperscript{\textdagger}}
\author[2]{Younes Farhangi Barooji}
\author[2]{Poul-Martin Bendix}
\author[3]{Mandar M. Inamdar\textsuperscript{\textdagger}}
\author[1]{Jakub Sedzinski\textsuperscript{\textdagger}}
\affil[1]{Novo Nordisk Foundation Center for Stem Cell Medicine (reNEW), Department of Biomedical Sciences, University of Copenhagen, 2200 Copenhagen, Denmark}
\affil[2]{Biocomplexity, Niels Bohr Institute, Faculty of Science, University of Copenhagen, 2200 Copenhagen, Denmark}
\affil[3]{Department of Civil Engineering, Indian Institute of Technology Bombay, Mumbai 400076, India}
\date{}

\begingroup
\renewcommand{\thefootnote}{\fnsymbol{footnote}}
\footnotetext{\textsuperscript{\textdagger}Corresponding authors: raghavan.thiagarajan@sund.ku.dk, minamdar@iitb.ac.in, jakub.sedzinski@sund.ku.dk}
\endgroup

\maketitle 

\textbf{\large{Abstract}}

Subcellular protein complexes and organelles exhibit diverse dynamic behaviors that reflect the mechanical constraints and organization of the intracellular environment. Although some structures follow classical Brownian motion, many display anomalous dynamics, including subdiffusion and superdiffusion, driven by viscoelasticity, molecular crowding, and cytoskeletal interactions. Transitions between these regimes are increasingly recognized as critical for subcellular organization, yet how they are regulated and influence pattern formation remains unclear. Here, we investigate the spatial arrangement of cilia on the apical surface of multiciliated cells (MCCs) in developing \textit{Xenopus laevis} embryos, where coordinated ciliary beating depends on the precise organization of hundreds of centriole-derived basal bodies (BBs). Using quantitative confocal, high-resolution and high-speed TIRF imaging together with theoretical modeling, we show that BB trajectories undergo time-resolved transitions between diffusive and anomalous motion, with distinct regimes that correlate with apical surface expansion. During the early stages, actin remodeling facilitates the dispersal of BBs by providing a permissive, low-confinement environment. As development progresses, the actin network becomes increasingly cross-linked, forming a dense meshwork that constrains BB movement and promotes uniform spacing across the apical domain. Disruption of $\alpha$-actinin-1, a major actin cross-linking protein, impairs the integrity of the apical actin meshwork, weakens BB confinement, and disrupts regular spatial patterning, ultimately compromising the spatial arrangement of BBs required for proper cilia alignment. Together, we show that progressive apical actin cross-linking coordinates BB positioning and regulates their dynamic state, guiding the shift from diffusive to confined motion. This transition in dynamics enables the emergence of a uniform BB pattern, which in turn ensures the aligned deployment of motile cilia necessary for effective directional fluid flow.\\

\section*{Introduction}

The subcellular environment is a dynamic and heterogeneous medium in which particles, ranging from membraneless macromolecular assemblies to membrane-bound organelles, exhibit motion that often deviates from classical Brownian behavior\cite{hofling_anomalous_2013, metzler_non-brownian_2016, agrawal_morphology_2022}. Such anomalous dynamics, including subdiffusion and superdiffusion, can arise from molecular crowding, spatial heterogeneity, or active forces generated by cytoskeletal networks\cite{golding_physical_2006, wirtz_particle-tracking_2009, bruno_transition_2009, etoc_non-specific_2018, woringer_protein_2018}. These non-Brownian modes of motion are functionally important, contributing to subcellular organization, cargo trafficking, and spatial patterning within cells\cite{hafner_spatial_2018, weiss2004anomalous, joshi_emergent_2024}.

The actin cytoskeleton, a dynamic and viscoelastic network, plays a central role in the shaping of intracellular dynamics by providing mechanical constraints and generating contractile and extensile forces\cite{lieleg_structure_2010, salbreux_actin_2012, prost_active_2015, gross_how_2017, banerjee_actin_2020}. Remodeling of actin filaments, cross-linking, and interactions with molecular motors enable transitions between diffusive and anomalous motion\cite{bruno_transition_2009, maelfeyt_anomalous_2019, anderson2021subtle}. Among the many subcellular structures regulated by actin mechanics, centrioles and their mature form, basal bodies (BBs), serve as a powerful model for investigating how cytoskeletal remodeling governs organelle positioning and collective pattern formation.

Centrioles are conserved microtubule-based organelles that function across diverse cell types to organize microtubule arrays, facilitate mitotic spindle formation, and establish planar cell polarity and directional migration\cite{marshall_centrioles_2001, nigg_centrosome_2011, conduit_centrosome_2015, carvajal2016centriole, etienne-manneville_integrin-mediated_2001, luxton_orientation_2011, kupfer_polarization_1982, higginbotham_centrosome_2007, tang_centrosome_2012, hannaford_positioning_2024}. In ciliated cells, the mother centriole matures into a BB that anchors and nucleates a cilium\cite{preble_basal_1999, kobayashi_regulating_2011, breslow_mechanism_2019, kumar_how_2021}. Multiciliated cells (MCCs), a specialized epithelial cell type, generate hundreds of such BB and associated motile cilia on their apical surface\cite{brooks_multiciliated_2014, boutin_biology_2019, mahjoub_development_2022}. These cilia drive directional fluid flow in various tissues: enabling mucociliary clearance in the airway\cite{fahy2010airway, knowles2002mucus}, transporting ova through the oviduct\cite{halbert1976egg, lyons2002fallopian}, and circulating cerebrospinal fluid in the brain ventricles\cite{worthington1963ependymal}.

The precise positioning of the BBs on the apical surface is essential for coordinated ciliary beating and effective fluid flow generation \cite{arata_coordination_2022, meunier_multiciliated_2016, bustamante-marin_cilia_2017, nawroth_multiscale_2020}. Disruption of BB patterning alters mucociliary clearance and contributes to ciliopathies that affect the respiratory, reproductive, and nervous systems\cite{ohata2014loss, kunimoto2012coordinated, park_dishevelled_2008, robinson2020camsap3, legendre_motile_2021, wallmeier2020motile}. Understanding how BBs are spatially organized during MCC development remains a central question in cell and developmental biology.

In particular, the BB arrangements within the apical domain of MCC are tissue specific: unipolar in the brain ependyma~\cite{mirzadeh_cilia_2010}, linear in the mammalian airway epithelium~\cite{herawati_multiciliated_2016, shiratsuchi_dual-color_2024}, and uniformly distributed in the \textit{Xenopus} embryonic epidermis~\cite{werner_actin_2011, turk2015zeta}. These diverse spatial patterns reflect an emergent organization governed by tissue-specific molecular and mechanical cues. BB positioning has been associated with planar cell polarity pathways, cytoskeletal dynamics, and structures associated with BB, such as the basal feet~\cite{park_dishevelled_2008, werner_actin_2011, antoniades_making_2014, herawati_multiciliated_2016}. In both \textit{Xenopus} embryonic MCCs and mammalian airway MCCs, the number of centrioles scales with the apical surface area, indicating a conserved coupling between organelle biogenesis and cell geometry~\cite{kulkarni_mechanical_2021, nanjundappa2019regulation}. However, the dynamic principles by which BBs self-organize within the apical domain remain poorly understood.

Emerging evidence suggests that the apical actin cytoskeleton plays a crucial role in regulating BB positioning. In \textit{Xenopus} MCCs, the assembly of the apical actin after centriole amplification drives the expansion of the apical surface and provides a mechanical and structural scaffold for the BB docking\cite{pan_rhoa-mediated_2007, sedzinski_emergence_2016, yasunaga_microridge-like_2022}. Proteins such as WDR5 and Filamin-A promote actin polymerization and cross-linking around BBs\cite{kulkarni_wdr5_2018, chevalier_mir-34449_2015}, while disruption of the apical actin meshwork disrupts the BB docking and spatial organization\cite{boisvieux-ulrich_cytochalasin_1990, werner_actin_2011}. These observations suggest that the apical actin network not only scaffolds BBs but also imposes mechanical constraints that influence their motility and final distribution.

Viewed as active particles embedded in an evolving viscoelastic medium, BBs offer a unique model for exploring how the cytoskeletal architecture shapes organelle positioning. Their behavior resembles that of active colloids, where spatial order arises from local interactions and mechanical feedback with the surrounding matrix\cite{bruot_realizing_2016, vernerey_biological_2019, bechinger_active_2016}. Modeling studies have shown that BBs can self-organize through cytoskeleton-dependent interactions\cite{herawati_multiciliated_2016, namba_cytoskeleton_2020}, yet how these processes are regulated by the dynamic and cross-linked architecture of the apical actin network remains largely unexplored.

Here, we demonstrate that remodeling of the apical actin cytoskeleton, particularly through cross-linking, governs the dynamic behavior and spatial organization of BB during the development of \textit{Xenopus laevis} MCCs. By combining quantitative confocal imaging with high-speed TIRF microscopy, we show that BBs exhibit in-plane motility characterized by anomalous diffusion, transitioning from free to confined dynamics as the apical domain expands. The perturbation of actin cross-linking through $\alpha$-actinin-1 KD disrupts both the confinement of the BB and regular spatial patterning by altering the actin meshwork, which normally forms pocket-like structures that embed and position the BBs within the apical domain. Theoretical modeling reveals that actin-mediated spatial bias and repulsive interactions imposed by the cross-linked network are required to reproduce the emergent BB distribution. Together, these findings establish a direct mechanistic link between the architecture of apical actin and BB dynamics, highlighting how cytoskeletal remodeling influences the positioning and patterning of subcellular organelles.

\section*{Results}

\subsection*{Basal body patterning is synchronized with apical surface expansion}

The formation of MCCs in the \textit{X. laevis} embryonic epidermis involves a coordinated morphogenetic program known as radial intercalation\cite{stubbs2006radial, ventura_multiciliated_2022}, during which the progenitors of MCCs insert into the outer epithelial layer and expand their apical domain in a process called apical emergence\cite{sedzinski_emergence_2016}. This apical surface is essential for the generation of motile cilia, which are nucleated from the BB after their ascent, docking, and distribution across the apical domain (\figref{figonea}).

To investigate BB dynamics during apical emergence, we performed early-stage injections in \textit{X. laevis} embryos to drive the MCC-specific expression of fluorescently tagged proteins. Chibby, a centriolar and BB-associated protein required for BB docking and ciliogenesis \cite{burke2014chibby}, was used to label BB, while LifeAct \cite{riedl2008lifeact}, an F-actin marker, was used to visualize the apical expansion. Both constructs were expressed under the control of the MCC-specific $\alpha$-tubulin promoter\cite{a1999two} (\figref{figoneb, figonec, sfigonea}, Movie 1). Confocal microscopy enabled 3D time-lapse imaging of MCCs throughout the entire BB ascent, docking, and spatial organization process (\figref{figoneb, figonec}, Movie 2). These "coarse-timed" recordings were acquired at intervals of 30 or 60 s.

We focus our analysis on the 2.5-hour window during which apical emergence occurs, a phase previously characterized by a near-linear increase in the apical area followed by a plateau (\sfigref{sfigoneb})\cite{sedzinski_emergence_2016}. During this period, BBs exhibited directed migration from the basal to the apical side of the cell (\figref{figonec}) and progressively docked at the apical membrane (\figref{figoneb}). Quantification of the appearance of the apical BBs revealed a linear accumulation over time (\figref{figoned, figonee}), while the density of the BBs remained constant throughout the expansion phase (\figref{figonef}). This indicated that BB delivery was closely coordinated with apical surface growth. Following docking, the BBs were redistributed throughout the apical domain (\figref{figoneb}), suggesting an additional level of spatial organization beyond the initial insertion on the apical side, prompting further investigation of their spatial distribution patterns. 

Previous studies examining the final spatial organization of BBs in \textit{X. laevis} embryos have shown that they adopt a characteristic spacing of approximately \SI{\sim 1}{\micro\meter} \cite{werner_actin_2011, chatzifrangkeskou2023jnk}. To understand how this pattern emerges over time, we quantified the dynamics of the BB distribution using Voronoi tessellation (\figref{figoneg}, Movie 3). At early time points, the tessellation pattern was heterogeneous, featuring a mixture of large and small areas. As the apical surface expanded and became progressively populated with BBs, the pattern transitioned to a more homogeneous configuration (\figref{figoneg, figoneh}). This transition was quantitatively reflected in a decreasing variance of the tessellation areas over time (\figref{figonei}). When analyzed as a function of apical surface area, the variance of the tesselation areas decreased during the early expansion phase (<\SI{150}{\micro\meter\squared}) and reached a plateau during later stages of expansion (>\SI{150}{\micro\meter\squared}) (\sfigref{sfigonec}).

To further characterize the spatial arrangement of BBs, we measured inter-BB distances using two complementary metrics: \textit{minimum distance}, which captures local packing precision, and \textit{pairwise distance}, which reflects overall spacing tendencies (\textit{see methods section 7.3.4 in SI}) (\sfigref{sfigoned}). The minimum distance (or nearest neighbor) showed a consistent decline over time (\figref{figonej}). When plotted against the apical area, it followed a similar trend: decreasing during the initial expansion phase and stabilizing beyond \SI{150}{\micro\meter\squared} (\sfigref{sfigonee}). This trend mirrored the behavior observed in the tessellation area variance. In particular, in the final stages of expansion, the nearest-neighbor distances converged to \SI{\sim 1}{\micro\meter} (\figref{figonej, sfigonee}), aligning with previous observations of the mature BB spacing. To confirm this, we computed the relative probability distribution of all nearest-neighbor distances across the entire apical expansion timeline. A dominant peak at \SI{\sim 1}{\micro\meter} was evident, reinforcing this value as the preferred spacing between BBs (\sfigref{sfigonef}).

Next, we analyzed pairwise distances to gain insight into how the global BB spacing evolved in relation to apical growth. This analysis revealed a gradual increase in the pairwise distance with the apical area, suggesting that the BBs became more widely distributed as the surface expanded (\sfigref{sfigoneg}). Despite this general spread, the dominant pairwise distance remained centered around \SI{\sim 6}{\micro\meter} (\sfigref{sfigoneh}), indicating that the most likely distance between any two BBs was consistently around \SI{\sim 6}{\micro\meter}.

Together, these findings demonstrate that the ascent, docking, and distribution of BB occur simultaneously with the apical expansion. The observation that BB density remains constant while both the tessellation area variance and nearest-neighbor distance decline from the onset suggests that MCCs initiate BB positioning with high spatial precision early in apical emergence. The plateauing of these parameters mid-way through expansion further implies that the final BB distribution is established well before the apical surface reaches its full size. Collectively, these results point to tight coordination between the mechanisms governing apical expansion and BB distribution.

\subsection*{Passive and active mechanisms shape basal body distribution}

To understand the mechanisms governing the distribution of BBs and their relationship to apical expansion, we characterized BB in-plane dynamics at the apical domain. Using time-lapse imaging, we tracked the movements of BB during apical surface expansion at coarse-timed intervals of 30 or 60 seconds (\figref{figtwoa}, Movie 4). This approach enabled us to examine the spatiotemporal properties of the BB entry and how their trajectories evolve toward a uniform distribution. BBs were found to insert into the apical domain at a steady rate of approximately 2 BBs per minute, with only minor temporal fluctuations (\sfigref{sfigtwoa}). 

Next, we asked whether there was a spatial bias in BB docking. To address this, we extracted the initial coordinates of the newly appearing BBs and classified them based on their position within the apical domain - either centrally or peripherally located (\sfigref{sfigtwob}; \textit{see methods section 7.3.6 in SI}). Interestingly, new BBs appeared preferentially in the periphery of the apical domain (\figref{figtwob, figtwoc}, Movie 5). Given that the apical periphery undergoes continuous expansion outward, this spatial preference suggests that the insertion of the BB is coupled to the extension of the apical domain. Thus, BB appearance and docking proceed at a constant rate, but occur preferentially within newly extended regions of the apical surface.

To further explore the spatial regulation of BB insertion, we quantified the positions of new BBs relative to: (i) BBs that had appeared in previous time points (\sfigref{sfigtwoc}) and (ii) pre-existing BBs present at the same time point (\sfigref{sfigthreea}), using both minimum and pairwise distance metrics. For the first case, we found that the minimum and pairwise distances between new BBs and those from preceding time points peaked at 5-\SI{6}{\micro\meter} (\sfigref{sfigtwod, sfigtwof}). Moreover, these distances increased progressively during apical expansion (\sfigref{sfigtwoe, sfigtwog}), indicating that BBs tend to appear in newly expanded, spatially unoccupied regions. For the latter case, the minimum distance between newly inserted BBs and the pre-existing BB population at the same time point, peaked at \SI{\sim 1}{\micro\meter} (\sfigref{sfigthreeb}) and plateaued at a similar value at the end of the expansion process (\sfigref{sfigthreec}). These findings are consistent with nearest-neighbor distance measurements among all BBs, which also peaked around \SI{\sim 1}{\micro\meter} (\figref{figonej, sfigonee}). In contrast, pairwise distances peaked at \SI{\sim 6}{\micro\meter} and exhibited a steady increase over time (\sfigref{sfigthreed, sfigthreee}). 

Together, these data suggest that BBs maintain a minimum inter-BB spacing of ~\SI{1}{\micro\meter}, while the global spacing increases toward a preferred distance of ~\SI{6}{\micro\meter} as the apical domain expands. This pattern is also evident in the increasing pairwise distances between all subpopulations of BBs, including pre-existing BBs (\sfigref{sfigoneg, sfigoneh}), newly inserted BBs from successive time points (\sfigref{sfigtwof, sfigtwog}), and new vs. pre-existing BBs (\sfigref{sfigthreed, sfigthreee}), suggesting an emergent spatial pattern in the organization of BBs. At the same time, the steady increase in pairwise distances supports the role of apical expansion as a driving force in the redistribution of BBs.

To understand how the BB distribution evolves over time, we quantified the characteristics of the BB trajectories. The contour lengths varied from approximately 5 to 150~\SI{}{\micro\meter} (\sfigref{sfigthreef}), with longer tracks reflecting complex, convoluted paths and shorter tracks suggesting limited mobility or transient presence. However, we ensured data integrity by implementing a strict tracking and filtering pipeline, that excluded spurious short trajectories thereby restricting the remaining short-lived events to genuine transient presence (\textit{see methods section 7.2.4 in SI}). Manual inspection further revealed that some BBs exited and re-entered the apical surface during expansion (\figref{figtwod, figtwoe}, Movie 6). These exit–entry events occurred multiple times 
though with substantial temporal variance (\sfigref{sfigthreei}). In particular, exit behavior was observed only after the apical domain exceeded an area of \SI{\sim 140}{\micro\meter\squared} and persisted thereafter, suggesting that the retention of BB becomes less stable beyond a certain domain size.

Interestingly, exiting BBs were often confined to specific subregions and exhibited minimal displacement (Movie 6). This observation prompted us to ask whether some BBs remained stationary throughout the expansion phase, potentially contributing to the population of trajectories with shorter end-to-end lengths (\sfigref{sfigthreeg}). In fact, analysis of step size distributions revealed a strong probability peak at \SI{\sim 0}{\micro\meter} (\sfigref{sfigthreeh}), which supports the presence of stationary BBs. At the same time, the observed distribution of tortuosity values was consistently greater than 1 (\figref{figtwof}), indicating that BB paths were non-linear, i.e., the contour lengths (\sfigref{sfigthreef}) consistently exceeded their end-to-end displacements (\sfigref{sfigthreeg}). To test for directional bias in the movement of the BBs, we computed the dot product of the displacement and position vectors, which revealed no preferred orientation of movement towards or away from the expanding periphery (\figref{figtwog, sfigthreej}). 

In summary, not all BBs are retained stably within the apical domain during expansion. They can display prolonged stationary behavior, undergo exit and re-entry cycles, or follow highly convoluted and non-directional trajectories. Consistent with these observations, the dynamics of BB exhibited mixed characteristics of both advective and diffusive behavior, as inferred from the alignment of the trajectory and the average of drift (\sfigref{sfigthreek, sfigthreel}) and the analysis of the mean square displacement (\sfigref{sfigthreem}), respectively. These results collectively indicate that, although apical expansion influences BB movements, it is not the sole determinant; The distribution of BB emerges through a combination of passive spreading and additional active regulatory mechanisms.

\subsection*{Basal body dynamics shift with apical domain expansion}

Apical expansion in MCCs is driven by the assembly of the actin network \cite{sedzinski_emergence_2016}. Simultaneously, the BBs ascend from the cell base, reach the apical surface, and become embedded within the actin cortex \cite{antoniades_making_2014}. At the end of the apical expansion, the BBs occupy only ~20\% of the apical area, while the remaining ~80\% is predominantly composed of actin (\sfigref{sfigonea, sfigfoura}). Based on this spatial distribution, we hypothesized that actin may play an active role in the regulation of BB dynamics and patterning.

Since actin remodeling occurs on timescales from seconds to tens of seconds \cite{higgs_regulation_1999, fritzsche_analysis_2013, kelkar_mechanics_2020}, we reasoned that key features of BB movement may be missed at the coarse-imaging intervals (30–60 s) used previously. To resolve BB behavior at finer temporal resolution, we used a TIRF setup equipped with a high-speed camera to acquire 30,000-frame movies (10.5 minutes duration) at 21 ms per frame (\figref{figthreea}, Movie 7).

These fast acquisition movies were collected across the entire range of apical expansion - from \SI{50}{\micro\meter\squared} to \SI{300}{\micro\meter\squared} - and used to calculate mean squared displacement (MSD) plots for each BB trajectory within each cell. To extract morphogenetically relevant features while minimizing pseudoreplication and intrinsic variability, we averaged BB trajectories by ensemble at the single cell level, resulting in one averaged MSD plot per cell (\figref{figthreea}; \figref{figthreeb}, \sfigref{sfigfourb}).

Across all apical areas and cells, the averaged MSD curves displayed similar qualitative behavior: At short timescales, they exhibited low slope values ($\alpha \ll 1$), indicating constrained motion; at longer timescales, the slopes increased and shifted into a range of subdiffusive ($\alpha < 1$), diffusive ($\alpha \approx 1$), or superdiffusive ($\alpha > 1$) regimes (\figref{figthreeb}, \sfigref{sfigfourb}). In particular, while short-timescale $\alpha$ values were invariant across apical expansion (\figref{figthreeb}, \sfigref{sfigfourd}), long-timescale $\alpha$ values varied with apical area (\figref{figthreeb, figthreec}), despite diffusion strength remaining constant (\sfigref{sfigfoure}).

Focusing on the long-timescale $\alpha$ values, we observed that BBs in smaller apical areas exhibited behavior close to free diffusion ($\alpha \approx 1$) (\figref{figthreec}). As the apical area increased, $\alpha$ gradually decreased to $\sim{0.6}$, indicating a transition to more restricted subdiffusive behavior. Piecewise linear regression revealed a breakpoint around \SI{145}{\micro\meter\squared} (\figref{figthreec}, \sfigref{sfigfourf}), indicating a change in slope between early (<\SI{150}{\micro\meter\squared}) and late (>\SI{150}{\micro\meter\squared}) phases of apical expansion (\figref{figthreec}, \figref{figthreed}). This behavioral transition was also evident in the morphology of the BB trajectories: in smaller areas, the trajectories appeared more elongated, while in larger areas they became increasingly convoluted (\figref{figthreea}, Movie 7). Consistently, end-to-end distances decreased with increasing area (\sfigref{sfigfourg}), while contour lengths remained constant (\sfigref{sfigfourh}), leading to increased tortuosity with apical expansion (\sfigref{sfigfouri}). Together, these data support a progressive change in the dynamics of the BB motility during the late phase of apical expansion.

Next, we examined the transition between short- and long-timescale MSD behaviors (\figref{figthreeb}, \sfigref{sfigfourc}). At short timescales, $\alpha$ values were near zero ($\alpha \ll 1$), consistent with strong confinement or “trapping” of BBs. Over longer timescales, $\alpha$ values spanned a broader range, reflecting either weakening of this confinement ($\alpha < 1$) or complete trap release ($\alpha \approx 1$ or $\alpha > 1$). To verify that this initial trapping was not a measurement artifact, we tracked fluorescent beads of a size similar to BBs (\SI{0.5}{\micro\meter}) fixed on the coverslip. The MSD of immobilized bead controls, was approximately an order of magnitude lower than that of BBs, confirming that the observed BB trapping reflects genuine biological motion (\sfigref{sfigfourj}, \textit{see methods section 8.3.6 in SI}).

We then quantified the time it took for the BBs to transition from trapped to released state (\sfigref{sfigfourc}) and found that this transition time was positively correlated with the apical area (\figref{figthreeb}, \figref{figthreee}). Specifically, cells in the late phase of expansion showed significantly longer transition times (\figref{figthreee}, \sfigref{sfigfourk}, \figref{figthreef}), suggesting that BBs remain confined for longer periods in larger apical domains. These findings imply that the interaction of BB with its local environment becomes increasingly restrictive as the apical domain expands.

Taken together, our results demonstrate that both the behavior of the trajectory of BB (as quantified by $\alpha$) and confinement dynamics (as measured by the transition time) evolve in parallel with the apical expansion. The consistent shift in both parameters near \SI{\sim 140}{\micro\meter\squared} (\figref{figthreec}, \figref{figthreee}) suggests a shared underlying mechanism driving this transition. The reduction in end-to-end displacement and the increase in tortuosity imply that BBs have greater spatial freedom during early expansion. In contrast, their movement becomes increasingly constrained in later stages. Moreover, the lengthening of the trap-release transition times supports the notion that BBs are held within confined regions for longer durations as expansion progresses.

\subsection*{In silico modeling of BB dynamics}

To gain further insight into the mechanisms driving the observed transitions in the dynamics of BBs and to explore the potential role of the actin network in this process, we next sought to replicate the behavior of BBs using theoretical modeling. 

First, to mechanistically explain the observed two-regime behavior in BB dynamics in the previous section, we implemented a stochastic model in which BBs diffuse within harmonic traps that themselves undergo anomalous motion (\figref{figthreeg}). In this framework, each BB is confined by a harmonic potential with diffusivity $D_w$ and trap relaxation time $\tau_w$, while the trap center follows fractional Brownian motion with diffusion strength $D$ and anomalous exponent $\alpha$~\cite{metzner_simple_2007}. The model equations were integrated using an Euler-Maruyama algorithm.  The Davis-Harte process corresponding to fractional Brownian motion (fBM)~\cite{munoz2021objective} was used to obtain the increments for anomalous diffusion of the trap (\textit{see Supplementary Information (SI) for complete mathematical formulation}). This framework successfully reproduced the experimentally observed MSD profiles: initial caging at short timescales followed by anomalous diffusion at longer timescales (\figref{figthreeh}). The physical basis for this model aligns with our experimental observations that BBs are embedded within dynamically remodeling traps that locally confine them while itself undergoing stochastic reorganization, thereby permitting long-range anomalous exploration on extended timescales.

The coarse-grained experimental imaging ($30~{\rm s}$ intervals) operates at timescales significantly longer than the trap-release transition time ($\tau \approx 5-15~{\rm s}$), thereby capturing only the anomalous diffusion regime without resolving the initial caging dynamics. Consequently, our computational model for the full spatiotemporal evolution of BB organization during apical emergence focuses on anomalous diffusion coupled with apical surface expansion, while explicitly accounting for the initial caging effect.

The model simulates a dynamically expanding circular apical domain with area prescribed by experimentally measured growth kinetics (\textit{see SI}). BBs are recruited at a constant rate matching experimental observations (Movie 8), with their initial positions sampled from spatially biased probability distributions that favor peripheral insertion, as observed experimentally (\textit{see SI}). Following insertion, each BB undergoes anomalous diffusion characterized by an exponent $\alpha$ randomly sampled from a Gaussian distribution ($\mu = 0.85$, $\sigma = 0.2$). Trajectories were generated using a piecewise fractional Brownian motion approximation based on the Davis-Harte algorithm: the full simulation interval of $150~{\rm min}$ was divided into three consecutive segments, each generated with a fixed Hurst exponent $H=\alpha/2$, with the exponent reduced by a factor of $3/4$ between successive segments and lower bounded by $\alpha = 0.4$. Segment-wise diffusion normalization was imposed using the same reference timescale $t_{\rm ref}=30~{\rm s}$.

Critical to reproducing the observed BB spacing was the incorporation of mechanical interactions mediated by traps surrounding each BB as was implicated by the fine-grained model for BB. We model these traps as soft repulsive zones of thickness $r_s$ and stiffness $k_s$ surrounding a hard BB core of radius $r_b$, with outer pocket radius $r_o=r_b+r_s$ (\textit{see SI}). Let $d$ denote the center-to-center distance between two BBs. Repulsion begins when $d<2r_o$: initially through elastic compression of the soft shells, and subsequently through a stiffer hard-core repulsion once the shells are fully compressed ($d<2r_b$). This two-stage interaction mechanism naturally prevents BB clustering beyond experimentally observed minimum spacings ($\sim{} 1~{\rm \mu m}$) while permitting dynamic spatial reorganization. BBs crossing the apical boundary were projected back into the apical domain at a small radial offset from the boundary.

Quantitative comparison of model outputs with experimental data confirmed successful recapitulation of key organizational metrics. The normalized variance of Voronoi tessellation areas decreased progressively over time (\figref{figthreei}), mirroring the experimental transition from scattered to organized BB distributions as density-dependent mechanical interactions drove spatial regularization. The average nearest-neighbor distance exhibited similar temporal evolution, decreasing steadily before plateauing at values consistent with experimental observations: $\sim1~{\rm \mu m}$ (\figref{figthreej}). These results demonstrate that the combination of peripherally biased BB insertion, anomalous diffusion dynamics, and soft-trap mediated mechanical interactions is sufficient to account for the emergent spatial patterning of BBs during apical expansion.

\subsection*{Progressive formation of an apical actin meshwork facilitates BB redistribution}

The theoretical model predicts the presence of soft traps that confine individual BBs which are necessary to reproduce the experimentally observed spacing. This raises a central question: do such trapping structures exist \textit{in vivo} and what molecular components give rise to them? Because BBs are embedded within a dense apical actin network, actin represents a strong candidate to provide the predicted trapping forces. To test this, we examined whether actin organization contributes to BB spatial patterning by acquiring high-resolution images of the apical surfaces of MCCs in embryos injected with Chibby-GFP, followed by fixation and actin (phalloidin) staining. These images revealed pronounced differences in actin organization as a function of apical domain size (\figref{figfoura}). In smaller apical domains, actin appeared disorganized and lacked a defined structure. In contrast, larger apical domains exhibited progressively more structured actin architectures, culminating in the emergence of a prominent meshwork organization (\figref{figfoura}).

During the early stages, when this actin meshwork was absent, BBs were arranged in clustered configurations (\figref{figfourb}). As the apical domain expanded, actin progressively assembled into a network composed of discrete pocket-like structures that surrounded individual BBs. The formation of interconnected actin pockets into a continuous meshwork across the apical surface correlated with the transition of BBs from a clustered to a more uniform spatial distribution (\figref{figfourb}).

To further characterize the dynamics of actin meshwork formation and its role in BB patterning, we quantified actin accumulation around the BBs. We compared the mean actin intensity at the BB positions (\sfigref{sfigfivea}) with that in the immediate surrounding regions (\sfigref{sfigfiveb}). Cumulative Distribution Function (CDF) plots of actin intensity for these two regions revealed overlapping distributions in small apical domains, which progressively diverged in larger domains (\figref{figfourc}). This indicated an increase in actin accumulation around the BBs relative to their immediate position. To further quantify these differences, we computed the surrounding-to-BB intensity ratios at the ${25}^{\text{th}}$ (\sfigref{sfigfived}), ${50}^{\text{th}}$ (\figref{figfourd}), and ${75}^{\text{th}}$ (\sfigref{sfigfivee}) percentiles of the CDF. Across all percentiles, the ratios increased with apical area, confirming enhanced actin enrichment in BB-surrounding regions in larger domains.

The CDF analysis compares how the full distribution of actin intensities differs between BB positions and surrounding regions, whereas frame averaging summarizes each frame by a single mean intensity value for each region. Thus, the CDF captures distribution-level shifts, while frame averaging provides a per-cell measure of average enrichment and reduces pseudoreplication. Consistent with the CDF based results, the ratio of frame-averaged actin intensity (surrounding / BB position) increased with apical area, further supporting enhanced actin accumulation around BBs in larger domains (\sfigref{sfigfivef}).

Notably, all CDF based measurements (\figref{figfourd, sfigfivec, sfigfived, sfigfivee}), as well as the frame-averaged intensity ratio (\sfigref{sfigfivef}), exhibited a common breakpoint around 170-\SI{185}{\micro\meter\squared}, beyond which actin enrichment increased markedly. Statistical analysis confirmed significant differences in these metrics before and after this threshold (\figref{figfoure, sfigfiveg, sfigfiveh, sfigfivei}).

Importantly, the actin-related breakpoint (\SI{\sim 175}{\micro\meter\squared}) occurs close in developmental time to the transition in BB dynamics previously identified at \SI{\sim 140}{\micro\meter\squared} (\figref{figthreec, figthreee}). Considering the continuous expansion of the apical surface over \SI{\sim 2.5}{\hour}, the proximity of these values likely represent the shared underlying developmental transition observed through different quantitative readouts, suggesting a strong mechanistic link between the two processes. In small apical domains lacking an organized actin network (\figref{figfoura, figfourb}), BBs exhibited elongated trajectories and short transition times (\figref{figthreea, figthreee}). As the actin meshwork assembly progressed with increasing apical area, the BB trajectories became more convoluted and transition times increased, reflecting greater confinement (\figref{figthreea, figthreec, figthreee}). Thus, the formation of the actin meshwork appears to progressively coordinate the dynamics of the distribution of BBs and ultimately guides the organization of BBs into a defined pocket-like architecture as indicated by our theoretical model. 

\subsection*{Disruption of the apical actin cross-linking impairs BB organization}

Actin meshwork formation results from the cross-linking of actin filaments by actin-binding proteins \cite{kadzik_f-actin_2020, lieleg_structure_2010, stricker_mechanics_2010}. To examine how this cross-linked actin network contributes to the distribution and patterning of BB, we knockdown (KD) $\alpha$-actinin-1, a key actin cross-linker \cite{tseng_how_2005, courson_actin_2010, ehrlicher_alpha-actinin_2015}, using a translation-blocking morpholino oligonucleotide targeting $\alpha$-actinin-1 mRNA. Embryos were injected at the 2–4 cell stage with 15 ng of morpholino, and apical expansion and BB behavior were imaged over 2.5 hours at 30–60~s intervals (\figref{figfivea, sfigsixa}, Movie 9). Despite $\alpha$-actinin-1 morpholino treatment, BBs were still able to migrate from the basal side to the apical surface (\figref{figfiveb}, Movie 10), and the apical domain expanded (\sfigref{sfigsixa}, Movie 9), with BBs docking to the apical surface during this process (\sfigref{sfigsixa}, Movie 9). However, the apical domain only reached an area of \SI{\sim 120}{\micro\meter\squared} at the end of the imaging (\figref{figfivec}), representing a more than twofold reduction in the expansion rate compared to the wild-type (WT) (\figref{figfived}). Furthermore, BB migration and docking were incomplete (\figref{figfiveb}), with approximately half as many BBs reaching the apical domain compared to WT (\figref{figfivee, figfivef}). These results underscore the importance of actin cross-linking for efficient apical expansion and BB docking.

Interestingly, in $\alpha$-actinin-1 knockdowns (KDs), apical expansion halted at a maximum of \SI{\sim 140}{\micro\meter\squared} (\figref{figfivec}), which corresponds closely to the area at which actin meshwork formation initiates in WT (\figref{figfour}), reinforcing the role of $\alpha$-actinin-1 in the assembly of actin networks. Despite these defects, we observed partial preservation of WT characteristics: new BBs still tended to appear near the periphery (\sfigref{sfigsixb}), and BB trajectories exhibited no directional bias, showing equal probability of movement toward or away from the expanding edge (\figref{figfivea, sfigsixc, sfigsixd}, Movie 11). These observations suggest that, while both the apical expansion and the BB docking were slowed, the coupling between these two processes was maintained. In fact, the density of BB in the apical domain remained comparable to that of WT (\sfigref{sfigsixe, sfigsixf}), indicating that expansion and BB insertion were proportionally delayed.

To determine whether $\alpha$-actinin-1 KD induced delay simply postponed BB organization or fundamentally disrupted it, we examined the apical domains of MCC at later developmental stages (Nieuwkoop and Faber (NF) stage\cite{nieuwkoop1994normal} 26). Although some MCCs eventually reached WT apical sizes, the organization of BBs remained perturbed (\sfigref{sfigsixg}). These defects were confirmed using a splice-blocking morpholino, which yielded similar phenotypes (\sfigref{sfigsixg, sfigsixh}). Together, these findings indicate that although actin cross-linking is not strictly required for apical expansion per se, it is essential for proper BB organization and spatial patterning.

Next, we assessed how loss of actin cross-linking affects the spatial distribution of BBs. The variance of the Voronoi tessellation areas, used here as a proxy of spatial regularity, was increased and more variable in the $\alpha$-actinin-1 KDs throughout expansion (\figref{figfiveg, figfiveh, figfivei}), resulting in a visibly incomplete distribution of BBs (\figref{figfiveg, figfivej}, Movie 12). To quantify these effects, we analyzed the minimum and pairwise distance metrics in three populations of BBs: (i) all BBs at each frame (\sfigref{sfigsevena}–\sfigref{sfigsevend}); (ii) newly inserted BBs relative to those from the previous frame (\sfigref{sfigsevene}–\sfigref{sfigsevenh}); and (iii) new BBs relative to pre-existing BBs within the same frame (\sfigref{sfigseveni}–\sfigref{sfigsevenl}). In all comparisons, the minimum and pairwise distances were consistently lower in $\alpha$-actinin-1 KDs than in WT (\sfigref{sfigsevena}–\sfigref{sfigsevenl}), reflecting the visually apparent clumping of BB (\figref{figfivea, sfigsixa}). These results confirm that actin cross-linking is required to maintain the appropriate spacing and prevent aggregation of BBs during their distribution.

In WT embryos, the apical area serves as a reliable proxy for morphogenetic progression. In contrast, in $\alpha$-actinin-1 KDs, this relationship is disrupted: developmental time (as measured by area) no longer maps linearly onto real time. To evaluate BB organization independently of experimental time, we reanalyzed spatial metrics relative to morphogenetic time, specifically, apical domain area (\sfigref{sfigeighta}–\sfigref{sfigeightn}). Up to the maximum apical area achieved by the $\alpha$-actinin-1 KDs (\SI{140}{\micro\meter\squared}), the variance of the tessellation (\sfigref{sfigeighta, sfigeightb}) and the minimum and pairwise BB distances (\sfigref{sfigeightc, sfigeighte, sfigeightg, sfigeighti, sfigeightk, sfigeightm}) showed qualitatively similar trends in both WT and $\alpha$-actinin-1 KD conditions. However, in almost all comparisons, the distributions shifted significantly (\figref{sfigeightd, sfigeightf, sfigeighth, sfigeightj, sfigeightl, sfigeightn}), indicating subtle but consistent increases in spatial heterogeneity.

In summary, these findings reveal that actin cross-linking mediated by $\alpha$-actinin-1 is not only required for timely apical expansion and BB recruitment, but is also critical to establishing spatial precision in BB organization. Although some aspects of BB positioning may emerge passively as the apical domain grows, the meshwork formed by cross-linked actin filaments is necessary for refining and stabilizing the final pattern.

\subsection*{$\alpha$-actinin-1 perturbation impairs transition in BB dynamics and actin meshwork formation}

The coarse-timed data (30-60 s interval) from the $\alpha$-actinin-1 KDs provided insights into the spatial heterogeneity of the BB distribution, arising from disrupted actin cross-linking. These observations were collected over 2.5 hours of experimental time, covering apical areas up to \SI{140}{\micro\meter\squared}, which corresponds to the maximum domain size reached in $\alpha$-actinin-1 KDs during this window and allows direct comparison with the WT condition. However, high-resolution end point imaging revealed that, despite delays, $\alpha$-actinin-1 KDs ultimately reached sizes of the apical domain comparable to those of WT at later developmental stages (\sfigref{sfigsixg}). Thus, the coarse-timed experiments captured the BB distribution and trajectory dynamics only during the early phase of apical expansion in the $\alpha$-actinin-1 KDs.

To overcome this limitation and examine the behavior of the BBs throughout the full range of apical expansion, including the late phase, we employed fine-timed imaging (21 ms interval), which allows sampling of a wide range of apical areas by capturing multiple MCCs over a 10.5 minute window (\figref{figthreea}). We hypothesized that this approach would reveal additional signatures of disrupted dynamics in $\alpha$-actinin-1 KDs (\figref{figsixa}, Movie 13).

We first analyzed the averaged MSDs of BBs across apical areas ranging from 50 to \SI{300}{\micro\meter\squared} (\figref{figsixb}). In $\alpha$-actinin-1 KDs, MSD plots showed constrained behavior ($\alpha \ll 1$) at short timescales, followed by a transition to primarily diffusive behavior ($\alpha \approx 1$) at longer timescales (\figref{figsixb}). This contrasts with WT trajectories, where BBs exhibit anomalous diffusion on long timescales (\figref{figthreeb}), reflecting evolving BB motions due to progressive actin meshwork formation. In $\alpha$-actinin-1 KDs, both short-timescale $\alpha$ values and diffusion strength remained constant across the range of apical areas (\sfigref{sfigninea, sfignineb}), and the long-timescale $\alpha$ values similarly showed no area-dependent trend (\figref{figsixc, sfigninee}), unlike in the WT, where these values decreased progressively with expansion (\figref{figthreec}). The absence of any breakpoint or shift in dynamics indicates that BB trajectories in $\alpha$-actinin-1 KDs are not remodeled during expansion. This is supported by the relatively constant end-to-end distances (\sfigref{sfigninec}) and tortuosity values (\sfigref{sfignined}), consistent with the elongated and unconstrained trajectories observed in $\alpha$-actinin-1 KDs (\figref{figsixa}).

We next focused on the transition in the $\alpha$ values between short and long timescales, which in WT reflects the shift from trapped to freely diffusing states. In $\alpha$-actinin-1 KDs, the BB trajectories followed a similar trend: low $\alpha$ values at short timescales and diffusive values at longer timescales (\figref{figsixb}). This transition also became more pronounced with increasing apical area (\figref{figsixd, sfigninef}), as observed in WT (\figref{figthreee}). However, the transition in $\alpha$-actinin-1 KDs was markedly delayed. In WT, transition times increased to \SI{\sim 12.5}{\second} at \SI{\sim 150}{\micro\meter\squared} (\figref{figthreee}), while in $\alpha$-actinin-1 KDs, the same transition time was only reached at \SI{\sim 300}{\micro\meter\squared} (\figref{figsixd}). This indicates that BBs in $\alpha$-actinin-1 KDs are released from confinement more rapidly and remain in a more dynamic state even in later stages of expansion, reflecting a failure of spatial refinement that normally occurs through actin cross-linking.

Altogether, these results show that in $\alpha$-actinin-1 KDs, the BB trajectories (\figref{figsixc}) and the trap-release dynamics (\figref{figsixd}) remain largely unchanged in the apical area range. This suggests that $\alpha$-actinin-1 mediated actin cross-linking is essential to regulate BB motion during apical expansion. In WT, this cross-linking begins around \SI{\sim 170}{\micro\meter\squared} (\figref{figfour}, \sfigref{sfigfive}), falling within the same apical area range as the transition in BB behavior. In $\alpha$-actinin-1 KDs, due to the lack of cross-linking, BBs retain diffusive trajectories and exhibit only brief trapping events even in large apical domains (\figref{figsixc, figsixd}). The elongated, unconstrained trajectories observed in the late stages (\figref{figsixa}, Movie 13) further support this. The delayed transition times in $\alpha$-actinin-1 KDs reflect an inability to stably confine BBs, thereby impairing their progressive spatial organization. These findings highlight the importance of the gradual buildup of actin cross-linking in refining BB dynamics and achieving proper patterning. Importantly, while the number of BBs is known to scale with the apical area \cite{nanjundappa2019regulation, kulkarni_mechanical_2021}, our $\alpha$-actinin-1 KD experiments reveal that the BB docking also critically depends on the actin architecture. Thus actin cross-linking introduces an additional layer of regulation in the relationship between the apical area and the BB number.

To confirm the impact of $\alpha$-actinin-1 KD treatment on actin organization, we performed high-resolution imaging of apical domains injected with Chibby-GFP and stained with phalloidin to visualize the actin network (\sfigref{sfignineg}). Across the full range of apical areas, $\alpha$-actinin-1 KDs displayed disorganized actin, lacking the meshwork architecture and characteristic actin pockets observed in WT cells, with BBs correspondingly arranged in clumped configurations (\figref{figsixe}). Quantitative analysis supported these observations: CDFs of actin intensity at BB positions and in their surrounding regions failed to diverge in large apical domains (\figref{figsixf}), in contrast to the behavior observed in WT cells, indicating an absence of actin enrichment around BBs. Consistently, the ratio of actin intensities (surrounding / BB position) at the ${50}^{\text{th}}$ percentile (\figref{figsixg, figsixh}) did not increase with apical area in $\alpha$-actinin-1 KDs.

Furthermore, frame-averaged actin intensities at BB positions and in the surrounding regions were consistently lower in $\alpha$-actinin-1 KDs throughout apical expansion compared to WT (\sfigref{sfignineh, sfigninek, sfigninei, sfigninel}), indicating a global reduction in actin accumulation in these regions. Accordingly, the ratio of frame-averaged actin intensity (surrounding / BB position) showed no increase in $\alpha$-actinin-1 KDs (\sfigref{sfigninej, sfigninem}), whereas in WT cells this ratio began to rise around \SI{178}{\micro\meter\squared} (\sfigref{sfigfivef, sfigninej}). Together, these results confirm the absence of actin meshwork formation in $\alpha$-actinin-1 KDs and establish $\alpha$-actinin-1 as a critical factor in actin meshwork assembly in MCCs.

Finally, we note that the absence of a breakpoint in actin intensity metrics (\figref{figsixg, sfigninej}) mirrors the lack of transition in the BB dynamics (\figref{figsixc, figsixd}). Together, these data demonstrate that in the absence of actin cross-linking, the actin meshwork fails to develop, preventing BB confinement and resulting in clumped, unpatterned distributions. These findings establish $\alpha$-actinin-1 dependent actin cross-linking as a key regulator of BB dynamics and spatial patterning in MCCs.

\subsection*{Heterogeneous distribution of the apical actin network leads to BB clumping}

The $\alpha$-actinin-1 KDs showed clear changes in BB dynamics and patterning, underscoring the importance of actin cross-linking in this process. To further investigate how BB disorganization manifests under these conditions, we quantified the spatial heterogeneity of BB distributions. Coarse-timed imaging was used to capture the chronological evolution of BB disorganization during apical expansion. Visually, two distinct clustering patterns were observed: (i) BBs aggregating into a single large cluster localized to one end of the apical domain (global cluster), or (ii) BBs forming multiple smaller clusters scattered across the surface (local cluster) (\figref{figsevena}).

To quantify heterogeneity in the BB distribution, we employed two clustering coefficients: a global clustering coefficient that incorporates both angular and radial variation and a local clustering coefficient based on the ratio of nearest-neighbor distance to the mean inter-BB distance (\textit{see methods section 7.3.12 in SI}). These metrics provided continuous and unbiased quantification of BB clustering throughout the apical domain, taking into account orientation and position. In WT cells, the global clustering coefficient (\figref{figsevenb}) and the local clustering coefficient (\figref{figsevend}) followed the expected pattern, consistent with other spatial organization metrics (\figref{figonei, figonej}): initially elevated, reflecting early-stage disorganization, followed by a gradual decrease as the BBs became more uniformly distributed.

Interestingly, $\alpha$-actinin-1 KDs also showed a decreasing trend in both coefficients (\figref{figsevenb, figsevend}), suggesting a trend towards spatial organization. However, the variance of these metrics was greater and their absolute values were significantly elevated compared to WT (\figref{figsevenc, figsevene}), indicating persistent disorganization.

To investigate the mechanism underlying the observed BB clustering, we turned to our theoretical model. For simulations replicating $\alpha$-actinin-1 KD, three key modifications were implemented to capture experimental perturbations: (i) 
BBs exhibited a strong spatial bias, with a higher probability of appearance within one region of the apical area compared to the rest of the domain (\textit{see SI}); (ii) diffusion strength was reduced by a factor of $5$ in non-preferred quadrants, creating localized confinement that restricts redistribution; and (iii) actin pocket thickness was reduced, weakening repulsive interactions and permitting closer BB approach (Movie 14). The temporal analysis of simulated clustering metrics revealed dynamics consistent with experimental observations (\sfigref{sfigtena, sfigtenb}). Both global and local clustering coefficients were initially elevated for wild-type (WT) and knockdown (KD) simulations, reflecting early-stage spatial disorganization when BB numbers were low. As time progressed, the global clustering coefficient decreased toward zero in WT simulations as BBs became scattered across the apical surface through stochastic peripheral appearance and anomalous diffusion. In the KD simulations, however, the global clustering coefficient remained saturated at relatively higher values even at later times, as BBs remained confined within the preferred quadrant. Similarly, local clustering decreased gradually in WT simulations but remained elevated in KD simulations due to reduced actin pocket sizes that permitted closer BB approach.

Notably, inducing clustering \textit{in silico} required simultaneous tuning of both parameters, i.e., BB repulsion distance and spatial bias in BB appearance; manipulating either alone was insufficient to drive clustering. This suggests that BB clustering in the experimental $\alpha$-actinin-1 KD arises from a combination of biased BB emergence and a lack of repulsive interactions, likely due to disrupted actin cross-linking, which normally enforces BB spacing. However, the origin of the preferential appearance could not be readily explained. Given that BB docking requires actin, we hypothesized that apical actin distribution may be involved in spatially biased BB docking.

Although clustering coefficients effectively capture heterogeneity in discrete BB distributions, they are not optimal for comparing spatially continuous signals such as actin intensity, which serves as a proxy for the underlying actin distribution. To address this, we developed a quadrant-based clumping factor that quantifies spatial heterogeneity for both discrete (BB positions) and continuous (actin intensity) distributions. The apical domain is divided into four equal angular sectors, and for each time frame, the number of BBs (or mean actin intensity) in each quadrant is determined. The clumping factor is then calculated as the normalized variance of these quadrant populations (\textit{see methods section 7.3.13 in SI}). If BBs are uniformly distributed across all quadrants, the clumping factor approaches 1; if all BBs are confined within a single quadrant, it approaches 4. This metric enables direct comparison of spatial heterogeneity between BB density and actin intensity distributions within a unified analytical framework (\sfigref{sfigtenc, sfigtend}, Movie 15).

In the experimental data, clumping factors for both BB density (\figref{sfigtene}, Movie 16) and actin intensity (\figref{figsevenf}, Movie 16) followed the same trend observed in clustering coefficients: initially elevated, then declining towards 1 over time (\figref{figsevenh,sfigteng}). This trend indicates a transition from spatial heterogeneity to homogeneity during normal apical expansion. In contrast, although the $\alpha$-actinin-1 KDs showed a similar downward trend (\figref{figsevenh, sfigteng}), their clumping factors remained significantly higher than those of WT, for both BBs (\figref{sfigtenf, sfigtenh}, Movie 16) and actin (\figref{figseveng, figseveni}, Movie 16), indicating persistent heterogeneity in both components. Consistent with the experimental observations, BB simulations incorporating preferential appearance and reduced repulsion exhibited higher clumping factors compared to simulations with regularly spaced BBs (\sfigref{sfigteni}).

Together, these analyses demonstrate that the $\alpha$-actinin-1 KDs consistently exhibit elevated spatial heterogeneity in the BB distribution, primarily in the form of BB clustering. These clusters tend to colocalize with regions of increased actin intensity, suggesting that BB disorganization reflects and is likely driven by the underlying heterogeneity of actin. These findings reinforce the critical role of actin cross-linking in the regulation of both BB dynamics and spatial patterning during apical expansion.

\section*{Discussion}

The actin cytoskeleton is a fundamental component and master regulator of cellular architecture and related functions \cite{rohn_actin_2010, salbreux_actin_2012, kelkar_mechanics_2020}. It can rapidly reorganize by assembling \cite{skau_specification_2015}, disassembling \cite{gautreau_nucleation_2022}, and forming higher-order structures such as filamentous networks \cite{tseng_how_2005, lieleg_structure_2010} and bundles \cite{furukawa_structure_1997}. These structures propagate forces across scales \cite{lappalainen_biochemical_2022, heisenberg_forces_2013}, facilitating processes ranging from intracellular transport \cite{khaitlina_intracellular_2014} and organelle positioning in the crowded cytoplasmic environment \cite{bornens_organelle_2008, kroll_principles_2024}, to cell migration \cite{schaks_actin_2019} and tissue morphogenesis \cite{munjal_actomyosin_2014}. For example, in plant cells, the actin cytoskeleton efficiently transports Golgi bodies, and their movement can be predicted from the global organization of the actin network \cite{breuer_system-wide_2017}.

Our quantitative analysis of the MCC apical domains in \textit{X. laevis} embryos reveals that the dynamic organization of the apical actin meshwork plays a decisive role in the emergent mobility and patterning of BBs. Importantly, we capture changes in meshwork formation that are acutely reflected in BB dynamics and patterning. This reinforces the idea that the dynamics of the BBs are tailored to their final patterning, which is rigorously orchestrated by actin remodeling. These findings align with the known role of actin dynamics in intracellular organization and significantly extend this established paradigm.

Previous studies have shown a correlative relationship between the size of the apical area and the BB count \cite{nanjundappa2019regulation, kulkarni_mechanical_2021}. Here, we demonstrate that this relationship is dynamic, and the BB count increases linearly with area. Interestingly, this correlation does not break down in the $\alpha$-actinin-1 KDs. Although the rate of expansion and BB docking is slower compared to that of the WT, the BB density is still preserved. This argues for a mechanism involving apical actin, which balances the expansion of the apical domain with the number of BBs docked. We propose that cross-linking-based meshwork formation maintains this balance, patterning the BBs while simultaneously providing mechanical stability to the expanding apical domain. Apical expansion then creates a new space that allows the recruitment of BBs. Interestingly, while cross-linking coordinates apical expansion and BB patterning, it also keeps these processes largely independent. If the processes were interdependent, then the apical expansion should already passively distribute the BBs. In contrast, the movement of BB trajectories in the direction opposing expansion argues for a decoupling mechanism between apical expansion and BB distribution. The correlation between actin remodeling and the shift in BB dynamics from diffusive (or even superdiffusive) to subdiffusive behavior implies that actin is the decoupling element. While actin polymerization pushes the boundary of the apical domain, the cross-linked actin meshwork presumably stabilizes the expanded area and simultaneously patterns the BBs. Thus, the link between apical expansion and BB patterning appears to be mediated through actin remodeling, although the direction of causality remains unclear.

The apical actin appears to take control of the BBs as soon as they are incorporated into the apical domain. In WT conditions, new BBs emerge near the periphery, close to the newly extended area, but then start migrating in a tortuous fashion until they settle among their pre-existing neighbors. This in-plane trafficking reflects the local activity of actin, which leads to mechanical push-pull within the apical cortex. In contrast, in $\alpha$-actinin-1 KDs with disrupted cross-linking, although new BBs appear peripherally, they wander randomly and ultimately accumulate into clusters rather than forming a structured arrangement. This is further confirmed by our minimum and pairwise distance metrics and clustering analysis, which show that actin integrity in the WT maintains uniform spacing. $\alpha$-actinin-1 KDs, on the other hand, exhibit higher clustering coefficients and greater heterogeneity in distance metrics. Although $\alpha$-actinin-1 KDs consistently show higher levels of heterogeneity compared to WT, a baseline level of heterogeneity is always present in the WT. This likely reflects intrinsic biological noise in the system. In this context, the actin meshwork can be viewed as a regulatory mechanism that refines the BB pattern against this inherent background noise. Disruption of actin cross-linking in $\alpha$-actinin-1 KDs compromises this regulatory mechanism, allowing intrinsic noise to become amplified and leading to disorganized spatial patterning of BB. Thus, actin cross-linking likely facilitates the coordinated movement of BBs into their final patterned positions. This is consistent with the current understanding of how actin networks and cross-linkers give rise to dynamic patterns across scales in diverse biological contexts, as well as their functional relevance \cite{bois_pattern_2011, prost_active_2015, heisenberg_forces_2013}.

Previous literature has suggested that BBs may bring actin polymerization factors, such as formins and Arp2/3, to the apical domain, thus promoting cross-linking \cite{yasunaga_polarity_2015, copeland_actin-based_2020}. However, blocking BB formation and apical docking does not impair the initial phase of apical expansion \cite{kulkarni_mechanical_2021}. These studies focused on endpoint effects, not time-evolution dynamics. Therefore, BBs might be influencing the apical expansion by helping the apical domain reach the critical amounts of actin-related proteins within a defined time frame for the formation of the meshwork. The results of the $\alpha$-actinin-1 KD in our study clearly show that the lack of actin cross-linking prevents the BB from docking to the apical domain, with no apparent effect on BB synthesis or migration. Specifically, the arrest of the apical expansion at \SI{\sim 120}{\micro\meter\squared} points to a mechanism other than cross-linking that may drive the expansion up to this range of the area. This is likely actin polymerization, which is then taken over by branching and cross-linking. If identified, this switch in branching and cross-linking could be considered a key developmental time point in MCC apical expansion. Thus, the results of this study further solidify the idea that the actin network provides both mechanical support and the dynamic rearrangements necessary for precise dynamics and organization of subcellular organelles, particularly in specialized cells such as MCCs that undergo extensive morphogenetic changes.

Perturbations in actin cross-linking significantly impair the spatiotemporal patterning of BBs, leading to observable defects in the expansion of the apical area, the count of BBs, the variance of the tessellation areas, and the minimum and pairwise distance metrics of different BB populations. These widespread disruptions highlight that actin cross-linking is not a passive process but an active organizer that enables the apical domain to facilitate BB arrangement. The fact that BB density, peripheral appearance, and trajectory direction remain similar to WT, while other parameters are disrupted, suggests a selective impairment of BB organization rather than a complete abolition of BB biogenesis or basal-to-apical migration. This implies that while initial BB appearance and general directional tendencies are preserved, the fine-tuning of their spatial relationships and overall patterning is highly dependent on an intact and cross-linked actin meshwork.

MSD analysis is a powerful tool for characterizing the nature of particle motion in complex biological environments \cite{wirtz_particle-tracking_2009, wang_principles_2021}. When a particle exhibits standard diffusion with Brownian motion, MSD scales linearly with time and the power-law exponent ($\alpha$) of the MSD-time relationship equals 1. However, deviations from this behavior indicate anomalous diffusion, including subdiffusion ($\alpha < 1$), superdiffusion ($\alpha > 1$), and ballistic motion ($\alpha \approx 2$) \cite{wirtz_particle-tracking_2009, wang_principles_2021}. Our findings from MSD analysis provide critical insight into BB dynamics and their strong correlation with actin meshwork formation. The transition of cortical actin into a cross-linked meshwork occurs concurrently with a shift in the BB dynamics, from freely diffusing or superdiffusive motion in small apical domains to confined, subdiffusive motion in expanded domains. This suggests that the actin meshwork lattices of larger domains effectively cage the BBs, restricting long-range movement. In contrast, when actin cross-linking is disrupted in $\alpha$-actinin-1 KDs, the meshwork fails to form, and BBs remain relatively unconstrained and diffusive throughout. This result is in line with studies that demonstrate that WDR5 \cite{kulkarni_wdr5_2018} and Filamin-A \cite{chevalier_mir-34449_2015} contribute to an F-actin lattice essential for uniform BB spacing. We show that, in the absence of this caging effect, BBs exhibit clumping and enhanced mobility. Further, our data directly link actin meshwork formation to BB single-particle dynamics: the meshwork imposes subdiffusive constraints, while in its absence, BBs behave in a nearly Brownian fashion. Thus, while coarse-timed experiments highlighted the role of cross-linking in the organization of BBs, it is the fine-timed data that reveal the underlying mechanism. These findings suggest that actin remodeling progressively primes the BB dynamics to achieve a final ordered state.

This conclusion is supported by a large body of biophysical literature on diffusion in complex media. In polymer networks, particles often exhibit anomalous behavior due to physical obstacles and active interactions \cite{golding_physical_2006, wirtz_particle-tracking_2009, bruno_transition_2009, etoc_non-specific_2018, woringer_protein_2018}. Depending on the dynamic state of the environment and the nature of particle–environment interactions, particles can exhibit complex combinations of anomalous and diffusive behaviors, separated by a characteristic timescale \cite{maelfeyt_anomalous_2019, etoc_non-specific_2018, luo_simulation_2023}. These types of behavior, including transient caging, have been observed in various systems, including nonbiological particles in colloidal fluids and glasses \cite{weeks_three-dimensional_2000}, granular materials under cyclic shear \cite{marty_subdiffusion_2005} or vertical shaking \cite{scalliet_cages_2015}, as well as biological contexts such as lipid bilayers \cite{golan_resolving_2017, molina-garcia_crossover_2018} and F-actin networks \cite{wong_anomalous_2004}.

Consistent with these findings, our MSD analysis of WT MCCs shows a transition from constrained to diffusive, or even superdiffusive, motion in BB trajectories. These transitions occur over a characteristic timescale (transition time), which increases from approximately 5 s to 15 s as the apical area expands. This observed transition time aligns with the timescale of actin remodeling reported in the literature \cite{higgs_regulation_1999, fritzsche_analysis_2013, kelkar_mechanics_2020}. The increase in transition time correlates with the formation of the meshwork, highlighting the role of the actin network in tuning the behavior of the BB trajectory. In contrast, in $\alpha$-actinin-1 KDs, this transition is severely delayed, suggesting that BBs behave as if in a more fluid, unconstrained environment. This observation agrees with studies in reconstituted cytoskeletal systems, where actin-microtubule cross-linked networks lead to subdiffusive motion of embedded tracer particles \cite{anderson2021subtle}. Thus, in WT, the actin meshwork likely cages BBs during late apical expansion, reducing their mobility and resulting in subdiffusive motion. In its absence, as in $\alpha$-actinin-1 KDs, BBs diffuse freely. The delay in the transition time in the $\alpha$-actinin-1 KDs indicates that, without reinforcement via cross-linking, the actin network never matures into a phase capable of constraining the BB motion.

While transitions between anomalous and diffusive behavior have been reported, to our knowledge, this is the first demonstration of a transition from one anomalous regime to a combination of anomalous and diffusive behaviors, spanning subdiffusive and superdiffusive motion, in the context of development. This suggests that the physics governing subcellular dynamics in developmental processes, such as MCC apical expansion in \textit{X. laevis}, may be more complex than currently appreciated. It also emphasizes the need for careful consideration in choosing appropriate models for intracellular trajectory behavior \cite{munoz2021objective}.

Recent advances in active gel theory provide a compelling framework for understanding the role of cytoskeletal remodeling in generating subcellular patterns, such as the BB organization described in this work \cite{lieleg_structure_2010, prost_active_2015, gross_how_2017, salbreux_actin_2012, banerjee_actin_2020}. In active gels such as the actin cytoskeleton, cross-linkers transiently bind and unbind to polymerizing actin filaments, tuning network connectivity and modulating viscoelastic properties \cite{wachsstock_affinity_1993, wachsstock_crosslinker_1994, falzone_actin_2013, foffano_dynamics_2016, lieleg_structure_2010, grooman_morphology_2012, ahmed_dynamic_2015, ehrlicher_alpha-actinin_2015, bueno_generalized_2022}. Our findings are consistent with this framework: the $\alpha$-actinin-1 based meshwork emerges during apical expansion and constrains BB mobility. This suggests that the meshwork acts as a spatially extended mechanical scaffold, regulating the positioning of the BB on the apical surface.

The concurrent shift in actin meshwork formation and BB dynamics, from fast, diffusive behavior to slower, subdiffusive motion, suggests that the surrounding cytoskeletal environment becomes increasingly confining. Notably, both shifts occur around similar expansion range: \SI{140}{\micro\meter\squared} for BB dynamics and \SI{170}{\micro\meter\squared} for meshwork formation. These observations suggest a possible percolation transition \cite{wachsstock_affinity_1993, lieleg_structure_2010, grooman_morphology_2012, ahmed_dynamic_2015, ehrlicher_alpha-actinin_2015, bueno_generalized_2022} around this area range, where the meshwork becomes sufficiently connected across the apical surface to behave as a constraining solid-like medium over timescales relevant for BB motion. This interpretation is consistent with previous work \cite{herawati_multiciliated_2016} that describes BBs as particles embedded in a viscoelastic medium. Our results further specify that this viscoelastic nature arises from actin polymerization, which drives activity, and $\alpha$-actinin-1 cross-linking, which provides mechanical integrity.

Thus, the apical actin network in MCCs is not only a passive stabilizer of apical domain expansion but also functions as a viscoelastic gel whose mechanical properties, regulated by cross-linkers such as $\alpha$-actinin-1, drive the emergent spatial organization of BBs. These findings are in line with observations in other systems: for example, apical actin organization is critical for anchoring hundreds of cilia in human airways \cite{pan_rhoa-mediated_2007}; perturbations of WDR5 \cite{kulkarni_wdr5_2018} and filamin \cite{chevalier_mir-34449_2015} disrupt BB spacing in \textit{Xenopus}; and actin-modulating drugs and actin-associated factors affect BB distribution and docking in \textit{Xenopus} and zebrafish MCCs, respectively \cite{werner_actin_2011, kumar_microvillar_2018}. Together with our data, these studies suggest a conserved mechanism by which actin meshwork dynamics actively dictate BB organization.

In conclusion, through a combination of imaging, quantitative analysis and perturbation studies, we establish the cytoskeletal organization of actin as a central driver of BB spatial patterning in \textit{X. laevis} MCCs. We anticipate that identifying this mechanistic link between cytoskeletal mechanics and BB dynamics will advance our understanding of how subcellular architecture emerges during development and how its dysregulation contributes to disease states.\\ \\

\clearpage
\majorheading{Methods}
\addcontentsline{toc}{section}{\hyperref[sec:Methods]{\large\bfseries Methods}}
The sections below describe the experimental procedures, and all materials used in these experiments are listed in tabular form in the List of Materials section in the Supplementary information.

\section{\textit{Xenopus laevis} embryo preparation and manipulation}
\subsection{Husbandry}
Adult \textit{Xenopus laevis} were obtained from NASCO (Fort Atkinson, WI, USA) or Xenopus 1 (Dexter, MI, USA). Animals were maintained in a centralized facility in accordance with guidelines provided by the National Xenopus Resource (Marine Biological Laboratory, Woods Hole, MA, USA), the European Xenopus Resource Center (University of Portsmouth, UK), and Xenbase (http://www.xenbase.org). All procedures were reviewed and approved by the Danish National Animal Ethics Committee (permit no. 2017-15-0201-01237).

\subsection{Embryo preparations}
Ovulation in adult female \textit{Xenopus laevis} was induced by injection of human chorionic gonadotropin (hCG; Chorulon; 500 IU per animal) on the day prior to experiments. Approximately 16 h later, eggs were collected by gentle abdominal squeezing of hormonally primed females. Eggs were fertilized in 1/3× Modified Ringer’s (MR) solution (33 mM \ce{NaCl}, 0.6 mM \ce{KCl}, 0.67 mM \ce{CaCl2}, 0.33 mM \ce{MgCl2}; pH 7.6) by addition of a small piece of crushed testis obtained from a sacrificed adult male. After \SI {\sim 2} h at \SI{20}{\degreeCelsius}, fertilized embryos were dejellied by incubation in 2.5–3\% (w/v) Cysteine solution (pH 7.8) for 8 min, followed by several rinses in 1/3× MR. Morphologically healthy embryos were then selected and transferred to fresh 1/3× MR for subsequent manipulations.

\subsection{Microinjections of plasmids and morpholinos}
For targeted fluorescent labeling of BBs and actin in multiciliated cells (MCCs), embryos were microinjected with the following plasmids driven by the $\alpha$-tubulin promoter: Chibby-GFP/RFP and LifeAct-GFP/RFP, at 5 ng/µL per blastomere. Microinjections were performed into the ventral blastomeres at the (Nieuwkoop and Faber (NF)\cite{nieuwkoop1994normal}) 2-cell or 4-cell stage, using either 1/3× MR solution or 2\% Ficoll in 1/3× MR. For $\alpha$-actinin-1 knockdown experiments, morpholino doses were carefully titrated to 15 ng per blastomere. Injection of morpholino at the 2–4-cell stage caused basal bodies to migrate from the basal to apical side, expansion of the apical domain, and docking of basal bodies into the expanding apical domain, while formation of the actin meshwork was impaired. Translation-blocking (ATG) or splice-blocking $\alpha$-actinin-1 morpholinos (Gene Tools; morpholino sequences are listed in Table 1) were co-injected with Chibby-GFP/RFP and LifeAct-GFP/RFP constructs for knockdown experiments. Prior to injection, the morpholino solution was heated to \SI{90}{\degreeCelsius} for 10 min, centrifuged, and the supernatant was subsequently mixed with the Chibby- and LifeAct-based constructs to obtain the desired concentrations. After injection, embryos were incubated at 18-\SI{23}{\degreeCelsius} in 1/3x MR until the desired stage (20-26) for live imaging or fixation.

\subsection{Immunostaining for high-resolution imaging}
For high-resolution Airyscan imaging of the actin meshwork, embryos were injected at the 2–4 cell stage with the $\alpha$-tubulin driven Chibby-GFP plasmid and fixed at the desired developmental stage for phalloidin staining. For knockdown experiments, $\alpha$-actinin-1 morpholino was co-injected with Chibby-GFP. Embryos were fixed between stages 22–26 to visualize actin meshwork formation across different BB distributions and apical domain sizes. At the desired stage, vitelline membranes were carefully removed, and embryos were rinsed once in MEMFA (0.1 M 3-(N-morpholino)propanesulfonic acid (MOPS), pH 7.4; 2 mM Ethylene glycol-bis($\beta$-aminoethyl ether)-N,N,$N'$,$N'$-tetraacetic acid (EGTA); 1 mM \ce{MgSO4}; 3.7\% formaldehyde), followed by fixation in fresh MEMFA on a rotator either for 2 h at room temperature (\SI{\sim22}{\degreeCelsius}) or overnight at \SI{4}{\degreeCelsius}. Following fixation, embryos were washed three times in Tris-buffered saline containing Triton X-100 (TBST; TBS: 155 mM \ce{NaCl}, 10 mM Tris-HCl, pH 7.4, supplemented with 0.1\% Triton X-100) for 5 min each. Blocking was performed by incubation in blocking buffer (10\% fetal bovine serum (FBS) and 5\% dimethyl sulfoxide (DMSO) in TBS) for 1 h, followed by washing in TBST for 1 h on a rotator. Embryos were then incubated with fluorescent phalloidin (555, 568, or 647; 1:500 dilution) for 2 h at room temperature or overnight at \SI{4}{\degreeCelsius}. After staining, embryos were washed three times in TBST over 30 min and stored in TBS at \SI{4}{\degreeCelsius} for up to 10 days. All fixed embryos were imaged within 10 days of staining.

\subsection{Embryo mounting for imaging}
\subsubsection{Live imaging}
Embryos at the desired developmental stage were manually removed from their vitelline membranes and briefly rinsed in drops (\SI{\sim{70}}{\micro\litre}) of 1.5–2\% low– or ultra-low–melting-point agarose preheated to \SI{65}{\degreeCelsius}. This rinsing step was repeated twice using fresh agarose drops to dilute residual MR solution coming from the embryo transfer. A fresh drop (\SI{\sim{70}}{\micro\litre}) of molten agarose was placed onto a coverslip mounted in the imaging chamber (Attofluor Cell Chamber for microscopy). Rinsed embryos (5–7 per drop) were transferred into this agarose drop and their orientations were adjusted so that the flank of the embryos faced the coverslip. Then excess agarose surrounding the embryos was carefully aspirated to slightly flatten the embryo surface against the coverslip. This increased the accessible surface area for imaging without excessive flattening. The agarose was allowed to set at \SI{13}{\degreeCelsius} for 3 min. A small volume (\SI{\sim{20}}{\micro\litre}) of agarose was then added to the top and sides of the set agarose to reinforce the mount and allowed to solidify for an additional 3 min at \SI{13}{\degreeCelsius}. This reinforcement step was repeated twice before gently filling the imaging chamber with fresh 1/3× MR solution. All agarose mounting steps were performed at a reduced ambient room temperature (\SI{\sim{20}}{\degreeCelsius}). Further, agarose was dispensed using pipettes with tips cooled by briefly holding them in hand prior to contact with embryos. These precautions minimized heat-induced damage to the embryonic surface. Agarose-mounted embryos were imaged for up to 5 h, after which a fresh batch of embryos was mounted.
\subsubsection{Fixed imaging}
Phalloidin-stained embryos stored in TBS were placed directly onto a coverslip mounted in the imaging chamber. Excess TBS was gently aspirated to allow the embryos to flatten against the coverslip, maximizing the available surface area for imaging. A small coverslip fragment, with silicone grease applied to its four corners, was gently pressed from the top of the embryos to stabilize their position. The imaging chamber was then carefully flooded with TBS and transferred for imaging.

\section{Microscopy}
\subsection{Imaging setups}
\subsubsection{Confocal microscopy}
All coarse-timed imaging were performed using either a Zeiss LSM880 or a Leica Stellaris line-scanning confocal microscope. Detailed acquisition specifications for each system are provided below. \textit{Zeiss LSM880}: Imaging was performed using 1–3\% laser power from Argon 25 mW (488 nm, 514 nm), diode-pumped solid-state (DPSS) 10 mW (561 nm), and helium–neon (HeNe) 5 mW (633 nm) laser lines; a 40x C-Apochromat water immersion objective (NA 1.2, Korr M27, WD 0.22 mm) or 60x C-Plan-Apochromat oil immersion objective (NA 1.4, DIC M27, WD 0.14 mm); conventional photomultiplier tube (PMT) detectors; and ZEN Black acquisition software. \textit{Leica Stellaris}: Imaging was performed using 1–3\% laser power from a freely tunable white-light laser (485–685 nm); a 40x HC PL APO CS2 oil immersion objective (NA 1.3, WD 0.24 mm); highly sensitive HyD S detectors; and LAS X acquisition software. For both systems, the frame size and zoom were adjusted to ensure a minimum of 2x Nyquist sampling in the xy plane ($\text{pixel size} \leq \Delta_{xy}/2$), where $\Delta_{xy}$ denotes the lateral resolution, estimated as $\Delta_{xy} \approx 0.61\lambda/NA$ based on the acquisition settings. Full three-dimensional image stacks of entire MCC were acquired at 30 s or 60 s intervals with an axial resolution (optical section thickness) of $\Delta_{z} \approx 2n\lambda/NA^{2}$. Image stacks were sampled with a z-step size of \SI{\sim{0.4}}{\micro\meter}, corresponding to an optical section of \SI{\sim{0.9}}{\micro\meter}, thereby satisfying the Nyquist criterion for axial sampling ($\text{z-step size} \leq \Delta_{z}/2$).

\subsubsection{High-speed imaging}
To capture fine BB movements at the MCC apical domain, imaging was performed using a custom-built total internal reflection fluorescence (TIRF) microscope based on an Olympus IX83 inverted platform. Imaging was carried out using 0.5–2\% laser power from Cobolt 200 mW laser lines (488 nm and 561 nm); a 150x universal apochromat Olympus TIRF oil-immersion objective (NA 1.45, WD 0.08 mm); a high-speed Hamamatsu ImagEM X2 electron-multiplying charge-coupled device (EMCCD) camera (106.667 nm per pixel for 150x 1.45 NA objective); and cellSens acquisition software. Using this setup, 30,000 images of BBs were acquired over 10.5 min at the rate of 1 frame every 21 ms. The frame size and zoom were adjusted to ensure 2x Nyquist sampling in the lateral (xy) plane for single-plane TIRF imaging, with all images acquired at a fixed focal plane throughout the time lapse acquisition. The excitation focus position was maintained by a motorized Olympus TIRF module, and axial drift was continuously corrected using an Olympus IX3-ZDC2 z-drift compensation system. Photobleaching was minimized using a real-time controller that synchronized camera exposure with an acousto-optic tunable filter (AOTF).

\subsubsection{High-resolution imaging}
High-resolution imaging of the actin meshwork was performed using the Airyscan super-resolution mode on a Zeiss LSM980 microscope. Imaging was carried out using 1–3\% laser power from diode 30 mW (488 nm), DPSS 25 mW (561 nm), DPSS 8 mW (594 nm), and diode 25 mW (639 nm) laser lines; a 63x Plan-Apochromat oil-immersion objective (NA 1.4, WD 0.19 mm); an Airyscan2 detector with a 32-channel gallium arsenide phosphide (GaAsP) detector array; and ZEN Blue 3 acquisition software. The frame size and zoom were adjusted to ensure a minimum of 2× Nyquist sampling in the lateral (xy) plane. Z-stacks were acquired to cover the entire apical cortex of MCCs with a z-step size of \SI{\sim{0.3}}{\micro\meter} using a piezo stage, corresponding to an optical section of \SI{\sim{0.9}}{\micro\meter}, thereby satisfying the Nyquist criterion for axial sampling.

\subsection{Pre-acquisition considerations for coarse-timed imaging}
During imaging of BB distribution and apical domain expansion, ongoing morphogenetic movements of the embryo caused continuous lateral (xy) drift of the region of interest (ROI). To facilitate downstream image analysis, the field of view (FOV) was continuously monitored throughout acquisition, and the ROI was manually repositioned by stage movement whenever drift was observed. Similarly, the z-stack range was adjusted as needed to compensate for axial drift, ensuring that the full height of the MCC remained within the defined imaging volume at all time points. Laser power and detector gain were adjusted during acquisition to maintain consistent visibility of BBs and apical actin throughout the imaging period.

\subsection{Pre-acquisition considerations for fine-timed imaging}
Fine-timed imaging was performed using a TIRF setup, which requires a planar sample for proper visualization of BBs moving across the apical plane. Therefore, only cells with a flat apical profile were selected to avoid losing BBs moving out of the TIRF plane. Using this setup, we found that BB fluorescence (Chibby-GFP) remained stable and trackable for 10.5 min at 1 frame every 21 ms. Apical areas of different sizes were imaged for 10.5 min to characterize BB trajectories across the full apical area range. To measure the apical area, we first acquired 400 frames in the actin channel, followed by 30,000 frames in the BB channel. To assess microscope stability, including lateral (xy) and axial (z) drift, and to provide reference objects for localization noise correction during particle detection and tracking, we imaged \SI{0.5}{\micro\meter} diameter beads ($\sim{}$BB size) attached to coverslips. These beads were imaged with the same settings (1 frame every 21 ms for 10.5 min), yielding 30,000 frames of bead data. Bead imaging was performed prior to BB experiments to validate the setup and provide reference data for downstream image analysis.

\section{Softwares} 
All image and data analyses were performed using established plugins and custom-written scripts in ImageJ \cite{schneider2012nih}, Fiji \cite{schindelin2012fiji}, and MATLAB R2023b (MathWorks, Natick, MA, USA). Analyses were carried out on a Windows 11-based workstation equipped with 192 GB of RAM, an 8 GB graphics card, and an Intel Xeon W-125 CPU (2.5 GHz). Statistical analyses were conducted using GraphPad Prism 11. Figures and schematics were generated using Adobe Illustrator 27 and Inkscape 1.4.

Detailed descriptions of all image analysis methods are provided in the Supplementary Information.

\section*{Acknowledgments}
We thank the John Wallingford group from the University of Texas at Austin for the alpha-Tubulin-LifeAct-GFP and alpha-Tubulin-LifeAct-RFP constructs and the Brian Mitchell group from the Northwestern University Feinberg School of Medicine for sharing the alpha-Tubulin-Chibby-GFP and alpha-Tubulin-Chibby-RFP constructs. We are grateful to the \textit{Xenopus} facility caretakers at the University of Copenhagen; Jutta Bulkescher and the reNEW Imaging Platform, and the Core Facility for Integrated Bioimaging (Faculty of Health and Medical Sciences, University of Copenhagen) for technical support and assistance with imaging. We thank Lene Oddershede and Akbar Samadi, formerly a postdoctoral researcher in her group at the Niels Bohr Institute, for technical assistance and discussions on fine-timed experiments. We thank Guilherme Bastos Ventura for critical reading of the manuscript and for valuable comments. We also thank members of the Sedzinski group for their valuable comments and suggestions.

\subsection*{Funding}
This work was supported by grants from Novo Nordisk Fonden (NNF21CC0073729, J.S.), European Research Council Consolidator Grant (ERC CoG 101125803 MechanoFate, J.S.). J.S. acknowledges the support of the Novo Nordisk Foundation (NNF22OC0076414, NNF19OC0056962) and the Leo Foundation (LF-OC-19-000219). P.M.B and Y.F.B acknowledge Novo Nordisk Foundation grant number NNF20OC0061176.

\subsection*{Author contributions}
R.T. conceived and designed the study, performed all experiments, analyzed the data, and wrote the manuscript with contributions from all other authors. Y.F.B. performed TIRF imaging experiments and TIRF data analysis. P.B. supervised TIRF experiments. M.M.I. conceived and designed the study, generated the theoretical model description, and helped with data analysis. J.S. conceived and designed the study, supervised the project, and acquired funding.

\subsection*{Use of Artificial Intelligence tools}
The authors acknowledge the use of ChatGPT, Gemini, Perplexity, and Claude for brainstorming, editing, coding and technical assistance, and proofreading during the preparation of this manuscript. All content generated with the assistance of these tools was subsequently reviewed and edited by the authors, who take full responsibility for the final version of the manuscript.

\subsection*{Data availability}
The data and materials supporting the findings of this study are available within the article and its supplementary information files. Additional information and relevant raw data are available from the corresponding authors upon request. 

\subsection*{Code availability}
The code used for data analysis and simulations is available in the GitHub repositories \href{https://github.com/RaghavanThiagarajan/Basal-Body-analysis.git}{BB analysis} and \href{https://github.com/minamdar/Basal-Body-Simulations}{BB simulations}, respectively.

\subsection*{Competing interests}
The authors declare no competing interests.


\section*{Main figures}

\begin{figure}[p]
  \centering
  \includegraphics[width=\textwidth]{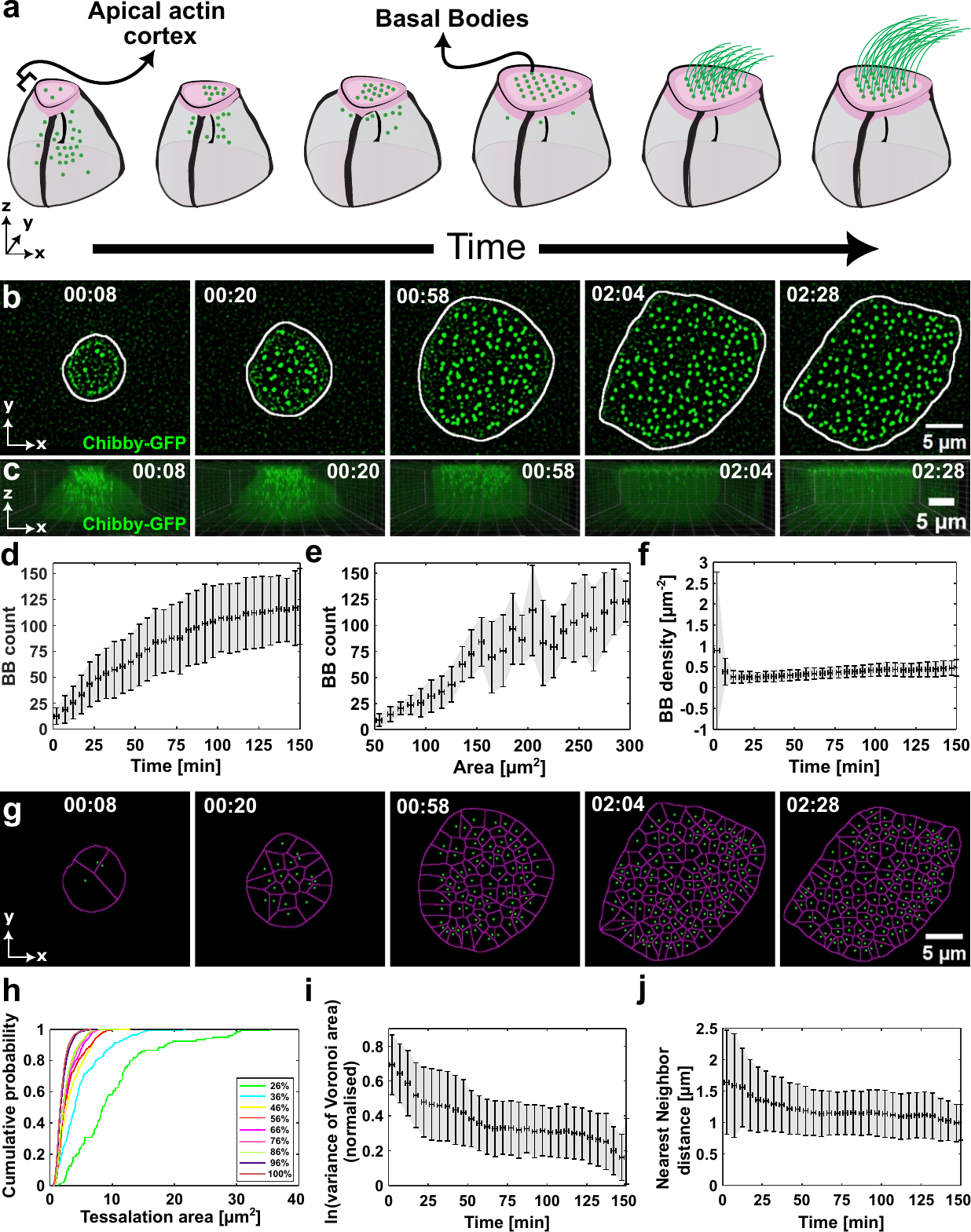}
  \refstepcounter{figure}
  \label{fig:fig1}
\end{figure}
\clearpage 
\noindent\textbf{Figure \ref{fig:fig1}. Basal body patterning is synchronized with apical surface expansion.}
All data correspond to coarse-timed imaging (frame interval = 30–60 s; total duration = 2.5 h). (\textbf{a}) Schematic showing the sequence of BB dynamics during multiciliated cell (MCC) morphogenesis. BBs ascend from the basal to the apical side, followed by distribution and patterning within the apical plane. BBs are shown in green, and the apical actin cortex is shown in pink. (\textbf{b}) Time-lapse sequence of BBs (Chibby-GFP) within the apical domain, with the apical periphery outlined by a white contour (see Movie 1). (\textbf{c}) Corresponding sequence showing the ascent of the BB from the basal to apical side of the MCC (see Movie 2). (\textbf{d}) Number of BBs plotted against time. (\textbf{e}) Number of BBs plotted against the apical area. (\textbf{f}) BB density over the apical area plotted against time. (\textbf{g}) Voronoi tessellations of BBs for the time-lapse sequence shown in (b) (see Movie 3). (\textbf{h}) Cumulative distribution functions (CDFs) of Voronoi tessellation areas for every 10\% \ increase in the apical area, for the data shown in (g). (\textbf{i}) Variance of Voronoi tessellation areas plotted against time. Variance values are natural log–transformed (ln) and min–max normalized. (\textbf{j}) The Nearest-neighbor distance of the BBs plotted against time. In (\textbf{d}), (\textbf{e}), (\textbf{f}), (\textbf{i}) and (\textbf{j}), error bars represent mean $\pm$ standard deviation (SD) of data averaged over 9 cells from 5 experiments, using 5 min (time) or \SI{10}{\micro\meter\squared} (area) bins. In (\textbf{b}), (\textbf{c}), and (\textbf{g}), time is in hh:mm.

\begin{figure}[p]
  \centering
  \includegraphics[width=\textwidth]{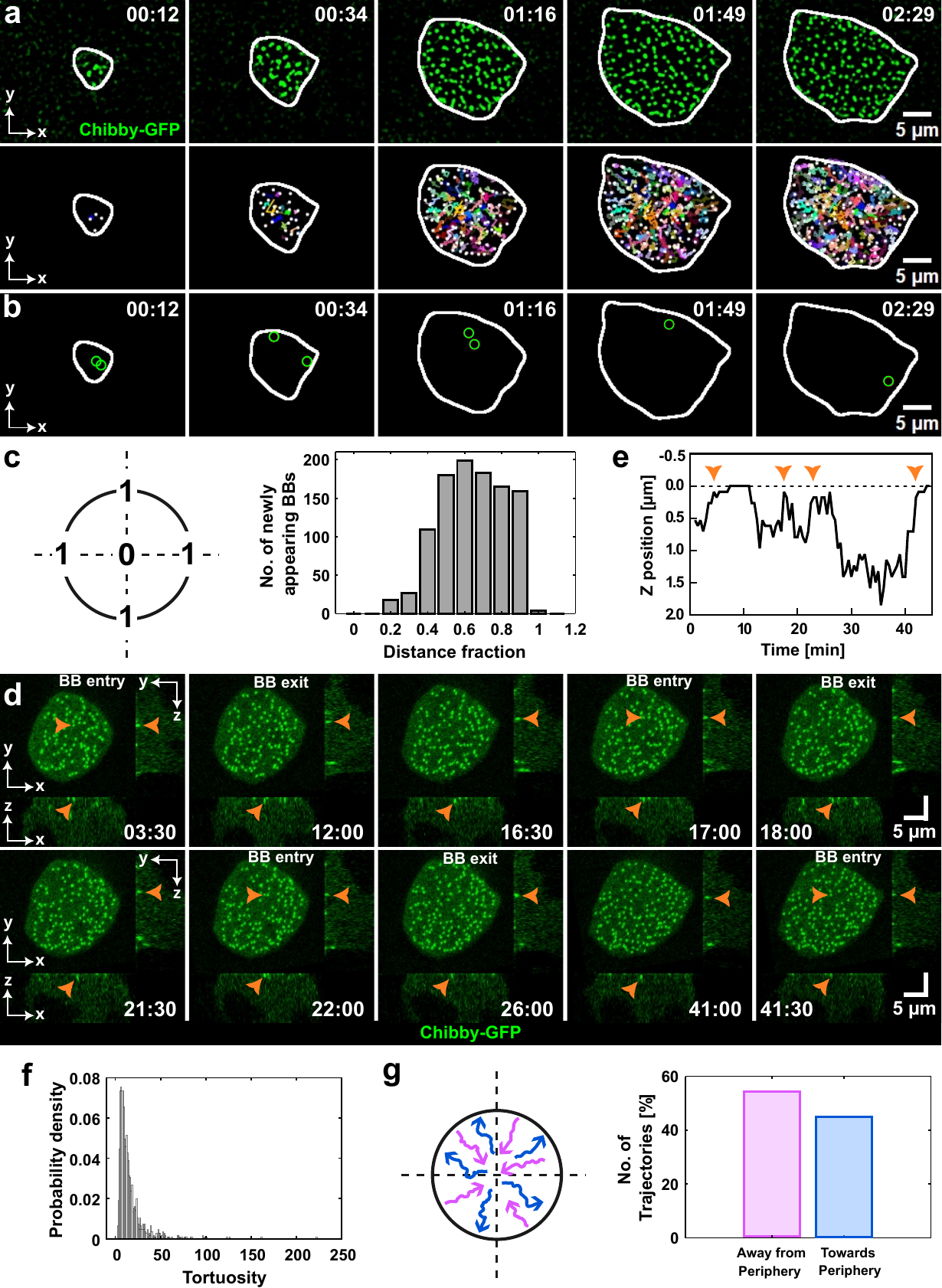}
  \refstepcounter{figure}
  \label{fig:fig2}
\end{figure}
\clearpage 
\noindent\textbf{Figure \ref{fig:fig2}. Passive and active mechanisms shape BB distribution.}
All data correspond to coarse-timed imaging (frame interval = 30–60 s; total duration = 2.5 h). (\textbf{a}) First row: time-lapse sequence of BBs (Chibby-GFP) in the apical domain. Second row: trajectories of BBs for the sequence above (see Movie 4). (\textbf{b}) Appearance of BBs near the apical periphery. The circles mark the BBs, and the white outline marks the apical boundary (see Movie 5). (\textbf{a}) and (\textbf{b}) share the same time-lapse sequence where the apical periphery is outlined by a white contour. (\textbf{c}) The schematic shows the distance fraction where 0 corresponds to the center and 1 to the periphery of the apical domain. Plot shows the distribution of newly appearing BBs across distance fractions within the apical domain. (\textbf{d}) Time-lapse sequence showing the entry and exit of a BB in the apical domain. Each panel shows views in xy (top), xz (bottom), and yz (right); the orange arrow highlights the tracked BB and the time corresponds to the plot in (e) and  Movie 6. (\textbf{e}) Plot showing the change in axial Z position of the BB shown in (d) over time. The dotted line at zero marks the apical surface. Arrows indicate the positions at which the BB approaches or reaches the apical surface during transient entry events before subsequently exiting. The final arrow marks the entry event after which the BB remains stably docked at the apical surface. (\textbf{f}) Probability density of trajectory tortuosity with a peak at 9.6. (\textbf{g}) The schematic illustrates BB trajectories directed towards (blue) and away (pink) from the periphery. The plot shows the percentage of trajectories moving in each direction. Data in (\textbf{c}), (\textbf{f}), and (\textbf{g}) are pooled from 9 cells across 5 experiments. In (\textbf{f}) and (\textbf{g}), a total of 1052 BB trajectories were analyzed across these cells. In (\textbf{a}) and (\textbf{b}), time is in hh:mm; in (\textbf{d}), time is in mm:ss.

\begin{figure}[p]
  \centering
  \includegraphics[width=\textwidth]{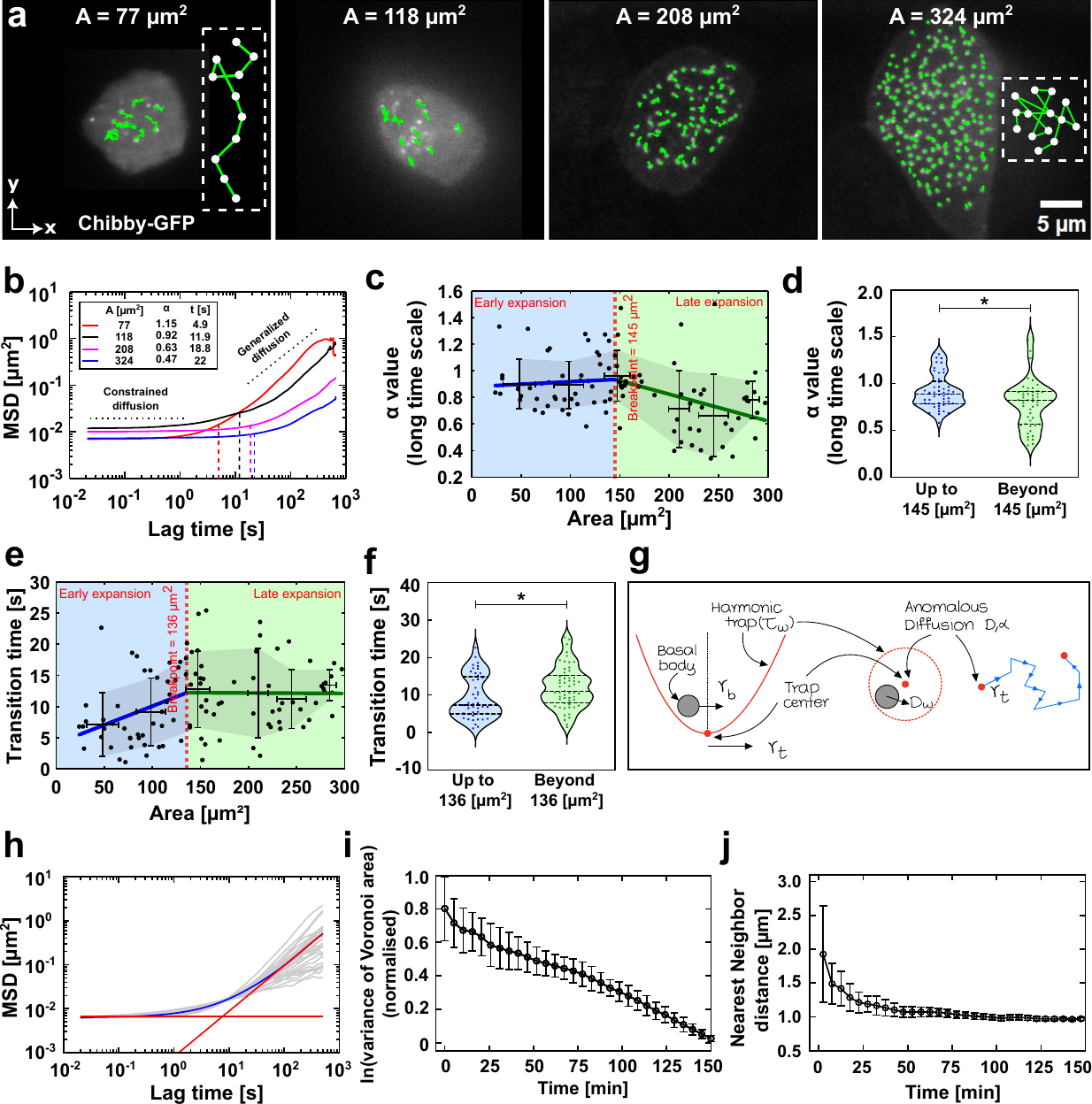}
  \refstepcounter{figure}
  \label{fig:fig3}
\end{figure}
\clearpage 
\noindent\textbf{Figure \ref{fig:fig3}. Basal body dynamics shift with apical domain expansion.} Data in (\textbf{a-f}) correspond to fine-timed imaging (frame interval = 21 ms; 30,000 frames; total duration = 10.5 min). (\textbf{a}) Images of four apical domains of different sizes, shown at the last time point of a 10.5 minute fine-timed image sequence. BBs (Chibby-GFP) are shown with their overlaid trajectories. The inset in the first image shows a schematic illustrating elongated trajectories in smaller apical domains, while the inset in the last image highlights convoluted trajectories in larger domains (see Movie 7). (\textbf{b}) MSDs of BBs from cells shown in (a), plotted against lag time. Each curve represents the ensemble-averaged MSD of trajectories within a single cell. The vertical dotted lines mark the transition times. Reported for each curve are the apical area, $\alpha$ value, and transition time. The black dotted lines above the curves indicate the phases of short- and long-timescales, depicting constrained diffusion at short times and a mix of sub-, diffusive, and superdiffusive behaviors at long times. (\textbf{c}) Long-term $\alpha$ values plotted against the apical area. (\textbf{d}) Comparison of $\alpha$ values before and after the area breakpoint in (c); exact P = 0.0144. (\textbf{e}) Transition time plotted against the apical area. (\textbf{f}) Comparison of transition times before and after the area breakpoint in (e); exact P = 0.0173. In (\textbf{c}) and (\textbf{e}), each dot represents the ensemble average per cell; error bars represent mean $\pm$ SD of averaged data, based on 4,604 trajectories from 100 cells across 17 experiments, using \SI{50}{\micro\meter\squared} bins. The red dotted line marks the breakpoint ($\alpha$ value - \SI{145}{\micro\meter\squared}; transition time – \SI{136}{\micro\meter\squared}) identified by piecewise linear regression (\textit{see methods section 8.3.5 in SI}) while blue and green lines are regression fits before and after the breakpoints. The blue and green shaded regions denote data before and after the breakpoints and serve as visual guides for the early (<\SI{150}{\micro\meter\squared}) and late (>\SI{150}{\micro\meter\squared}) phases of apical expansion. Data for (\textbf{d}) include 837 trajectories from 50 cells (\(\leq\) \SI{145}{\micro\meter\squared}) and 3767 from 50 cells (>\SI{145}{\micro\meter\squared}), and for (\textbf{f}), 675 trajectories from 45 cells (\(\leq\) \SI{136}{\micro\meter\squared}) and 3929 from 55 cells (>\SI{136}{\micro\meter\squared}), all from 17 experiments. In (\textbf{d} and \textbf{f}), the central line represents the median, and the upper and lower lines represent the $75^\text{th}$ and $25^\text{th}$ percentiles, respectively. (\textbf{g-j}) Model simulations. (\textbf{g}) Schematic of the particle trap arrangement in simulations showing initial caging followed by anomalous diffusion. (\textbf{h}) MSDs of simulated BBs: grey - individual particles; blue - ensemble average; red lines - fitted regimes; their intersection defines the transition time. Data obtained from 20 independent runs of the single-particle simulation. (\textbf{i}) Variance of Voronoi tessellation areas plotted against time. Variance values are natural log–transformed (ln) and min–max normalized. (\textbf{j}) The distance from the Nearest-neighbor plotted against time. In (\textbf{i} and \textbf{j}), error bars represent mean $\pm$ SD of data averaged over 20 simulations, using 5 min bins.

\begin{figure}[p]
  \centering
  \includegraphics[width=\textwidth]{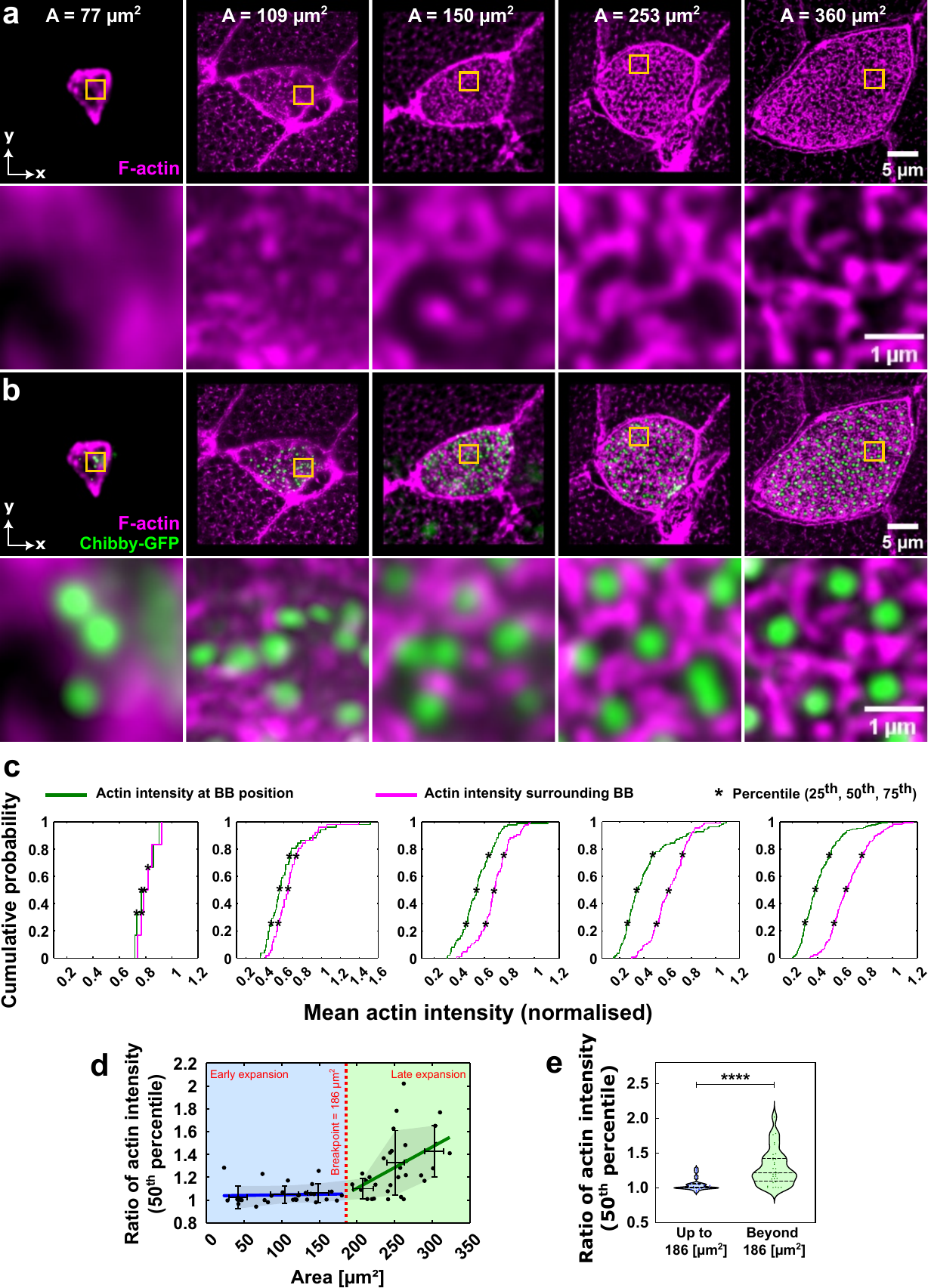}
  \refstepcounter{figure}
  \label{fig:fig4}
\end{figure}
\clearpage 
\noindent\textbf{Figure \ref{fig:fig4}. Progressive formation of an apical actin meshwork facilitates BB redistribution.}
All data from high-resolution Airyscan imaging. (\textbf{a–b}) Images of increasing apical domain areas showing the formation of an actin meshwork. (\textbf{a}) Actin (Phalloidin). (\textbf{b}) Actin and BBs (Chibby-GFP). The second rows in (a) and (b) show zoom-ins of the regions highlighted by orange squares. (\textbf{c}) CDFs of actin intensity at the BB positions (green) and in the surrounding regions (pink), corresponding to (a) and (b). Black markers indicate actin intensity at the ${25}^{\text{th}}$, ${50}^{\text{th}}$, and ${75}^{\text{th}}$ percentiles. (\textbf{d}) Ratio of actin intensity (surrounding / BB position) at the ${50}^{\text{th}}$ percentile of CDF, plotted against the apical area. Each dot represents a single cell, and the error bars represent mean $\pm$ SD of data averaged over 64 cells across 4 experiments, using \SI{50}{\micro\meter\squared} bins. The red dotted line marks the breakpoint (\SI{186}{\micro\meter\squared}) of the piecewise linear regression while blue and green lines represent the regression fits before and after the breakpoint. The blue and green shaded regions denote data before and after the breakpoint and serve as visual guides for the early (<\SI{150}{\micro\meter\squared}) and late (>\SI{150}{\micro\meter\squared}) phases of apical expansion. (\textbf{e}) Comparison of actin intensity ratios at ${50}^{\text{th}}$ percentile of CDF before and after the breakpoint in (d); P < 0.0001. The central line represents the median, and the upper and lower lines represent the $75^\text{th}$ and $25^\text{th}$ percentiles, respectively. Data: 32 cells for each category (\(\leq\)\SI{186}{\micro\meter\squared}) and (>\SI{186}{\micro\meter\squared}), from 4 experiments.

\begin{figure}[p]
  \centering
  \includegraphics[width=\textwidth]{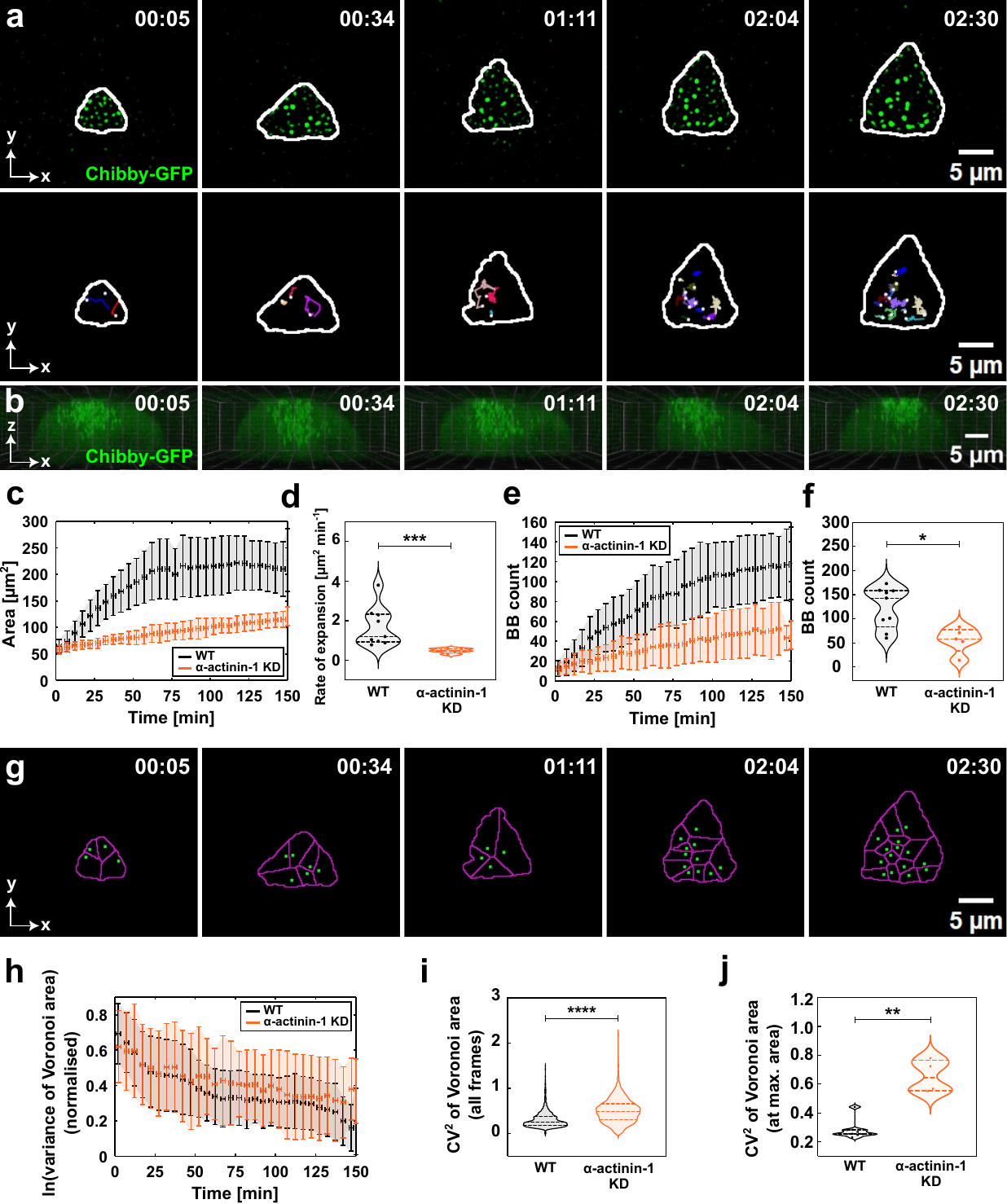}
  \refstepcounter{figure}
  \label{fig:fig5}
\end{figure}
\clearpage 
\noindent\textbf{Figure \ref{fig:fig5}. Disruption of apical actin cross-linking impairs BB organization.} All data correspond to coarse-timed imaging (frame interval = 30–60 s; total duration = 2.5 h). All plots compare WT (black) and $\alpha$-actinin-1 translation-blocking morpholino (orange) conditions. (\textbf{a,b,g}) Time-lapse sequences of an $\alpha$-actinin-1 translation-blocking morpholino injected cell. (\textbf{a}) First row: BBs (Chibby-GFP) within the apical domain; second row: corresponding BB trajectories; with the apical periphery outlined by a white contour (see Movies 9 and 11). (\textbf{b}) Sequence showing BB ascent from the basal to the apical side of the cell (see Movie 10). (\textbf{c}) Apical area plotted against time. (\textbf{d}) Comparison of apical expansion rates; exact P = 0.001. (\textbf{e}) BB number plotted against time. (\textbf{f}) BB number at the maximum apical area reached within 2.5 h; exact P = 0.0028. (\textbf{g}) Voronoi tessellations of BBs in an $\alpha$-actinin-1 KD MCC corresponding to (a) (see Movie 12). (\textbf{h}) Variance of voronoi tessellation areas plotted against time. Variance values are natural log–transformed (ln) and min–max normalized. (\textbf{i-j}) Comparison of variability in Voronoi tessellation areas: (\textbf{i}) pooled across all frames (2.5 h); P < 0.0001; (\textbf{j}) at the maximum apical area reached within 2.5 h; exact P = 0.0028. Variability is quantified as $CV^{2}$, the squared coefficient of variation (variance divided by the square of the mean). Data in (\textbf{c-f}) and (\textbf{h-j}) are from 9 WT cells across 5 experiments and from 5 KD cells across 4 experiments. For (\textbf{i}), data were pooled from 2611 frames (WT) and 1589 frames (KD). In (\textbf{c}), (\textbf{e}), and (\textbf{h}), error bars represent mean $\pm$ SD of binned data (5 min bins). In (\textbf{d}), (\textbf{f}), (\textbf{i}), and (\textbf{j}), the central line represents the median, and the upper and lower lines represent the $75^\text{th}$ and $25^\text{th}$ percentiles, respectively.

\begin{figure}[p]
  \centering
  \includegraphics[width=\textwidth]{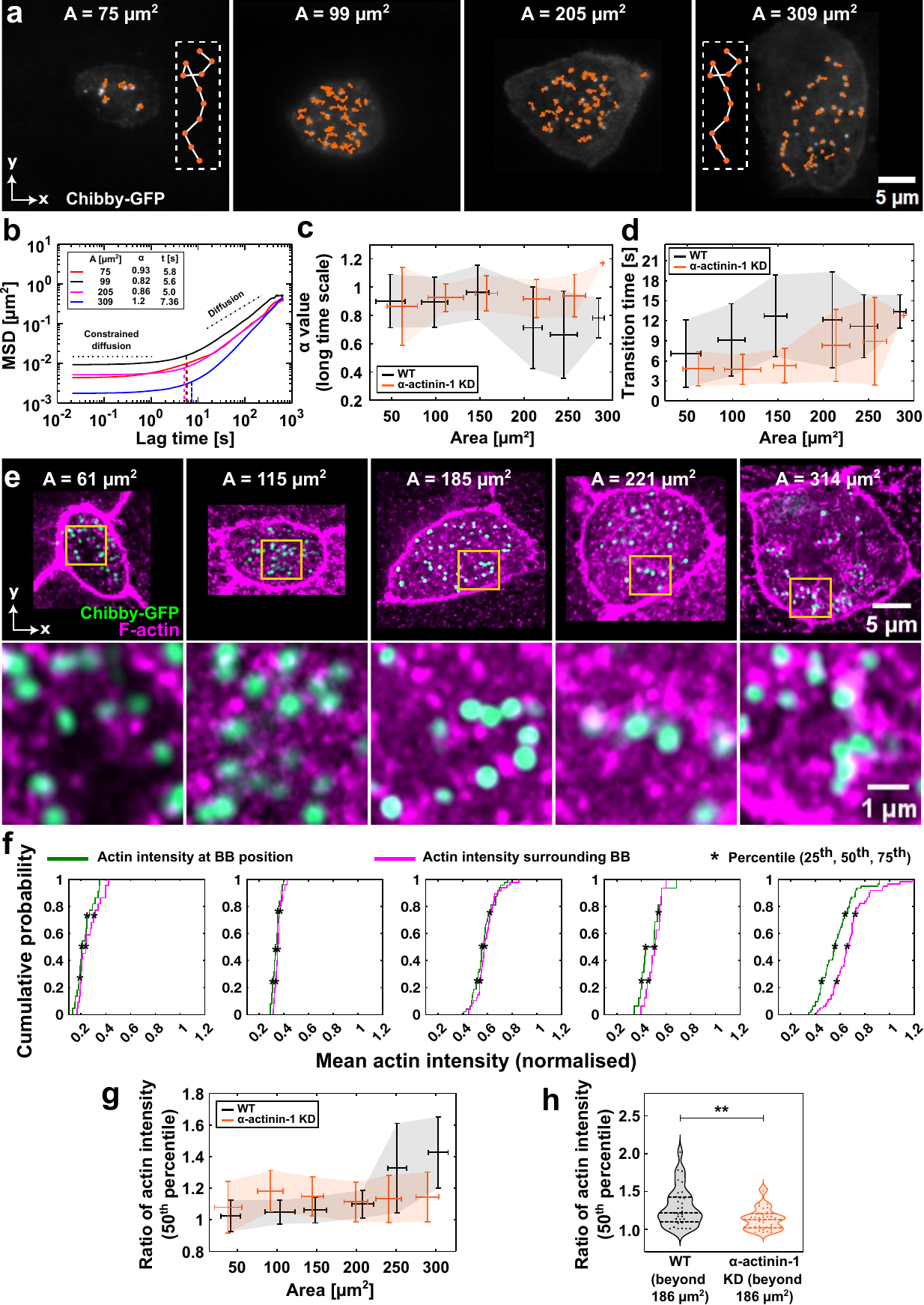}
  \refstepcounter{figure}
  \label{fig:fig6}
\end{figure}
\clearpage 
\noindent\textbf{Figure \ref{fig:fig6}. $\alpha$-actinin-1 perturbation impairs transition in BB dynamics and actin meshwork formation.} Data in (\textbf{a-b}) and (\textbf{e-f}) correspond to $\alpha$-actinin-1 translation-blocking morpholino condition; (\textbf{c-d}, \textbf{g-h}) compare WT (black) and $\alpha$-actinin-1 translation-blocking morpholino (orange) conditions.
Data in (\textbf{a-d}) correspond to fine-timed imaging (frame interval = 21 ms; 30,000 frames; total duration = 10.5 min). (\textbf{a}) Images of four apical domains of different sizes, shown at the last time point of a 10.5 min sequence. BBs (Chibby-GFP) are shown with overlaid trajectories (see Movie 13). The insets in the first and last images show schematics illustrating elongated trajectories in smaller and larger apical domains, respectively. (\textbf{b}) MSDs of BBs from the cells in (a), plotted against lag time. Each curve represents the ensemble-averaged MSD of trajectories within a single cell. Vertical dotted lines mark transition times. Reported for each curve are the apical area, $\alpha$ value, and transition time. (\textbf{c}) Long-time $\alpha$ values plotted against apical area. (\textbf{d}) Transition times plotted against apical area. In (\textbf{c-d}), $\alpha$ values and transition times are ensemble averages of all BB trajectories per cell. Error bars represent mean $\pm$ SD of averaged data over 4604 trajectories from 100 cells across 17 experiments (WT) and 2029 trajectories from 46 cells across 5 experiments (KD), using \SI{50}{\micro\meter\squared} bins.
Data in (\textbf{e-h}) correspond to high-resolution Airyscan imaging. (\textbf{e}) Images of apical domains of increasing size showing the lack of actin meshwork formation. Top row: actin (Phalloidin) and BBs (Chibby-GFP). Bottom row: zoom of regions highlighted in orange squares in the top row. (\textbf{f}) CDFs of actin intensity at BB positions (green) and surrounding regions (pink), corresponding to (e). Black markers indicate actin intensity at the ${25}^{\text{th}}$, ${50}^{\text{th}}$, and ${75}^{\text{th}}$ percentiles. (\textbf{g}) Ratio of actin intensity (surrounding / BB position) at the ${50}^{\text{th}}$ percentile of CDF, plotted against the apical area. Error bars represent mean $\pm$ SD of averaged data over 64 cells across 4 experiments (WT) and 70 cells across 3 experiments (KD), using \SI{50}{\micro\meter\squared} bins. (\textbf{h}) Comparison of actin intensity ratios at ${50}^{\text{th}}$ percentile of CDF between WT and KD, beyond \SI{186}{\micro\meter\squared}; (exact P = 0.0059). Data from 32 cells across 4 experiments (WT) and 36 cells across 3 experiments (KD). The central line represents the median, and the upper and lower lines represent the $75^\text{th}$ and $25^\text{th}$ percentiles, respectively.

\begin{figure}[p]
  \centering
  \includegraphics[width=\textwidth]{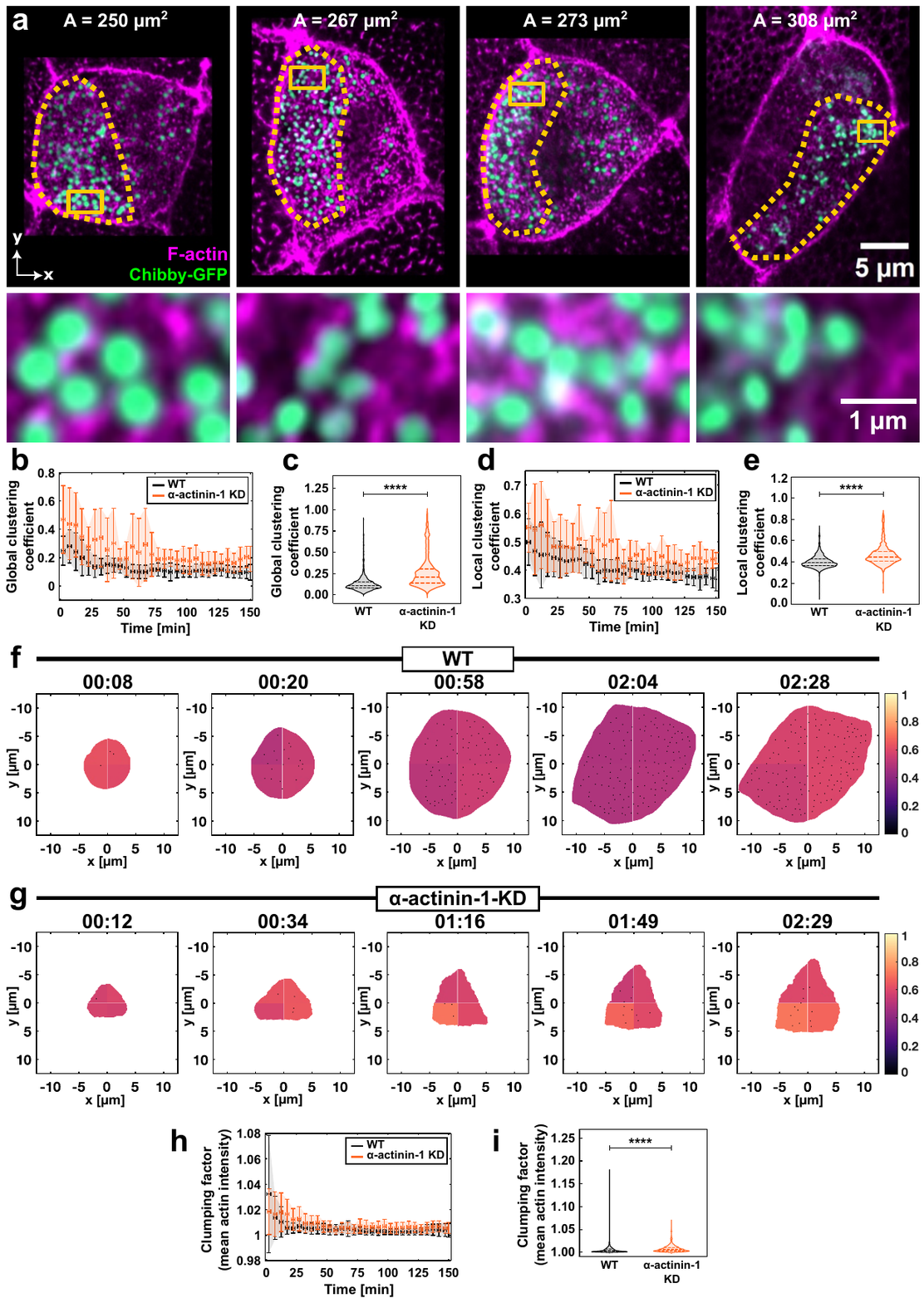}
  \refstepcounter{figure}
  \label{fig:fig7}
\end{figure}
\clearpage 
\noindent\textbf{Figure \ref{fig:fig7}. Heterogeneous distribution of the apical actin network leads to BB clumping.} All data from coarse-timed imaging (frame interval = 30–60 s; total duration = 2.5 h). Data in (\textbf{b-e}) and (\textbf{h-i}) compare WT (black) and $\alpha$-actinin-1 translation-blocking morpholino (orange) conditions. (\textbf{a}) Images showing the global and local clustering in $\alpha$-actinin-1 translation-blocking morpholino condition. Areas outlined by dotted orange lines indicate global clustering, whereas the zoom-ins of regions outlined by solid lines show local clustering of BBs. (\textbf{b, d}) Time evolution of the global (\textbf{b}) and local (\textbf{d}) clustering coefficients from experiments. (\textbf{c, e}) Comparisons of global (\textbf{c}) and local (\textbf{e}) clustering coefficients between WT and KD. (\textbf{f-g}) Time-lapse sequences of apical domains discretized into four quadrants and color-coded for mean actin intensity: (\textbf{f}) WT; (\textbf{g}) KD (see Movie 16). (\textbf{h}) Time evolution of the clumping factor for actin. (\textbf{i}) Comparison of clumping factors between WT and KD for actin. In (\textbf{b}), (\textbf{d}) and (\textbf{h})), error bars represent mean $\pm$ SD of binned data (5 min bins). Statistics for comparisons: (\textbf{c}) 2,554 frames (WT) vs. 1,455 frames (KD), P < 0.0001; (\textbf{e}) 2,552 frames (WT) vs. 1,455 frames (KD), P < 0.0001; (\textbf{i}) 1,670 frames (WT) vs. 1,182 frames (KD), P < 0.0001. In (\textbf{c}), (\textbf{e}) and (\textbf{i}), the central line represents the median, and the upper and lower lines represent the $75^\text{th}$ and $25^\text{th}$ percentiles, respectively. Data in (\textbf{b-e}) and (\textbf{h-i})) were obtained from 9 cells across 5 experiments (WT) and 5 cells across 4 experiments (KD).

\clearpage

\begin{suppfigure} 
  \centering
  \refstepcounter{supfig}
  \label{fig:sfig1}
\end{suppfigure}

\begin{suppfigure} 
  \centering
  \refstepcounter{supfig}
  \label{fig:sfig2}
\end{suppfigure}

\begin{suppfigure} 
  \centering
  \refstepcounter{supfig}
  \label{fig:sfig3}
\end{suppfigure}

\begin{suppfigure} 
  \centering
  \refstepcounter{supfig}
  \label{fig:sfig4}
\end{suppfigure}

\begin{suppfigure} 
  \centering
  \refstepcounter{supfig}
  \label{fig:sfig5}
\end{suppfigure}

\begin{suppfigure} 
  \centering
  \refstepcounter{supfig}
  \label{fig:sfig6}
\end{suppfigure}

\begin{suppfigure} 
  \centering
  \refstepcounter{supfig}
  \label{fig:sfig7}
\end{suppfigure}

\begin{suppfigure} 
  \centering
  \refstepcounter{supfig}
  \label{fig:sfig8}
\end{suppfigure}

\begin{suppfigure} 
  \centering
  \refstepcounter{supfig}
  \label{fig:sfig9}
\end{suppfigure}

\begin{suppfigure} 
  \centering
  \refstepcounter{supfig}
  \label{fig:sfig10}
\end{suppfigure}

\bibliography{literature_refs}

\end{document}


\title{\textbf{\LARGE{}{Supplementary Information} \\ [0.6em]
\Large{Actin cross-linking organizes basal body patterning through anomalous diffusion transitions}}}

\author[1]{Raghavan Thiagarajan\textsuperscript{\textdagger}}
\author[2]{Younes Farhangi Barooji}
\author[2]{Poul-Martin Bendix}
\author[3]{Mandar M. Inamdar\textsuperscript{\textdagger}}
\author[1]{Jakub Sedzinski\textsuperscript{\textdagger}}
\affil[1]{Novo Nordisk Foundation Center for Stem Cell Medicine (reNEW), Department of Biomedical Sciences, University of Copenhagen, 2200 Copenhagen, Denmark}
\affil[2]{Biocomplexity and Biophysics Section, Niels Bohr Institute, University of Copenhagen, 2200 Copenhagen, Denmark}
\affil[3]{Department of Civil Engineering, Indian Institute of Technology Bombay, Mumbai 400076, India}
\date{}

\maketitle 
\tableofcontents
\clearpage

\phantomsection
\majorheading{Supplementary figures and Movies}
\addcontentsline{toc}{section}{\large\bfseries Supplementary figures and Movies}
\label{sec:suppfigs}

\section{Supplementary figures}

\begin{suppfigure}[H]
  \centering
  \includegraphics[width=\textwidth]{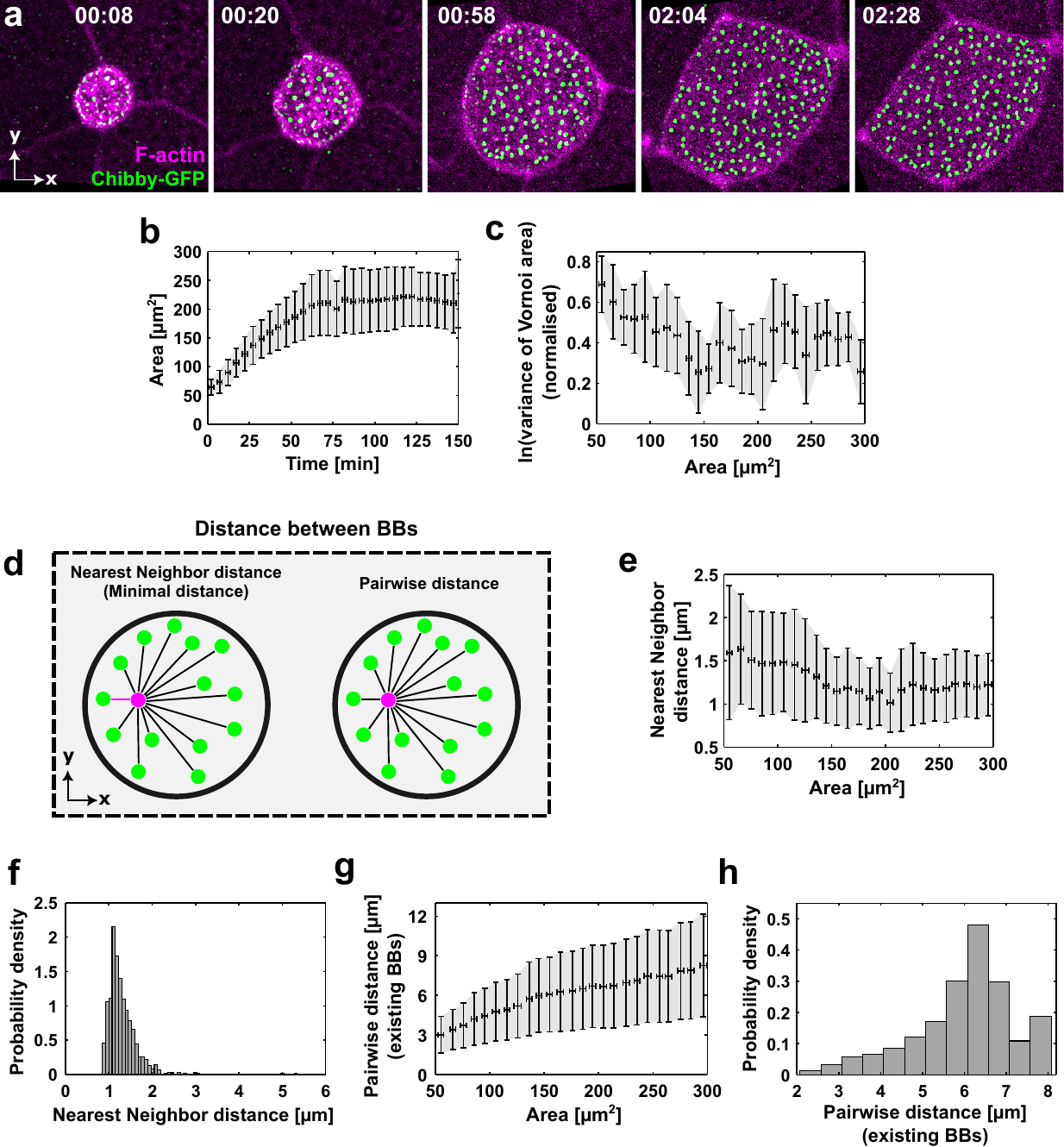}
  \refstepcounter{supfig}
  \label{fig:sfig1}
\end{suppfigure}
\clearpage 
\noindent\textbf{Figure \thesupfig. Basal Body (BB) patterning is synchronized with apical surface expansion.}
All data correspond to coarse-timed imaging (frame interval = 30–60 s; total duration = 2.5 h). (\textbf{a}) Time-lapse sequence of BBs (Chibby-GFP, green) embedded within the actin cortex (LifeAct-RFP, pink) of the expanding apical domain, corresponding to \figref{figonea} (see Movie 1). Time is in hh:mm. (\textbf{b}) Apical area plotted against time. (\textbf{c}) Variance of Voronoi tessellation areas plotted against apical area. Variance values are natural log–transformed (ln) and min–max normalized. (\textbf{d}) Schematic explaining the calculation of minimal (nearest-neighbor) and pairwise distances between BBs. The reference BB (pink) is compared with cohabiting BBs (green); pink line marks the nearest neighbor and the black lines mark the pairwise distances \textit{(see methods section 7.3.4 in SI)}. (\textbf{e}) Nearest-neighbor distance plotted against apical area. (\textbf{f}) Probability density of nearest-neighbor distances with a peak at \SI{1.23}{\micro\meter\squared}. (\textbf{g}) Pairwise distance plotted against apical area. (\textbf{h}) Probability density of pairwise distances with a peak at \SI{6.5}{\micro\meter}. Distances in (\textbf{f}) and (\textbf{h}) were obtained from 9 cells across 5 experiments, covering 1487 frames and $\sim97,000$ BBs (f) and $\sim1,048,000$ BBs (h). Distances were computed for all BBs in each frame, then frame-averaged and pooled over the full 2.5 h time-lapse sequence. In (\textbf{b}), (\textbf{c}), (\textbf{e}), and (\textbf{g}), error bars represent mean $\pm$ standard deviation (SD) of data averaged over 9 cells from 5 experiments, using 5 min (time) or \SI{10}{\micro\meter\squared} (area) bins.

\begin{suppfigure}[p]
  \centering
  \includegraphics[width=\textwidth]{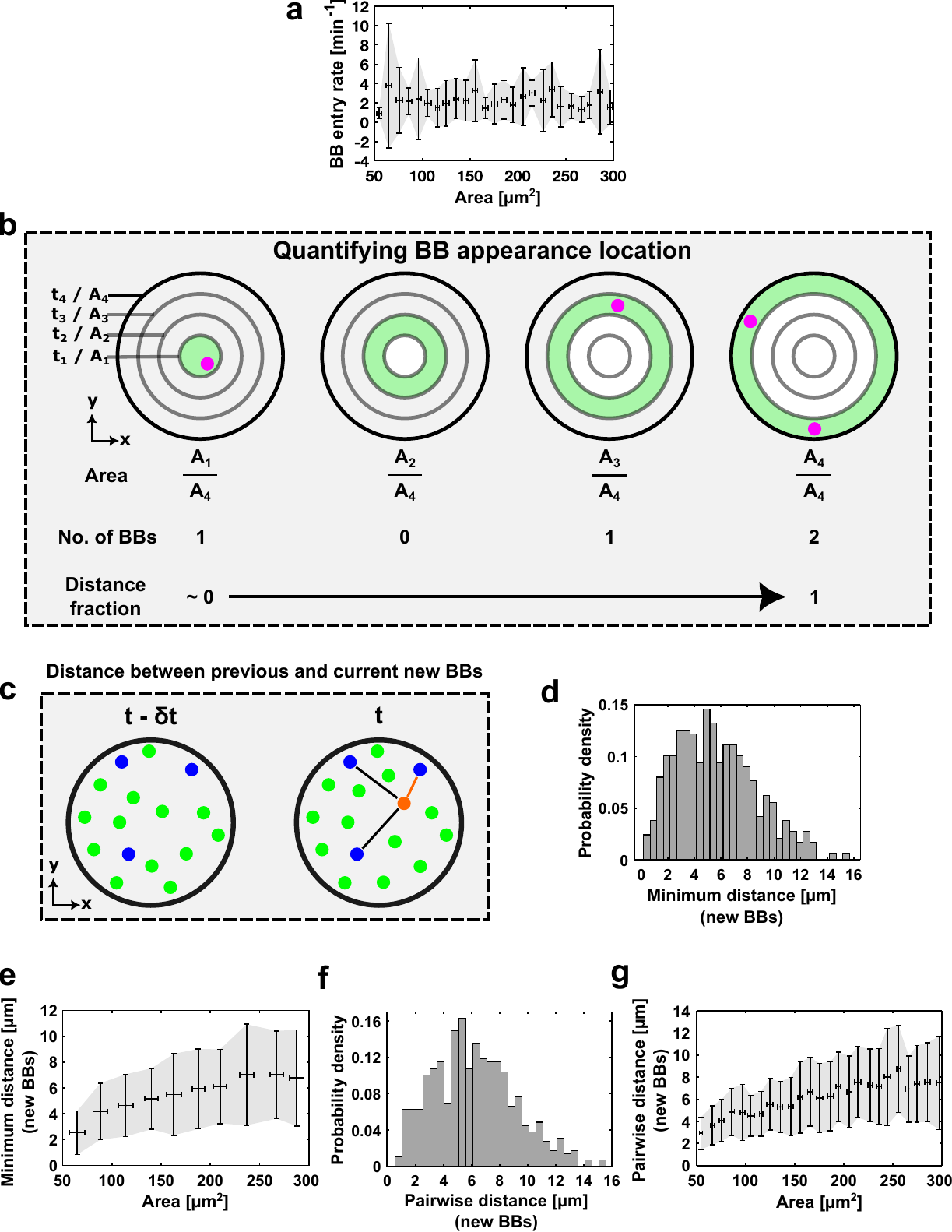}
  \refstepcounter{supfig}
  \label{fig:sfig2}
\end{suppfigure}
\clearpage 
\noindent\textbf{Figure \thesupfig. Passive and active mechanisms shape BB distribution.} All data correspond to coarse-timed imaging (frame interval = 30–60 s; total duration = 2.5 h). (\textbf{a}) Entry rate of BBs plotted against apical area. (\textbf{b}) Schematic showing calculation of BB appearance location, corresponding to \figref{figtwoc}. Black contours represent apical boundaries; the solid black contour marks the boundary at the current time point, whereas semi-transparent contours mark boundaries at previous time points. The green donut-shaped regions between successive contours indicate the newly formed apical area. Pink dots mark BBs. The area, BB count and distance fraction for the current time point are given below the schematic (\textit{see methods section 7.3.6 in SI for description}). (\textbf{c-g}) Distances between newly appearing BBs at consecutive time points. (\textbf{c}) Schematic showing minimal and pairwise distances between new BB at time \(t\) (orange) and at the previous time point $t - \delta t$ (blue). Lines (black and orange) indicate pairwise distances; orange highlights the minimal distance; green shows pre-existing BBs (\textit{see methods section 7.3.4 in SI for description}). (\textbf{d}) Probability density of minimal distances with a peak at \SI{5.3}{\micro\meter}. (\textbf{e}) Minimal distance plotted against apical area. (\textbf{f}) Probability density of pairwise distances with a peak at \SI{6}{\micro\meter}. (\textbf{g}) Pairwise distance plotted against apical area. In (\textbf{a}) and (\textbf{d-g}), data are pooled from 9 cells across 5 experiments. In (\textbf{d-g}), distances were calculated for all BBs in each frame and then frame-averaged before pooling across the full 2.5 h time-lapse sequences of all cells. The plots in (\textbf{d}) and (\textbf{f}) were computed from 851 minimal and 1221 pairwise distances for 575 frames. In (\textbf{a}), (\textbf{e}) and (\textbf{g}), error bars represent mean $\pm$ SD of binned data (\SI{10}{\micro\meter\squared} bins). 

\begin{suppfigure}[p]
  \centering
  \includegraphics[width=\textwidth]{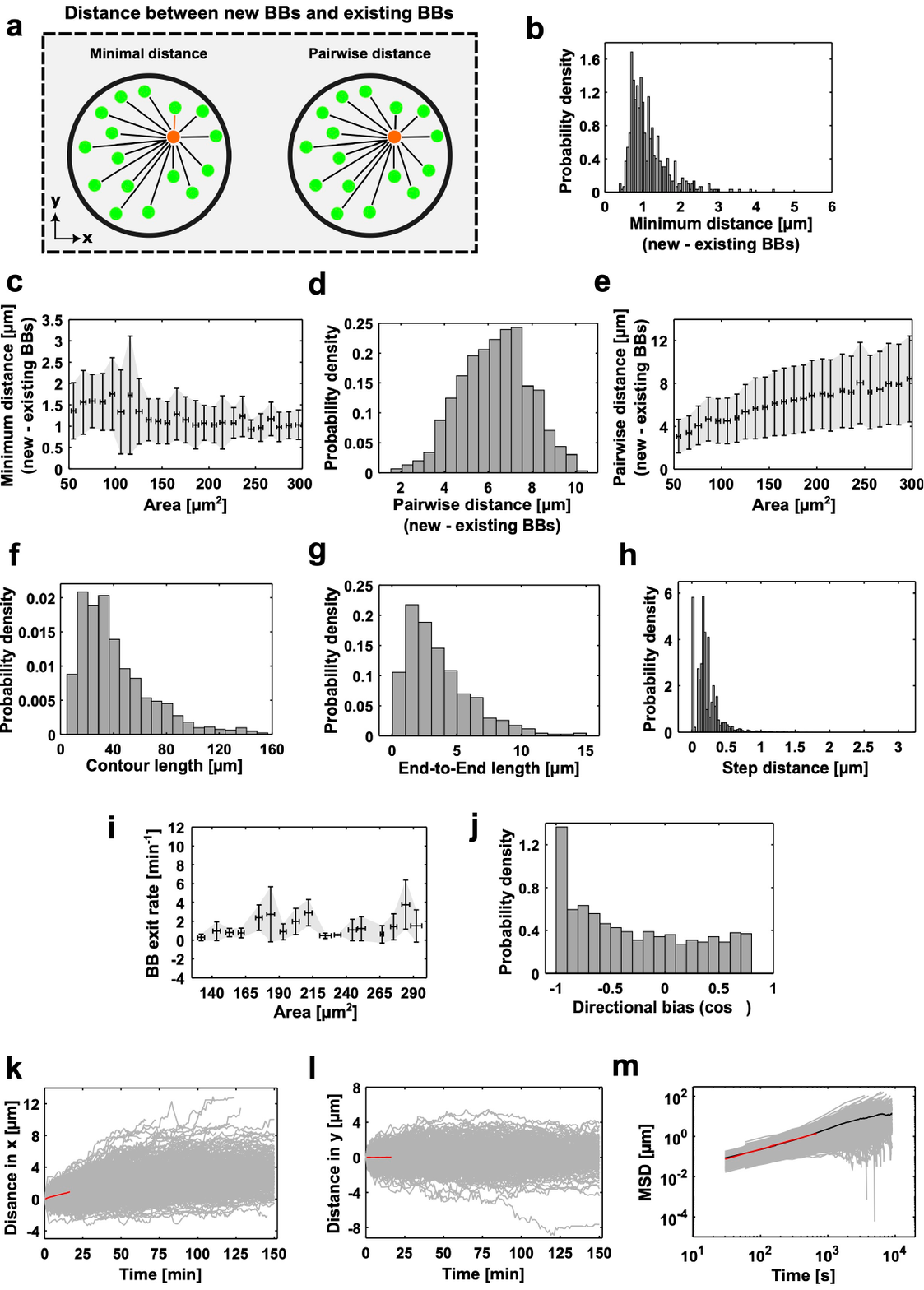}
  \refstepcounter{supfig}
  \label{fig:sfig3}
\end{suppfigure}
\clearpage 
\noindent\textbf{Figure \thesupfig. Passive and active mechanisms shape BB distribution.} All data correspond to coarse-timed imaging (frame interval = 30–60 s; total duration = 2.5 h). (\textbf{a-e}) Distances between new and pre-existing BBs. (\textbf{a}) Schematic showing distances between new BBs (orange) and pre-existing BBs (green). Black lines indicate pairwise distances, and orange highlights the minimal distance (\textit{see methods section 7.3.4 in SI for description}). (\textbf{b}) Probability density of minimal distances with a peak at \SI{1}{\micro\meter}. (\textbf{c}) Minimal distance plotted against apical area. (\textbf{d}) Probability density of pairwise distances with a peak at \SI{6.6}{\micro\meter}. (\textbf{e}) Pairwise distance plotted against apical area. In (\textbf{b-e}), data are pooled from 9 cells across 5 experiments. The distances were calculated for all BBs in each frame and then frame-averaged before pooling across the full 2.5 h time-lapse sequences of all cells. The plots in (\textbf{b}) and (\textbf{d}) were computed from 898 minimal and 39326 pairwise distances for 593 frames. In (\textbf{c}) and (\textbf{e}), error bars represent mean $\pm$ SD of binned data (\SI{10}{\micro\meter\squared} bins). (\textbf{f}) Probability density of contour lengths of BB trajectories. (\textbf{g}) Probability density of end-to-end displacements of BB trajectories. (\textbf{h}) Probability density of BB step sizes. (\textbf{i}) Exit rate of BBs from the apical domain as a function of apical area. The error bars represent mean $\pm$ SD of data averaged over 9 cells from 5 experiments, using \SI{10}{\micro\meter\squared} bins. (\textbf{j}) Probability density of the normalized dot product values quantifying trajectory direction relative to the periphery of the apical domain. (\textbf{k}) Mean x-displacement of all trajectories (grey: coherent drift component) and their ensemble average (red), plotted against time. (\textbf{l}) Mean y-displacements (grey: stochastic component; red: ensemble average), plotted against time. (\textbf{m}) Mean squared displacement (MSD) of trajectories against lag time. Grey curves: individual MSDs; black: ensemble average; red: fit line. A total of 1052 trajectories from 9 cells across 5 experiments were analyzed for (\textbf{f-h}) and (\textbf{j-m}). 

\begin{suppfigure}[p]
  \centering
  \includegraphics[width=\textwidth]{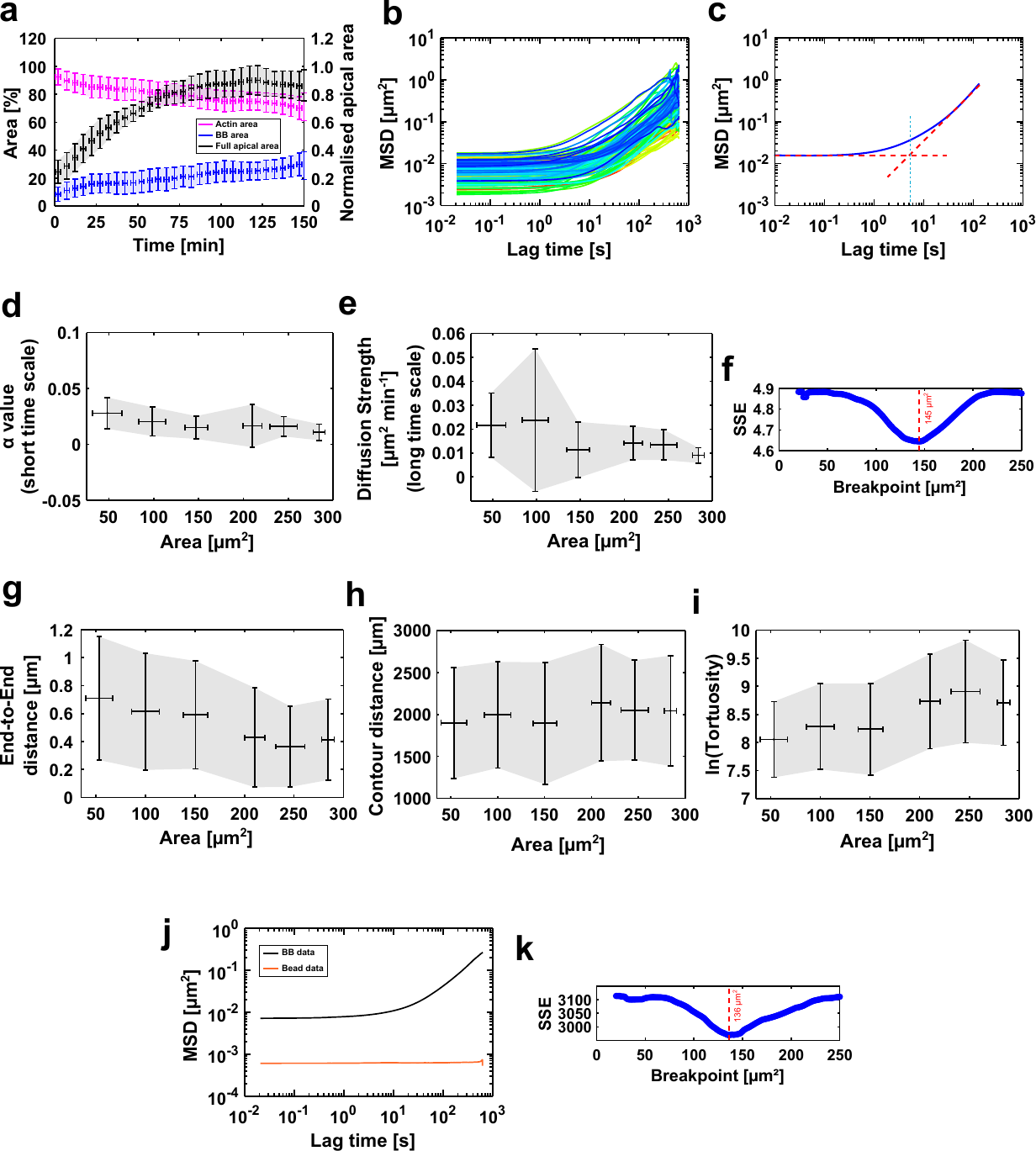}
  \refstepcounter{supfig}
  \label{fig:sfig4}
\end{suppfigure}
\clearpage 
\noindent\textbf{Figure \thesupfig. BB dynamics shift with apical domain expansion.} (\textbf{a}) Apical area percentages plotted against time. Black: total apical area (see \sfigref{sfigoneb}). Blue: area fraction occupied by BBs. Pink: area fraction occupied by actin. Data from coarse-timed imaging (frame interval = 30–60 s; total duration = 2.5 h). Error bars represent mean $\pm$ SD of data averaged over 9 cells across 5 experiments, using 5 min bins. (\textbf{b-k}) Data from fine-timed imaging (frame interval = 21 ms; 30,000 frames; total duration = 10.5 min). (\textbf{b}) Ensemble-averaged MSDs of BB trajectories from 100 cells across 17 experiments, plotted against lag time. (\textbf{c}) Schematic illustrating transition time extraction from MSD: blue - MSD curve; red dotted lines - short- and long-time fits; blue dotted line - intersection (transition time) (\textit{see methods section 8.3.3 in SI}). (\textbf{d}) Short-time $\alpha$ values plotted against apical area. (\textbf{e}) Long-time diffusion strength plotted against apical area. (\textbf{f}) Sum of Squared Errors (SSE) of long-time $\alpha$ values fits (from \figref{figthreec}), plotted across apical area; the red dotted line marks the breakpoint area (lowest SSE). (\textbf{g}) End-to-end displacement plotted against area. (\textbf{h}) Contour length plotted against area. (\textbf{i}) Tortuosity (natural log–transformed) plotted against area. (\textbf{j}) Ensemble-averaged MSDs of BBs (black; 4,604 trajectories from 100 cells) and \SI{0.5}{\micro\meter} beads (orange; 47 beads) (\textit{see methods section 8.3.6 in SI}). (\textbf{k}) SSE of transition time fits (from \figref{figthreee}), plotted across apical area; the red dotted line marks the breakpoint area (lowest SSE). In (\textbf{d, e, g-i}), error bars represent mean $\pm$ SD of data averaged over 4,604 trajectories from 100 cells across 17 experiments, using \SI{50}{\micro\meter\squared} bins.

\begin{suppfigure}[p]
  \centering
  \includegraphics[width=\textwidth]{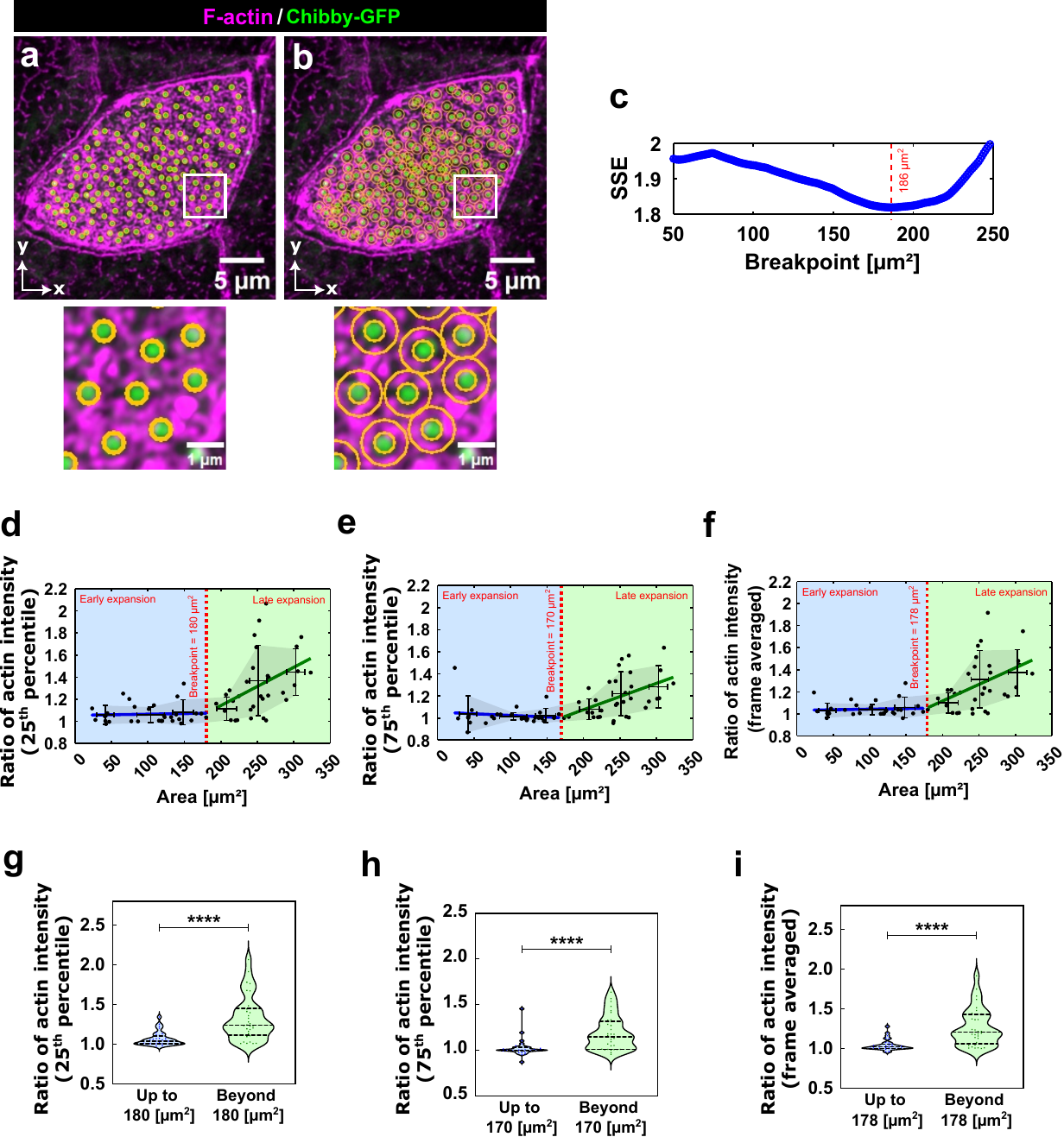}
  \refstepcounter{supfig}
  \label{fig:sfig5}
\end{suppfigure}
\clearpage 
\noindent\textbf{Figure \thesupfig. Progressive formation of an apical actin meshwork facilitates BB redistribution.} All data from high-resolution airyscan imaging. (\textbf{a-b}) Quantification of actin intensity at BB positions (\textbf{a}) and surrounding regions (\textbf{b}); zoomed-in ROIs from white areas are shown below. Orange circular ROI: position of the BB used for quantifying actin intensity at the BB; orange donut ROI: annular region used to quantify actin intensity in the surrounding regions. (\textbf{c}) SSE of actin intensity ratios at $50^{\text{th}}$ percentile of CDF (from \figref{figfourd}), across apical area. Red dotted line highlights the breakpoint area (lowest SSE). (\textbf{d-e}) Ratios of actin intensities (surrounding / BB positions) plotted against apical area at $25^{\text{th}}$ (\textbf{d}) and $75^{\text{th}}$ (\textbf{e}) percentiles of CDF. (\textbf{f}) Ratio of frame-averaged surrounding actin intensity to frame-averaged actin intensity at BB positions. (\textbf{d, e, f}) Each dot represents a single cell and error bars represent mean $\pm$ SD of data averaged over 64 cells from 4 experiments, using \SI{50}{\micro\meter\squared} bins. Red dotted lines mark the breakpoints identified by piecewise linear regression: $25^{\text{th}}$ percentile of CDF - \SI{180}{\micro\meter\squared}; $75^{\text{th}}$ percentile of CDF - \SI{170}{\micro\meter\squared}; frame-averaged - \SI{178}{\micro\meter\squared}, while blue and green lines are regression fits before and after the breakpoints. The blue and green shaded regions denote data before and after the breakpoints and serve as visual guides for the early (<\SI{150}{\micro\meter\squared}) and late (>\SI{150}{\micro\meter\squared}) phases of apical expansion. (\textbf{g-i}) Plots comparing ratios before vs. after breakpoints, using data from 63 cells across 4 experiments. (\textbf{g}) $25^{\text{th}}$ percentile of CDF, corresponding to (d): P < 0.0001; 32 cells (\(\leq\)\SI{180}{\micro\meter\squared}), 32 cells (>\SI{180}{\micro\meter\squared}). (\textbf{h}) $75^{\text{th}}$ percentile of CDF, corresponding to (e): P < 0.0001; 30 cells (\(\leq\)\SI{170}{\micro\meter\squared}), 34 cells (>\SI{170}{\micro\meter\squared}). (\textbf{i}) frame-averaged, corresponding to (f): P < 0.0001; 31 cells (\(\leq\)\SI{178}{\micro\meter\squared}), 33 cells (>\SI{178}{\micro\meter\squared}). In (\textbf{g-i}), the central line represents the median, and the upper and lower lines represent the $75^\text{th}$ and $25^\text{th}$ percentiles, respectively.

\begin{suppfigure}[p]
  \centering
  \includegraphics[width=\textwidth]{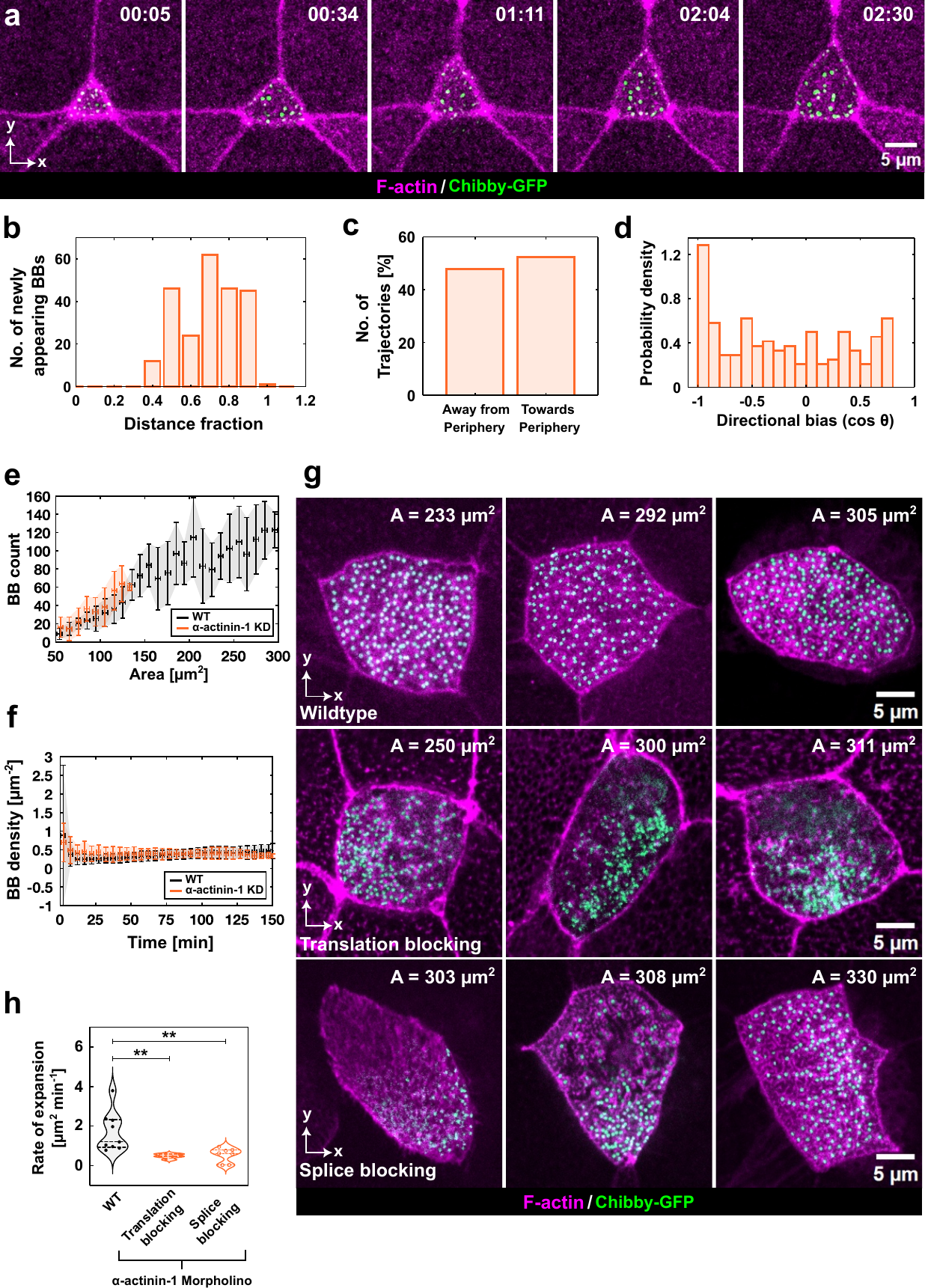}
  \refstepcounter{supfig}
  \label{fig:sfig6}
\end{suppfigure}
\clearpage 
\noindent\textbf{Figure \thesupfig. Disruption of apical actin crosslinking impairs BB organization.} Data in (\textbf{a-f}) and (\textbf{h}) correspond to coarse-timed imaging (frame interval = 30–60 s; total duration = 2.5 h). (\textbf{a-d}) Analyses of $\alpha$-actinin-1 translation-blocking morpholino injected cells. (\textbf{a}) Time-lapse sequence of BBs (Chibby-GFP, green) embedded in apical actin (LifeAct-RFP, pink), corresponding to \figref{figonea}. Time is in hh:mm (see Movie 9). (\textbf{b}) Distribution of newly appearing BBs across distance fractions, where 0 corresponds to the center and 1 to the periphery of the apical domain. (\textbf{c}) Percentage of trajectories moving towards or away from the expanding periphery. (\textbf{d}) Probability density of normalized dot product values quantifying trajectory direction relative to the apical periphery. (\textbf{e-f}) Comparison of WT (black) and $\alpha$-actinin-1 translation-blocking morpholino (orange)–injected cells. Error bars represent mean $\pm$ SD of binned data: \SI{10}{\micro\meter\squared} (area bins) or 5 min (time bins).(\textbf{e}) Number of BBs plotted against time. (\textbf{f}) BB density over the apical area plotted against time. (\textbf{g}) BB distributions in large apical domains for WT (top row), $\alpha$-actinin-1 translation-blocking morpholino (middle row), and $\alpha$-actinin-1 splice-blocking morpholino (bottom row) injected cells. (\textbf{h}) Apical expansion rates compared between WT, $\alpha$-actinin-1 translation-blocking morpholino (exact P = 0.0018), and $\alpha$-actinin-1 splice-blocking morpholino (exact P = 0.0097) conditions. The central line represents the median, and the upper and lower lines represent the $75^\text{th}$ and $25^\text{th}$ percentiles, respectively. In (\textbf{b-d}), data are from 5 cells across 4 experiments. In (\textbf{e-f}) and (\textbf{h}), data are from 9 cells across 5 experiments for WT, 5 cells across 4 experiments for $\alpha$-actinin-1 translation-blocking morpholino, and 7 cells across 3 experiments for $\alpha$-actinin-1 splice-blocking morpholino conditions.

\begin{suppfigure}[p]
  \centering
  \includegraphics[width=\textwidth]{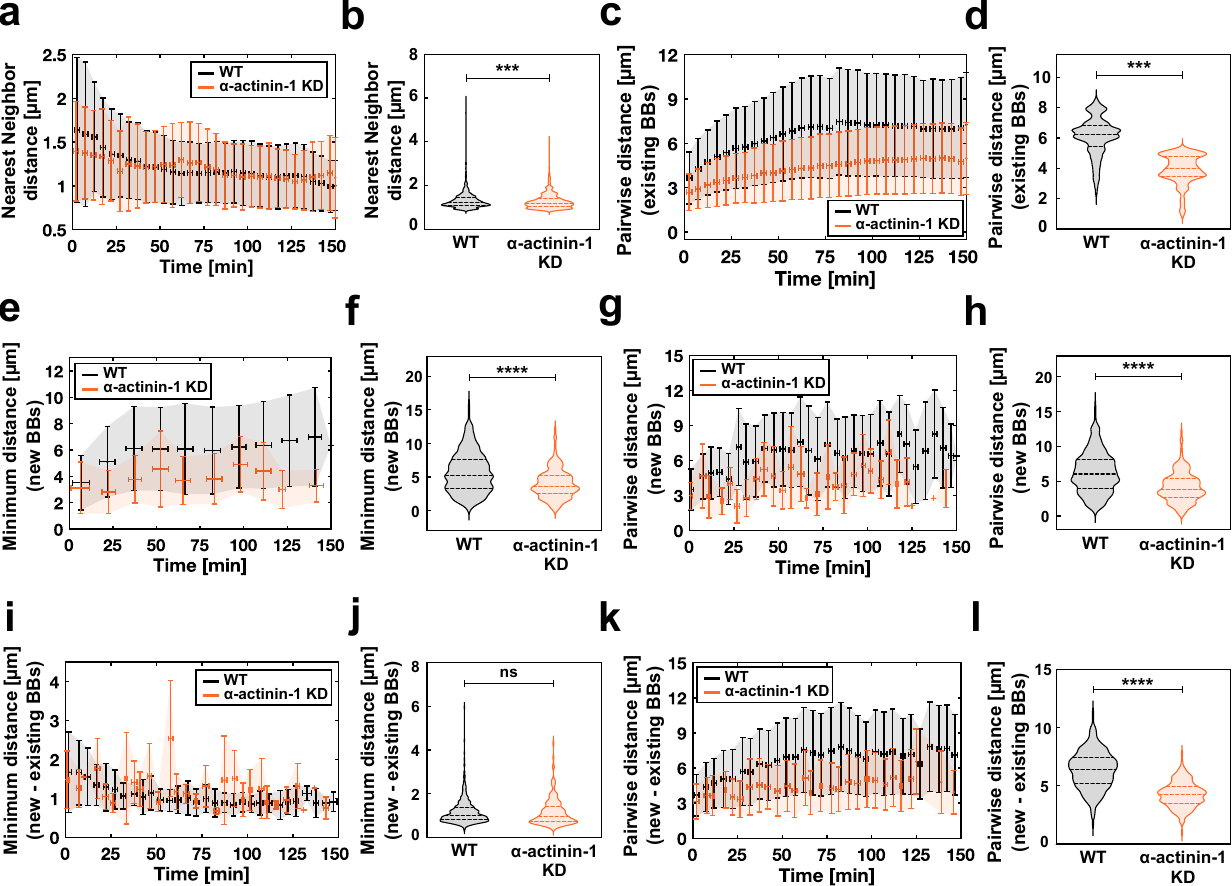}
  \refstepcounter{supfig}
  \label{fig:sfig7}
\end{suppfigure}
\clearpage 
\noindent\textbf{Figure \thesupfig. Disruption of apical actin crosslinking impairs BB organization.} All data correspond to coarse-timed imaging (frame interval = 30–60 s; total duration = 2.5 h). WT (black) and $\alpha$-actinin-1 translation-blocking morpholino (orange)–injected cells are compared and plotted for an experimental time of 2.5 h. (\textbf{a-d}) Distances between BBs in the current frame: (\textbf{a}) nearest neighbor distance evolution; (\textbf{b}) frame-averaged minimal distances; (\textbf{c}) pairwise distance evolution; (\textbf{d}) frame-averaged pairwise distances. (\textbf{e-h}) Distances between newly appearing BBs at consecutive frames: (\textbf{e}) minimal distance evolution; (\textbf{f}) frame-averaged minimal distances; (\textbf{g}) pairwise distance evolution; (\textbf{h}) frame-averaged pairwise distances. (\textbf{i-l}) Distances between new and pre-existing BBs in the current frame: (\textbf{i}) minimal distance evolution; (\textbf{j}) frame-averaged minimal distances; (\textbf{k}) pairwise distance evolution; (\textbf{l}) frame-averaged pairwise distances. In (\textbf{a}), (\textbf{c}), (\textbf{e}), (\textbf{g}), (\textbf{i}) and (\textbf{k}), error bars represent mean $\pm$ SD of binned data (5 min bins). Statistics for violin plots: (\textbf{b}) $\sim97,000$ distances from 1,487 frames (WT) vs. $\sim12,000$ distances from 1,131 frames (KD), P $\approx 0.0004$; (\textbf{d}) $\sim1,050,000$ distances from 1,487 frames (WT) vs. $\sim670,000$ from 1,131 (KD), P < 0.0001; (\textbf{f}) 851 distances from 575 frames (WT) vs. 204 from 177 (KD), P < 0.0001; (\textbf{h}) 1,221 distances from 575 frames (WT) vs. 239 from 177 (KD), P < 0.0001; (\textbf{j}) 898 distances from 593 frames (WT) vs. 146 from 177 (KD), P = 0.1964; (\textbf{l}) 39,326 distances from 593 frames (WT) vs. 4,831 from 177 (KD), P < 0.0001. In all the violin plots, the central line represents the median, and the upper and lower lines represent the $75^\text{th}$ and $25^\text{th}$ percentiles, respectively. Data were obtained from 9 cells across 5 experiments (WT) and 5 cells across 4 experiments (KD).

\begin{suppfigure}[p]
  \centering
  \includegraphics[width=\textwidth]{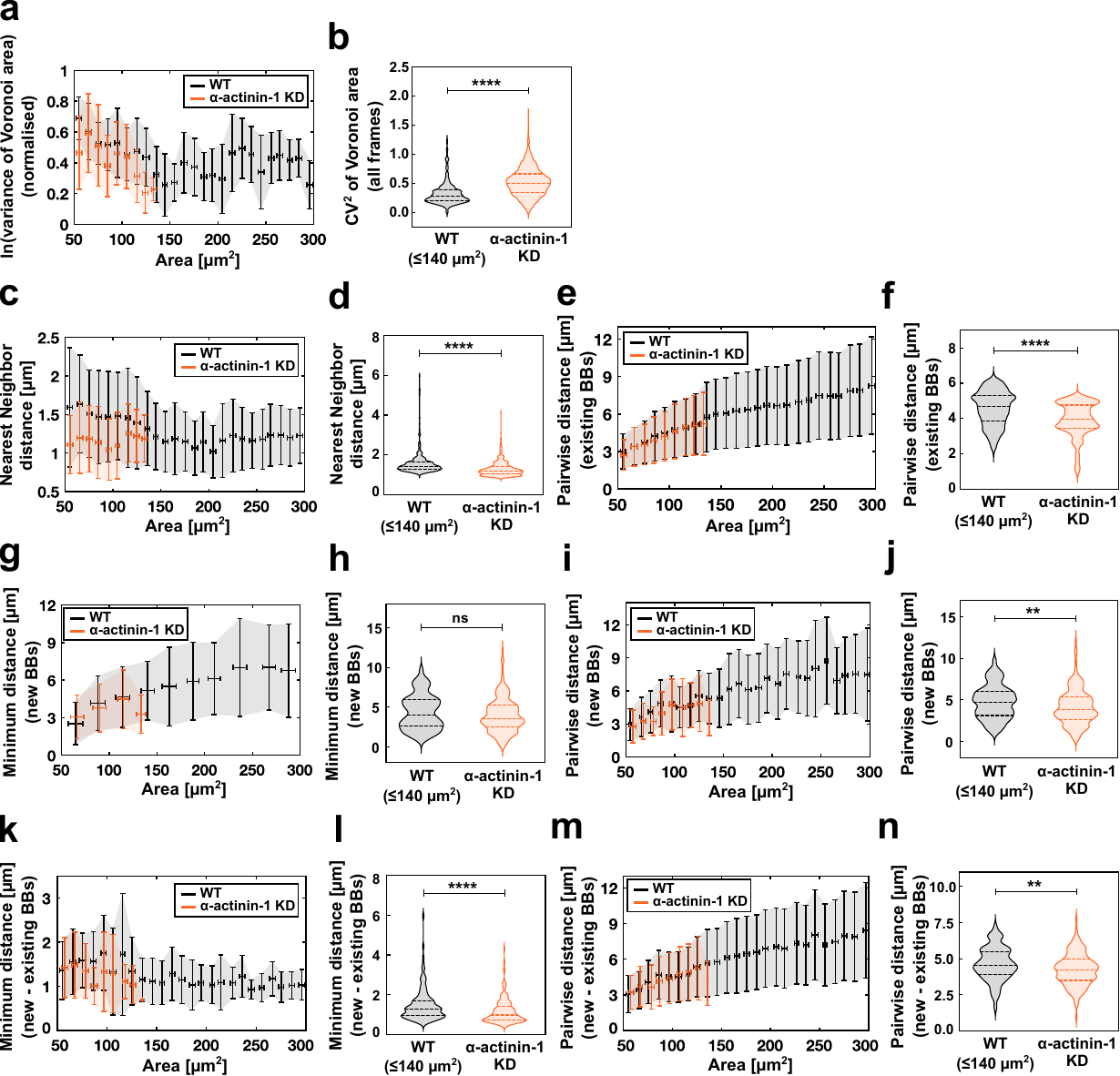}
  \refstepcounter{supfig}
  \label{fig:sfig8}
\end{suppfigure}
\clearpage 
\noindent\textbf{Figure \thesupfig. Disruption of apical actin crosslinking impairs BB organization.} All data correspond to coarse-timed imaging (frame interval = 30–60 s; total duration = 2.5 h). WT (black) and $\alpha$-actinin-1 translation-blocking morpholino (orange)–injected cells are compared as a function of morphogenetic time (apical area). Data for WT in the violin plots are plotted up to the maximum apical area reached in KD (\SI{140}{\micro\meter\squared}). (\textbf{a}) Evolution of variance. Variance values are natural log–transformed (ln) and min–max normalized. (\textbf{b}) Comparison of variability in Voronoi tessellation areas pooled across all frames (2.5 h). Variability is quantified as $CV^{2}$, the squared coefficient of variation (variance divided by the square of the mean). (\textbf{c-f}) Distances between BBs in the current frame: (\textbf{c}) minimal distance evolution, (\textbf{d}) frame-averaged minimal distances, (\textbf{e}) pairwise distance evolution, (\textbf{f}) frame-averaged pairwise distances. (\textbf{g-j}) Distances between newly appearing BBs across consecutive frames: (\textbf{g}) minimal distance evolution, (\textbf{h}) frame-averaged minimal distances, (\textbf{i}) pairwise distance evolution, (\textbf{j}) frame-averaged pairwise distances. (\textbf{k-n}) Distances between new and pre-existing BBs in the current frame: (\textbf{k}) minimal distance evolution, (\textbf{l}) frame-averaged minimal distances, (m) pairwise distance evolution, (\textbf{n}) frame-averaged pairwise distances. In (\textbf{a}), (\textbf{c}), (\textbf{e}), (\textbf{g}), (\textbf{i}), (\textbf{k}), and (\textbf{m}), error bars represent mean $\pm$ SD of binned data (\SI{10}{\micro\meter\squared} bins). Statistics for violin plots: (\textbf{b}) variances from 428 frames (WT) and 1176 frames (KD). (\textbf{d}) $\sim7,000$ distances from 414 frames (WT) vs. $\sim12,000$ distances from 1129 frames (KD), P < 0.0001; (\textbf{f}) $\sim148,000$ distances from 414 frames (WT) vs. $\sim670,000$ from 1129 (KD), P < 0.0001; (\textbf{h}) 178 distances from 137 frames (WT) vs. 204 from 177 (KD), P $\approx 0.0897$; (\textbf{j}) 294 distances from 137 frames (WT) vs. 239 from 177 (KD), P $\approx 0.0049$; (\textbf{l}) 184 distances from 137 frames (WT) vs. 146 from 177 (KD), P < 0.0001; (\textbf{n}) 2992 distances from 137 frames (WT) vs. 4831 from 177 (KD), P = 0.0025. In all the violin plots, the central line represents the median, and the upper and lower lines represent the $75^\text{th}$ and $25^\text{th}$ percentiles, respectively. Data were obtained from 9 cells across 5 experiments (WT) and 5 cells across 4 experiments (KD).

\begin{suppfigure}[p]
  \centering
  \includegraphics[width=\textwidth]{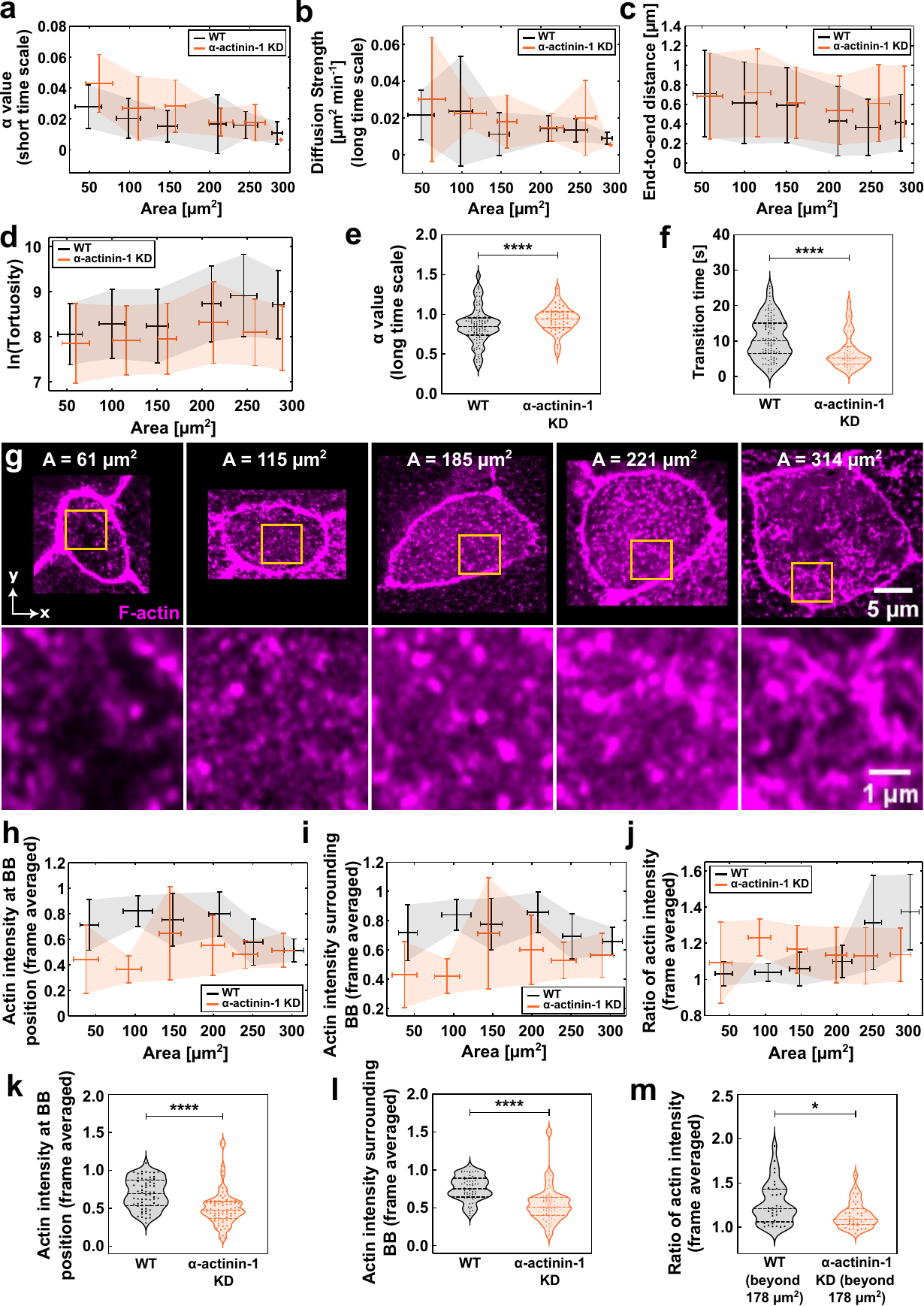}
  \refstepcounter{supfig}
  \label{fig:sfig9}
\end{suppfigure}
\clearpage 
\noindent\textbf{Figure \thesupfig. $\alpha$-actinin-1 perturbation impairs transition in BB dynamics and actin meshwork formation.} Data in (\textbf{a-f}) and (\textbf{h-m}) compare WT (black) and $\alpha$-actinin-1 translation-blocking morpholino (orange) conditions.
Data in (\textbf{a-f}) correspond to fine-timed imaging (frame interval = 21 ms; 30,000 frames; total duration = 10.5 min). (\textbf{a}) Short-time $\alpha$ values versus apical area. (\textbf{b}) Long-time diffusion strength versus apical area. (\textbf{c}) End-to-end displacement versus apical area. (\textbf{d}) Tortuosity (natural log–transformed) versus apical area. (\textbf{e}) Comparison of long-time $\alpha$ values, corresponding to \figref{figsixc}; exact P = 0.0071. (\textbf{f}) Comparison of transition times, corresponding to \figref{figsixd}; P < 0.0001. Error bars in (\textbf{a-d}) represent mean $\pm$ SD of binned data (\SI{50}{\micro\meter\squared} bins). Data in (\textbf{a-f}) include 4,604 trajectories from 100 cells across 17 experiments (WT) and 2,029 trajectories from 46 cells across 5 experiments (KD). Data in (\textbf{g-m}) correspond to high-resolution Airyscan imaging. (\textbf{g}) Images of apical domains of increasing size showing the lack of actin meshwork formation, corresponding to \figref{figsixe, figsixf}. Top row: actin (Phalloidin). Bottom row: zoom of regions highlighted in orange squares in the top row. (\textbf{h-m}) compare the frame-averaged actin intensity at BB position (\textbf{h, k}), frame-averaged surrounding actin intensity (\textbf{i, l}), and ratio of frame-averaged actin intensity between surrounding and BB position (\textbf{j, m}). Data are from 64 cells across 4 experiments (WT) and 72 cells across 3 experiments (KD). (\textbf{h-j}) Compare the evolution across apical area and error bars represent mean $\pm$ SD of binned data (\SI{50}{\micro\meter\squared} bins). (\textbf{k-m}) Violin plots for statistical comparison of: (\textbf{k}) frame-averaged actin intensities at BB position (P < 0.0001); (\textbf{l}) in the surrounding (P < 0.0001); and (\textbf{m}) their ratios beyond the breakpoint (\SI{178}{\micro\meter\squared}, see \sfigref{sfigfivef}), (exact P = 0.0123); data from 33 cells across 4 experiments (WT) and 37 cells from 3 experiments (KD). In all the violin plots, the central line represents the median, and the upper and lower lines represent the $75^\text{th}$ and $25^\text{th}$ percentiles, respectively. 

\begin{suppfigure}[p]
  \centering
  \includegraphics[width=\textwidth]{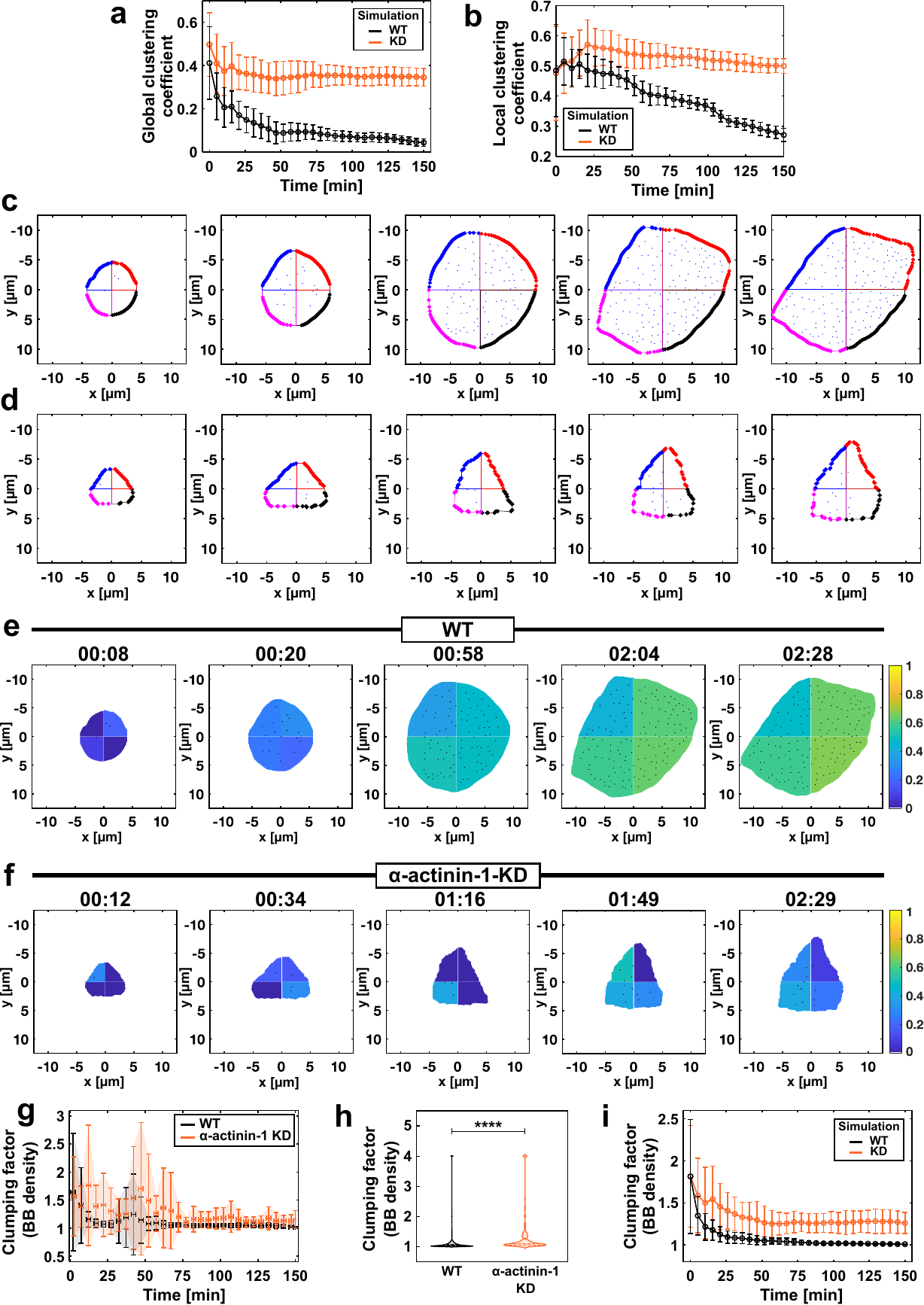}
  \refstepcounter{supfig}
  \label{fig:sfig10}
\end{suppfigure}
\clearpage 
\noindent\textbf{Figure \thesupfig. Heterogeneous distribution of the apical actin network leads to BB clumping.} (\textbf{a,b,i}) Data from the model, comparing WT (black) and KD (orange) simulations. (\textbf{c-h}) Data from coarse-timed imaging (frame interval = 30–60 s; total duration = 2.5 h), comparing WT and $\alpha$-actinin-1 translation-blocking morpholino conditions. (\textbf{a-b}) Time evolution for global (\textbf{a}) and local (\textbf{b}) clustering simulations. (\textbf{c-d}) Time-lapse sequences of apical domains showing quadrant discretization and BBs within each quadrant for WT (\textbf{c}) and KD conditions (\textbf{d}), corresponding to (\figref{sfigtene, figsevenf}) (WT) and (\figref{sfigtenf, figseveng}) (KD) (see Movie 15). (\textbf{e-f}) Time-lapse sequences of apical domains discretized into four quadrants and color-coded for BB density: WT (\textbf{e}) and KD (\textbf{f}) (see Movie 16). (\textbf{g}) Time evolution of the clumping factor for BBs. Error bars represent mean $\pm$ SD of binned data (5 min bins). (\textbf{h}) Comparisons of clumping factors between WT and KD for BBs; data obtained from 1,704 frames (WT) and 1,182 frames (KD), P < 0.0001. The central line represents the median, and the upper and lower lines represent the $75^\text{th}$ and $25^\text{th}$ percentiles, respectively. Data in (\textbf{g}) and (\textbf{h}) were obtained from 9 cells across 5 experiments (WT) and 5 cells across 4 experiments (KD).(\textbf{i}) Time evolution of the clumping factor for BBs from the simulations. In (\textbf{a}), (\textbf{b}) and (\textbf{i}), the error bars represent mean $\pm$ SD of data averaged over 20 simulations, using 5 min bins.


\begin{figure} 
  \centering
  \refstepcounter{figure}
  \label{fig:fig1}
\end{figure}

\begin{figure} 
  \centering
  \refstepcounter{figure}
  \label{fig:fig2}
\end{figure}

\begin{figure} 
  \centering
  \refstepcounter{figure}
  \label{fig:fig3}
\end{figure}

\begin{figure} 
  \centering
  \refstepcounter{figure}
  \label{fig:fig4}
\end{figure}

\begin{figure} 
  \centering
  \refstepcounter{figure}
  \label{fig:fig5}
\end{figure}

\begin{figure} 
  \centering
  \refstepcounter{figure}
  \label{fig:fig6}
\end{figure}

\begin{figure} 
  \centering
  \refstepcounter{figure}
  \label{fig:fig7}
\end{figure}

\section{Movie captions}

\textbf{Movie 1.} Left: BB (Chibby-GFP, green); Right: actin (LA-RFP, pink) and BB, showing the BB docking and distribution during the apical domain expansion in a WT MCC. Time in hh:mm.

\textbf{Movie 2.} 3D visualization of BB (Chibby-GFP, green) ascent from basal to apical side of a WT MCC, coupled with docking to the apical domain during expansion. Time in hh:mm.

\textbf{Movie 3.} Voronoi tessellations of BBs on the apical surface of a WT MCC. Left: data showing BBs (Chibby-GFP, green) used for constructing tessellations; the apical periphery is shown as a white contour. Right: extracted tessellations in pink with BBs as green dots. Time in hh:mm.

\textbf{Movie 4.} BB tracking and trajectory evolution in a WT MCC. The apical periphery is shown as a white contour. Left: data showing BBs (Chibby-GFP, green) used for tracking. Right: white dots represent BBs, and trajectories are shown in diverse colors. Time in hh:mm.

\textbf{Movie 5.} Spatial position of newly appearing BBs in a WT MCC. The apical periphery is shown as a white contour. Left: data showing BBs (Chibby-GFP, green). Right: circles mark the positions where new BBs appear. Time in hh:mm.

\textbf{Movie 6.} Exit and entry events of a BB (Chibby-GFP, green) in a WT apical domain shown from three different views: xy (top), xz (bottom), and yz (right). Yellow arrows in the xz and yz views continuously track the BB both while docked at the apical surface and after exiting beneath the apical domain. The yellow arrow in the xy view appears only when the BB is docked at the apical surface. The animated plot below shows the corresponding Z-position variation of the BB; '0' on the y-axis marks the apical surface. Time in mm:ss.

\textbf{Movie 7.} Fine-timed imaging sequences (frame interval = 21 ms; 30,000 frames; total duration = 10.5 min) of BBs (Chibby-GFP, green) overlaid with trajectories for four different apical domain sizes (WT). For visualization and file-size compatibility, maximum intensity projection was applied to every 30 frames, and the movie was downsized to 1000 frames, resulting in an apparent frame interval of 630 ms over a total duration of 10.5 min. Time in mm:ss.

\textbf{Movie 8.} WT model simulation of basal body distribution during apical domain expansion. BBs are shown as green dots within the expanding circular apical domain boundary (magenta circle). Simulation units are converted to experimental time and spatial scales to enable direct comparison with experiments. Time is shown in min.

\textbf{Movie 9.} Left: BB (Chibby-GFP, green); Right: actin (LA-RFP, pink) and BB, showing the BB docking and distribution during the apical domain expansion of an MCC injected with $\alpha$-actinin-1 translation-blocking morpholino. Time in hh:mm.

\textbf{Movie 10.} 3D visualization of BB (Chibby-GFP, green) ascent from basal to apical side, coupled with docking to the apical domain during expansion in an MCC injected with $\alpha$-actinin-1 translation-blocking morpholino. Time in hh:mm.

\textbf{Movie 11.} BB tracking and trajectory evolution in an MCC injected with $\alpha$-actinin-1 translation-blocking morpholino. The apical periphery is shown as a white contour. Left: data showing BBs (Chibby-GFP, green) used for tracking. Right: white dots represent BBs, and trajectories are shown in diverse colors. Time in hh:mm.

\textbf{Movie 12.} Voronoi tessellations of BBs on the apical surface of an MCC injected with $\alpha$-actinin-1 translation-blocking morpholino. Left: data showing BBs (Chibby-GFP, green) used for constructing tessellations; the apical periphery is shown as a white contour. Right: extracted tessellations in pink with BBs as green dots. Time in hh:mm.

\textbf{Movie 13.} Fine-timed imaging sequences (frame interval = 21 ms; 30,000 frames; total duration = 10.5 min) of BBs (Chibby-GFP, green) overlaid with trajectories for four different apical domain sizes ($\alpha$-actinin-1 translation-blocking morpholino condition). For visualization and file-size compatibility, maximum intensity projection was applied to every 30 frames, and the movie was downsized to 1000 frames, resulting in an apparent frame interval of 630 ms over a total duration of 10.5 min. Time in mm:ss.

\textbf{Movie 14.} KD model simulation of basal body distribution during apical domain expansion. BBs are shown as green dots within the expanding circular apical domain boundary (magenta circle). Simulation units are converted to experimental time and spatial scales to enable direct comparison with experiments. Time is shown in min.

\textbf{Movie 15.} Quadrant discretization of apical domains during expansion, showing BB docking and distribution for WT and $\alpha$-actinin-1 translation-blocking morpholino conditions. Each quadrant is marked by a distinct color, and BBs are shown as blue dots. Time in hh:mm.

\textbf{Movie 16.} Discretized quadrants of apical domains color-coded for BB density and mean actin intensity. Top and bottom rows correspond to WT and KD conditions, while left and right columns correspond to BB density and mean actin intensity, respectively. BBs are shown as black dots. Time in hh:mm.

\clearpage

\clearpage
\phantomsection
\majorheading{Theoretical Model}
\addcontentsline{toc}{section}{\large\bfseries Theoretical Model}
\label{sec:theory}

\section{Dynamics of Basal Body movements at fine-grained time-scales}

The basal bodies (BB) exhibit stochastic movements as per experimental tracking. The tracking was experimentally done in two modes. In the fine-grained mode, the time-interval  between two frames was $\Delta t_{\rm fg} = 20~{\rm ms}$  and the total duration of measurement was $20~{\rm min}$. In the coarse-grained mode, the time-interval was $\Delta t_{\rm cg} = 30~{\rm s}$ and the measurement for done for the entire duration $\approx 150~{\rm min}$ of a typical experiment. 

In the fine-mode, the following features of BB movements were observed. As a function of time-lag $\Delta t$, the mean square displacement of a particular BB could be approximately represented as:
\begin{align}
\mathrm{MSD}(\Delta t) \approx
\begin{cases}
\Delta R^2_0, & \Delta t < \tau,\\
4D\Delta t^\alpha, & \Delta t > \tau.
\end{cases}
\end{align}
This indicated that the basal bodies are caged within a region of size $\Delta R_0$ for time-lags $\Delta t$ smaller than $\tau$, and they undergo anomalous diffusion with exponent $\alpha$ on longer time-scales. The typical time-scales for $\tau$ were experimentally found to be of the order $5-15~{\rm s}$ and the exponent $\alpha $ approximately in the range $0.4 - 1.4$, thus exhibiting both subdiffusion and superdiffusion. 

\begin{figure}[htbp]
    \centering
    \includegraphics[width=0.8\linewidth]{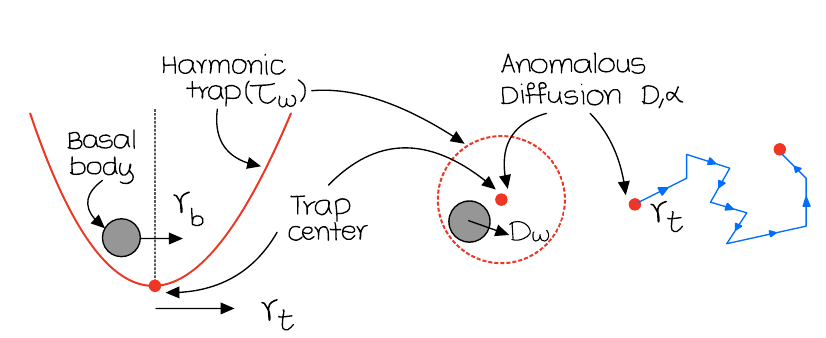}\par
    \refstepcounter{modelfigure}
    \textbf{Figure \themodelfigure.} 
    \justifying  {In the fine-grained model, the early time caged movements of the basal body (BB), with position ${\mathbf r}_b$, within the actin pocket is modeled as a harmonic trap of relaxation time $\tau_w$. The BB diffuses within the trap with diffusivity $D_w$ . The trap center, with position ${\mathbf r}_t$, itself undergoes anomalous diffusion with exponent $\alpha$ and strength $D$.}
    \label{fig:fabrymodel}
\end{figure}

To model this process, we adapted the framework developed by Metzner et al.~\cite{metzner_simple_2007}. We assumed that a BB $i$ is caged inside a harmonic trap with a diffusivity $D_w$ of the BB within the cage with a trap relaxation time of $\tau_w$. The trap itself is modeled to undergo anomalous diffusion with strength $D$ and exponent $\alpha$ (Figure~\ref{fig:fabrymodel}). The equations of motion for the position $\mathbf{r}_{bi}$ and $\mathbf{ r}_{ti}$ of the BB and the center of the harmonic trap are, respectively, given as\\

\FloatBarrier
\begin{align}
    {d \over dt }\mathbf{r}_{bi} &= -{1 \over \tau_w}({\mathbf r}_{bi} - \mathbf{r}_{ti}) + \boldsymbol{\xi}_i(t) \label{eq:diff},\\
    {d \over dt }\mathbf{r}_{ti} &= \boldsymbol{\eta}_i(t) \label{eq:anom},
\end{align}
where $\boldsymbol{\xi}_{i}(t)$  is standard Gaussian noise with zero mean $\langle \boldsymbol{\xi}_i\rangle = \mathbf{0}$ and delta correlation  $\langle \boldsymbol{\xi}_i(t)\cdot\boldsymbol{\xi}_i(t')\rangle = 4D_w \delta(t-t')$ and $\boldsymbol{\eta}_i$ is non-Gaussian noise with zero mean $\langle \boldsymbol{\eta}_i \rangle = \mathbf{0}$ and power-law correlation $\langle \boldsymbol{\eta}_{i}(t)\cdot\boldsymbol{\eta}_{i}(t')\rangle = 4D|t-t'|^\alpha$. We use standard Euler-Maruyama algorithm to integrate these equations as
\begin{align}
    \mathbf{r}_{bi}^{t + dt} &= \mathbf{r}_{bi}^t - {1 \over \tau_w}(\mathbf{r}_{bi}^t - \mathbf{r}_{ti}^t) + \sqrt{2D_wdt}\Delta G(t),\\
    \mathbf{r}_{ti}^{t + dt} &= \mathbf{r}_{ti}^t  + \sqrt{2D}{dt}^H\Delta N(t),
\end{align}
where $H = \alpha/2$ is the so-called Hurst exponent and $dt$ is the time-step of the simulation~\cite{munoz2021objective}. $\Delta G$ is a sequence of random numbers from Gaussian distribution with zero mean and unit standard deviation. $\Delta N(t)$ are the increments obtained from the Davis-Harte process, corresponding to fractional Brownian motion (fBM) that is used to model the anomalous diffusion of the actin trap. Increments for fBM were generated using the circulant embedding method~\cite{davies1987, wood1994}, as implemented via the \texttt{fbm} Python package \cite{flynn2019fbm} and adapted for MATLAB. 

\begin{figure}[htb]
    \centering
    \includegraphics[width=0.5\linewidth]{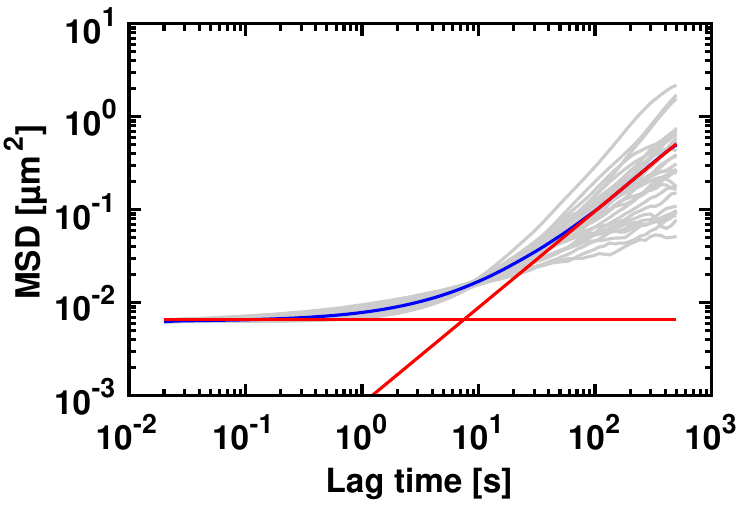}\par
    \refstepcounter{modelfigure}
    \textbf{Figure \themodelfigure.} 
    \justifying MSDs of simulated BBs: grey, individual particles; blue, ensemble average; red lines, fitted regimes; their intersection defines the transition time. Data obtained from $30$ independent runs of the single-particle simulation. The moving trap model described in the text and shown in Figure~\ref{fig:fabrymodel} exhibits commensurate MSD with an initial caging regime followed by anomalous diffusion. The fitted ensemble-averaged MSD yields a plateau $\Delta R_0^2 \sim 10^{-2}~{\rm \mu m^2}$, consistent with $4D_w\tau_w$, and a power-law regime with $D_{\rm eff} \approx 0.014~{\rm \mu m^2\,min^{-\alpha}}$ and $\alpha \approx 1.04$, close to $D_{\rm raw}\cdot t_{\rm ref}^{1-\alpha} \approx 0.017~{\rm \mu m^2\,min^{-\alpha}}$. The intersection of the two fitted red lines gives the caging or elbow time $\tau \approx 7.6~{\rm s} \sim 10~{\rm s}$, consistent with the analytical estimate $\tau = \left(D_w\tau_w/D_{\rm eff}\right)^{1/\alpha} \approx 6.9~{\rm s}$ and the experimentally observed range $5$--$15~{\rm s}$. See Table~\ref{tab:params_fg} for the full list of model parameters.
    \label{fig:MSD_trap}
\end{figure}

To capture the experimentally observed BB-to-BB heterogeneity in anomalous dynamics,  the exponent $\alpha$ is sampled independently for each realisation from a Gaussian distribution with mean $\mu = 0.85$ and standard deviation $\sigma = 0.2$, 
clipped to the range $[0.4, 1.4]$ consistent with the experimentally observed spread. Since $D$ carries units ${\rm \mu m^2\,s^{-\alpha}}$ that depend on $\alpha$, using 
a fixed numerical value across realisations with different $\alpha$ would be physically  inconsistent. We therefore normalise $D$ per realisation as $D = D_{\rm raw} \cdot t_{\rm ref}^{1 - \alpha},$
where $D_{\rm raw} = 0.0161~{\rm \mu m^2/min}$ is the mean diffusion strength fitted  from fine-timed experimental trajectories and $t_{\rm ref} = 10~{\rm s}$ is the mean  elbow transition time measured experimentally. This ensures  ${\rm MSD}(t_{\rm ref}) = 4D_{\rm raw}\,t_{\rm ref}$ is identical across all realisations regardless of $\alpha$. A sample mean square displacement (MSD) obtained using this model is shown in Figure~\ref{fig:MSD_trap}.

The model admits simple analytical insights. In the short-time 
limit $\Delta t \ll \tau_w$, the BB displacement within the trap saturates, 
giving a plateau $\Delta R_0^2 = 4D_w\tau_w$. Equating this with the long-time 
anomalous diffusion MSD $4D\tau^\alpha$ gives the elbow time
\begin{equation}
    \tau = \left(\frac{D_w\tau_w}{D}\right)^{1/\alpha},
    \label{eq:elbow}
\end{equation}
at which the BB transitions from caged to anomalous diffusion. 
The numerical values of these quantities are discussed in the caption of 
Figure~\ref{fig:MSD_trap}.
\FloatBarrier
The mechanism described in Figure~\ref{fig:fabrymodel} is applicable to our system. This is supported by our experimental observations showing that basal bodies are surrounded by an actin network that can effectively trap them. Moreover, the actin network undergoes remodeling as the apical surface evolves and can be effectively thought to undergo stochastic movements that statistically show signature of anomalous diffusion.

\section{Coarse-grained basal body dynamics}
In the coarse-grained experimental mode of BB tracking, the time-interval between two frames,  $\Delta t_{\rm cg} = 30~{\rm s} > \tau \approx 5-15~{\rm s}$, the trapping time-scale observed experimentally. Hence, at this experimental resolution, we do not observe the initial trapping. However, in this case the MSD fits reasonably well with
\begin{equation}
    {\rm MSD}(\Delta t) = 4D \Delta t^\alpha,
\end{equation}
where all the terms have identical interpretation as earlier. A majority of the phenomena reported in our experiments was quantified using coarse-grained imaging in which the initial trapping behavior is not captured. Hence, while modeling the complete BB dynamics, in conjunction with the apical surface growth, we do not incorporate the initial trapping dynamics, but instead focus on the anomalous diffusion BB dynamics. The complete model is detailed below. 

\begin{figure}[htb]
    \centering
    \includegraphics[width=0.4\linewidth]{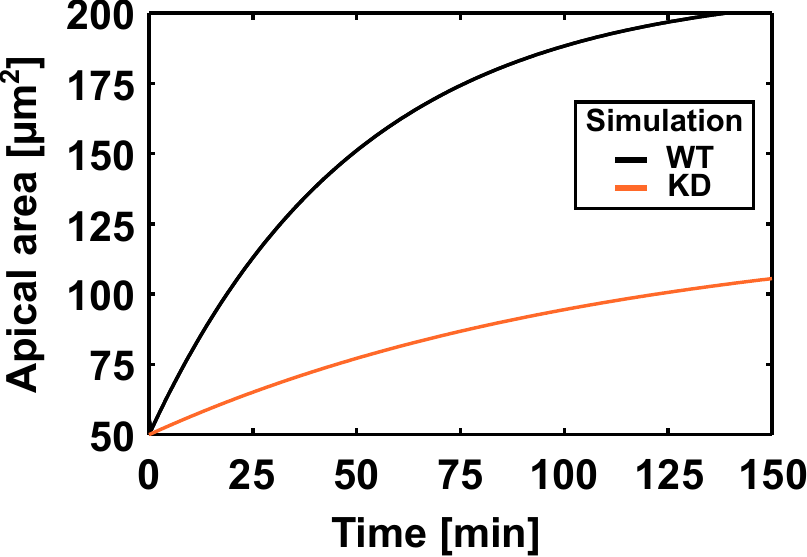}\par
    \refstepcounter{modelfigure}
    \textbf{Figure \themodelfigure.} 
    \justifying The growth of apical surface area (${\rm \mu m^2}$) as a function of time (in mins) for the WT (black) and KD (orange).
    \label{fig:Apical}
\end{figure}

\begin{figure}[htb]
    \centering
    \includegraphics[width=0.4\linewidth]{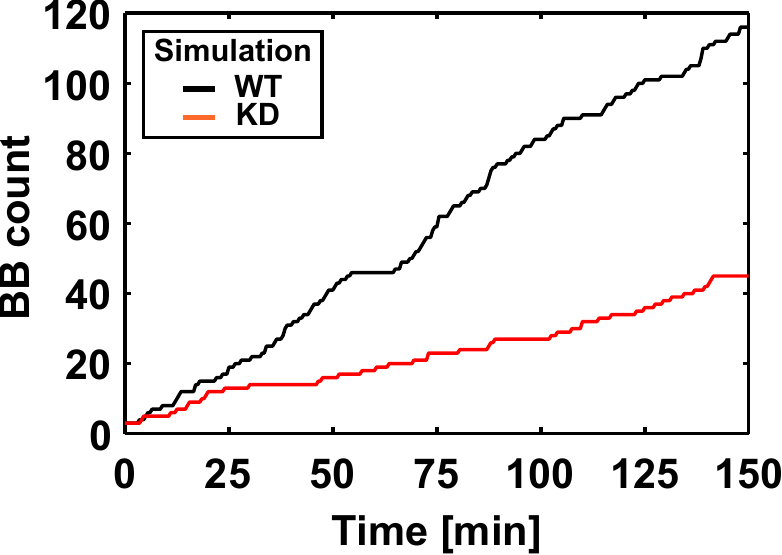}\par
    \refstepcounter{modelfigure}
    \textbf{Figure \themodelfigure.} 
    \justifying The basal bodies steadily appear in the apical surface from the basal surface. The black and red curves, respectively, corresponds to WT and KD.
    \label{fig:BB}
\end{figure}

\begin{figure}[H]
    \centering
    \includegraphics[width=0.8\linewidth]{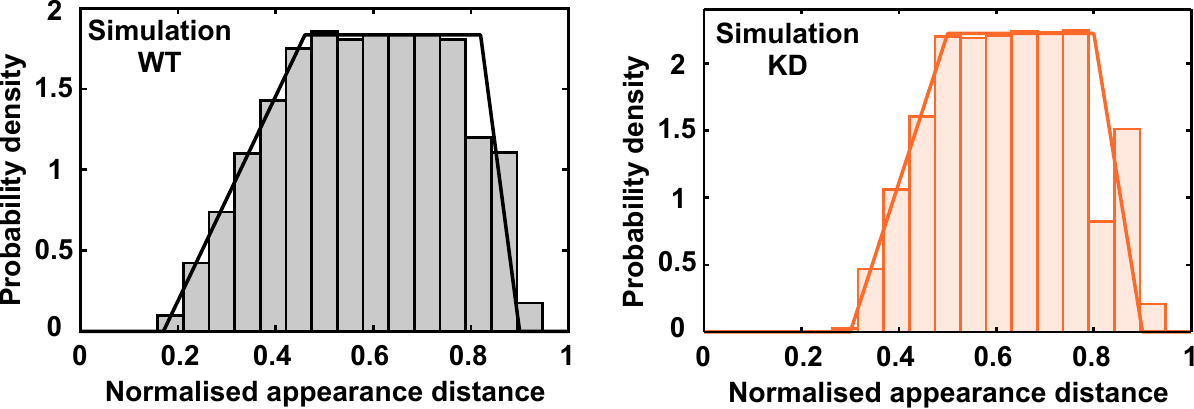}\par
    \refstepcounter{modelfigure}
    \textbf{Figure \themodelfigure.} 
    \justifying {The probability of appearance of basal bodies in the apical surface as a function of normalized radius of the apical surface for WT (left) and KD (right) simulations using a model that closely resembles the experimentally observed basal body appearance. The distribution is trapezoidal: zero for $\rho < a$, rising linearly from $\rho = a$ to $\rho = b$, uniform between $\rho = b$ and $\rho = c$, falling linearly to zero at $\rho = d$, and zero for $\rho > d$. For WT, $(a, b, c, d) = (0.17, 0.46, 0.82, 0.90)$ and for KD, $(a, b, c, d) = (0.30, 0.50, 0.80, 0.90)$, where $\rho$ is the normalised radial distance from the cell centre.}
    \label{fig:Appearence}
\end{figure}

Our computational model simulates the spatiotemporal dynamics of basal body organization in ciliated epithelial cells. The model incorporates several key biological features that govern basal body positioning and clustering behavior during apical emergence. We model that the apical surface of the cell grows dynamically while maintaining its circular geometry throughout the simulation. The surface area increases according to a prescribed time-dependent function that reflects the natural expansion patterns observed experimentally. Figure~\ref{fig:Apical} demonstrates this growth pattern, showing how the apical surface area evolves over time for wild-type (WT) and knockdown (KD) simulations.
\FloatBarrier
Basal bodies appear at the apical surface following a steady recruitment process that occurs at a constant rate throughout the simulation period. This continuous appearance of new basal bodies mimics the temporal dynamics of basal body insertion observed experimentally. As shown in Figure~\ref{fig:BB}, the cumulative number of basal bodies increases linearly with time, reflecting the steady-state recruitment mechanism implemented in our model. The basal bodies do not appear uniformly, but their appearance is biased as observed experimentally. 
\FloatBarrier
The appearance of basal bodies follows a spatially dependent probability distribution that varies with the normalized distance from the cell center. Figure~\ref{fig:Appearence} shows the probability density functions for basal body appearance, revealing the spatial preferences that govern where new basal bodies are most likely to emerge during the simulation.
\FloatBarrier
After appearance, the basal bodies undergo anomalous diffusion as observed in the experiments. To mimic this, each basal body in the simulation is characterized by an anomalous exponent $\alpha$ that is randomly sampled from a Gaussian distribution with mean $\mu = 0.85$ and standard deviation $\sigma = 0.2$ for WT and $\mu = 0.72$ and $\sigma = 0.2$  for KD simulations. This parameter controls the dynamics of movement of basal bodies within the apical surface.  The fBM increments are generated using the Davis-Harte method as described above for the fine-grained case.  
The raw diffusion strength $D_r$ is normalised using a physical reference time $t_{\rm ref} = 30~{\rm s}$ (the coarse-grained frame interval) such that ${\rm MSD}(t_{\rm ref}) = 4D_r\,t_{\rm ref}$ is consistent across all realisations, i.e., $D = D_r \cdot t_{\rm ref}^{1-\alpha}$. Note that $t_{\rm ref} = 30 s$ here is the coarse-grained frame interval, distinct from $t_{\rm ref} = 10 s$ used in the fine-grained model which was the mean elbow transition time. Both serve the same normalisation purpose but are chosen to match the characteristic timescale of each model. When the time-dependent reduction in anomalous dynamics is used, each BB trajectory is generated as three consecutive Davis--Harte segments, each with a fixed Hurst exponent $H=\alpha/2$. The Hurst exponent is reduced between successive segments by a factor of $3/4$, with a lower bound $\alpha_{\min}=0.4$, and the diffusion-strength normalisation is recomputed separately for each segment using the same reference time $t_{\rm ref}$. This provides a piecewise approximation to the progressive confinement observed experimentally, rather than a continuously varying fractional process. 

\begin{figure}[htb]
    \centering
    \includegraphics[width=0.5\linewidth]{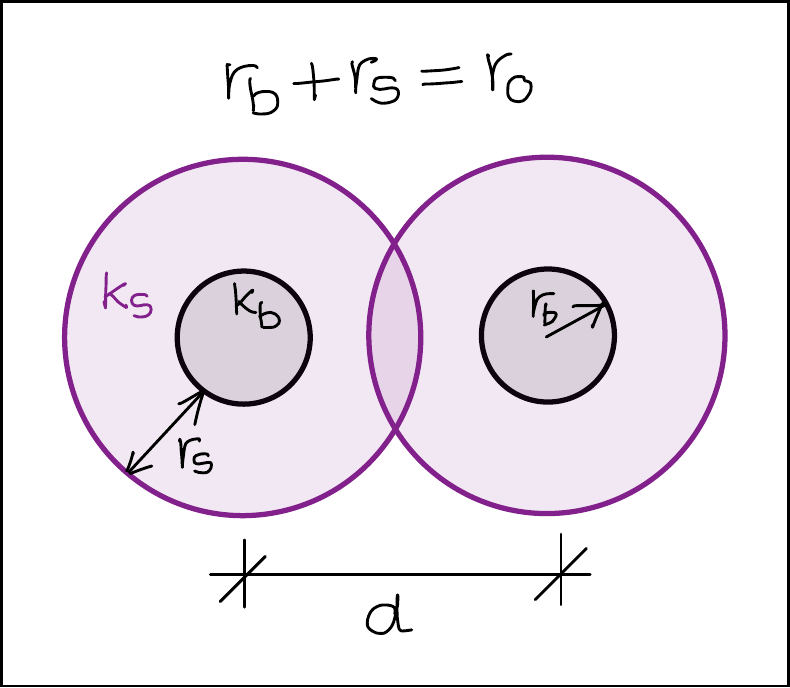}\par
    \refstepcounter{modelfigure}
    \textbf{Figure \themodelfigure.} 
    \justifying Mechanical interaction between two basal bodies. The radius of the dark basal body is $r_b$. The effective thickness of the surrounding actin pocket is $r_s$, so that the outer radius of the pocket is $r_o = r_b + r_s$. The basal bodies interact mechanically with each other when the center-to-center spacing satisfies $d < 2r_o$. The stiffness of the softer actin pocket and the stiffer basal body is $k_s$ and $k_b$, respectively.
    \label{fig:ActinPocket}
\end{figure}

The model incorporates explicit basal body interactions through a self-repulsion mechanism mediated by deformable actin pockets surrounding each basal body. These actin-rich regions create localized zones of altered mechanical properties that influence the positioning and movement of neighboring basal bodies. Figure~\ref{fig:ActinPocket} illustrates the geometric configuration of these interactions, showing how the actin pocket radius and basal body separation distance determine the strength of repulsive forces between adjacent structures. Denoting by $d = |\mathbf{r}_{ij}|$ the center-to-center distance between basal bodies $i$ and $j$, and by $r_o = r_b + r_s$ the outer radius of the actin pocket around a single basal body, the force exerted by basal body $j$ on basal body $i$ is calculated as:
\[
\mathbf{F}_{ij} = 
\begin{cases}
\mathbf{0} & \text{if } d \ge 2r_o \\[1.5ex]
k_{s} (2r_o - d) \dfrac{\mathbf{r}_{ij}}{|\mathbf{r}_{ij}|} & \text{if } 2r_b \le d < 2r_o \\[1.5ex]
\left[2k_{s} r_s + k_{b} (2r_b - d) \right]\dfrac{\mathbf{r}_{ij}}{|\mathbf{r}_{ij}|} & \text{if } 0 \le d < 2r_b
\label{eq:BB_interact}
\end{cases}
\]
\FloatBarrier
The idea is that when $2r_b \le d < 2r_o$, the softer actin pocket first resists the approach of the basal bodies. When the actin pocket of thickness $r_s$ is fully compressed, the stronger self-exclusion resistance between the stiffer basal bodies is activated, with the hard-core branch containing the maximum soft-shell force plus the additional hard-core contribution. Hence, the equation of motion of basal body $i$ is given as:
\begin{align}
    \frac{d}{dt}\mathbf{r}_i = \sum_{j} \mathbf{F}_{ij} + \boldsymbol{\eta}_i(t),
\end{align}
where the anomalous noise is implemented to be the same as in Eq.~\ref{eq:anom}, but here used for the BB directly instead of the trap as was done earlier. ${\mathbf F}_{ij}$ is the force between basal bodies $i$ and $j$ due to interaction shown in Figure~\ref{fig:ActinPocket} and described in Eq.~\ref{eq:BB_interact}. We note that without the actin pockets, the basal bodies can approach each other at distances corresponding to $r_b \approx 0.25~{\rm \mu m}$. This would lead to the observed spacing between the basal bodies to be much lesser than that experimentally observed indicating the presence of an additional steric cushion that naturally maps to actin pockets in our model. 

Basal bodies are constrained to remain within the apical surface. If a basal body strays outside the apical surface of radius $R(t)$, we project it back just inside the current domain, at a small radial offset from the boundary along the local radial direction.

While most of the simulation procedures for the KD condition are similar, the experimental condition introduces a few key modifications to the base model that reflect the altered cellular environment created by $\alpha$-actinin-1 morpholino treatment. These changes capture the essential features of $\alpha$-actinin-1 morpholino induced disruptions to normal basal body organization and positioning.

\begin{figure}[htb]
    \centering
    \includegraphics[width=0.3\linewidth]{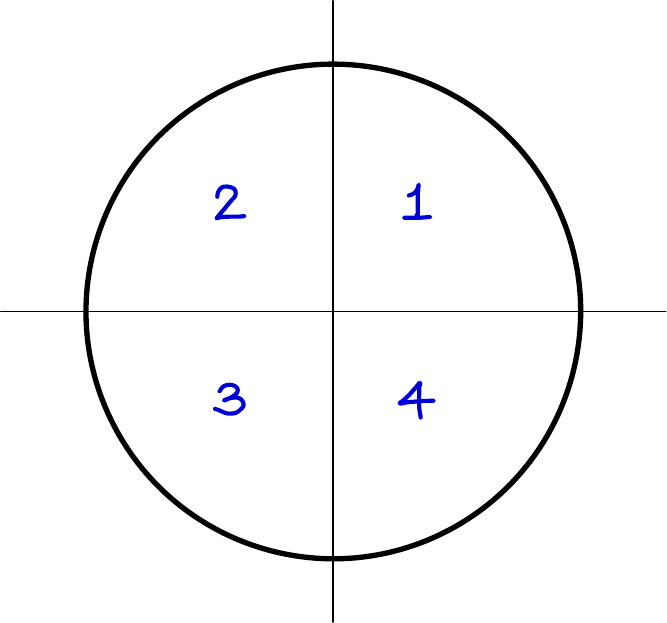}\par
    \refstepcounter{modelfigure}
    \textbf{Figure \themodelfigure.} 
    \justifying {For the morpholino case, we model that the basal bodies appear in quadrant $1$ at any instant with a probability of $0.7$, and with a probability of $0.5$ in the rest of the three quadrants combined.}
    \label{fig:Quadrant}
\end{figure}

Firstly, in the KD simulation, basal bodies exhibit a strong spatial bias in their appearance pattern, preferentially emerging within a single quadrant of the apical surface rather than distributing uniformly across the entire area. This quadrant-specific recruitment pattern reflects the polarized disruption of cellular organization that occurs following $\alpha$-actinin-1 morpholino treatment. Figure~\ref{fig:Quadrant} illustrates this spatial bias by dividing the circular apical surface into four equal sectors and highlighting the preferred region-$1$ of basal body emergence.
\FloatBarrier
Similarly, to accurately capture the localization of basal bodies observed in $\alpha$-actinin-1 morpholino treated cells, the diffusion strength parameter for the basal bodies is reduced by a factor of $5$ throughout the rest of the apical surface, with the exception of the preferred quadrant-$1$. This modification creates a form of mini-confinement that restricts basal body mobility and prevents the normal redistribution processes that would otherwise occur in WT conditions.

Finally, the thickness of actin pockets surrounding individual basal bodies is reduced in the KD model compared to the WT case. This reduction reflects the altered cytoskeletal organization that accompanies $\alpha$-actinin-1 morpholino treatment and affects the local mechanical environment experienced by each basal body and allow the basal bodies to cluster more near each other as compared to the WT. The modified actin pocket geometry influences the strength of inter-basal body interactions. The full list of coarse-grained simulation parameters is given in Table~\ref{tab:params}.

\section{Results}

Our analysis focuses on several key metrics that capture different aspects of basal body organization and dynamics as was done in the experiments. These measurements provide quantitative insights into the differences between the wild-type (WT) and knockdown (KD) conditions. The quantification of various quantities from the simulations closely follows their experimental counterparts. 

\begin{figure}[htb]
    \centering
    \includegraphics[width=0.5\linewidth]{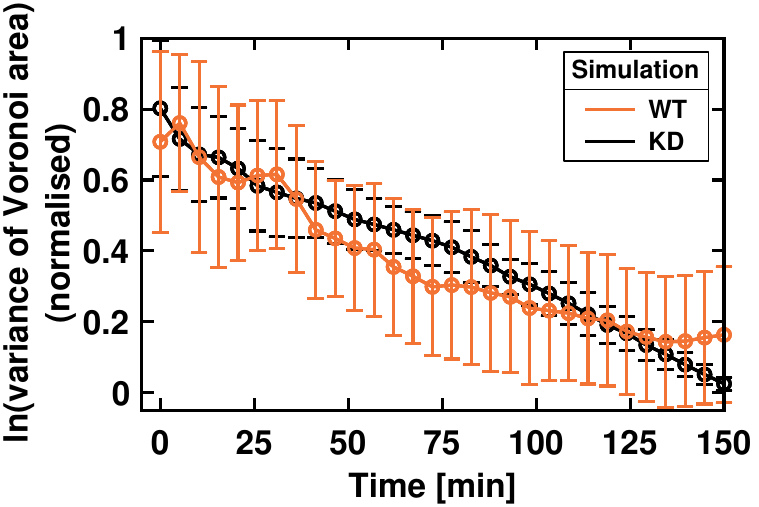}\par
    \refstepcounter{modelfigure}
    \textbf{Figure \themodelfigure.} 
    \justifying Normalized variance in area associated with basal bodies in the apical surface as a function of time for WT and KD cases. The error bars represent mean $\pm$ SD of data averaged over 20 simulations, using 5 min bins. 
    \label{fig:AreaVar}
\end{figure}

\begin{figure}[htb]
    \centering
    \includegraphics[width=0.4\linewidth]{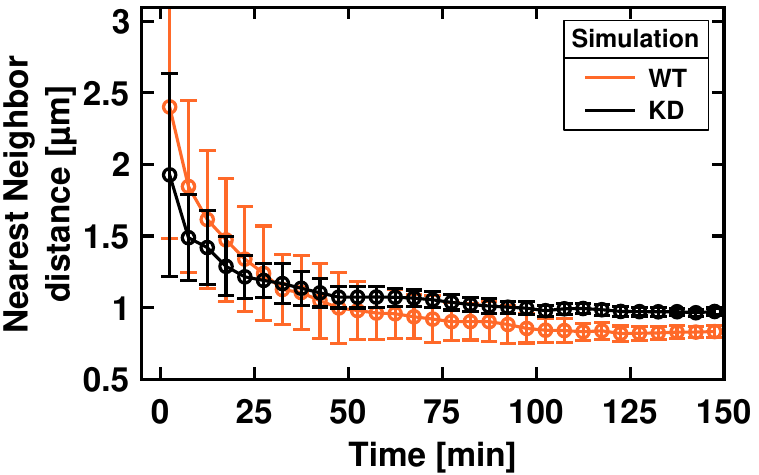}
     \includegraphics[width=0.35\linewidth]{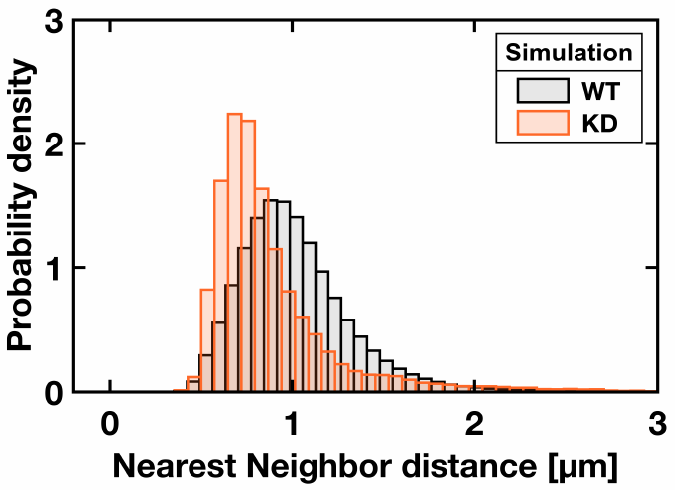}\par
    \refstepcounter{modelfigure}
    \textbf{Figure \themodelfigure.} 
    \justifying (Left) Average minimum spacing $d_{\rm min}$ between basal bodies as a function of time for WT and KD simulations. The error bars represent mean $\pm$ SD of data averaged over 20 simulations, using 5 min bins. (Right) Histogram for $d_{\rm min}$. As expected, $d_{\rm min}$ decreases with time and mimics the experimental observations. Data were obtained from 20 simulation runs.
    \label{fig:dmin}
\end{figure}

\begin{figure}[htb]
    \centering
    \includegraphics[width=0.4\linewidth]{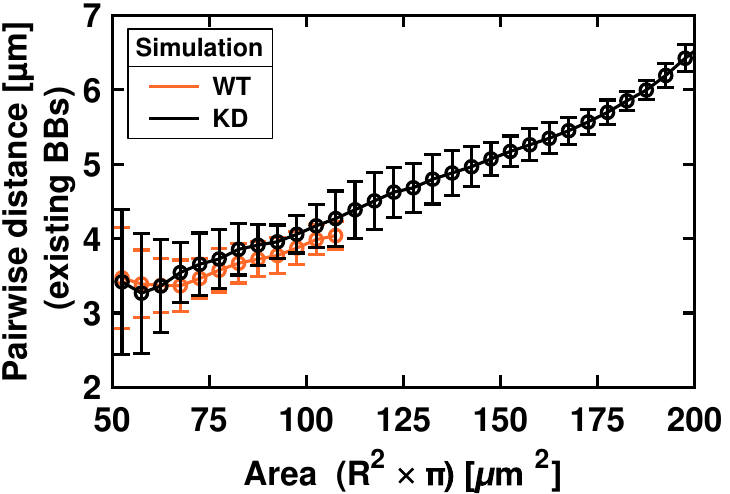}
      \includegraphics[width=0.4\linewidth]{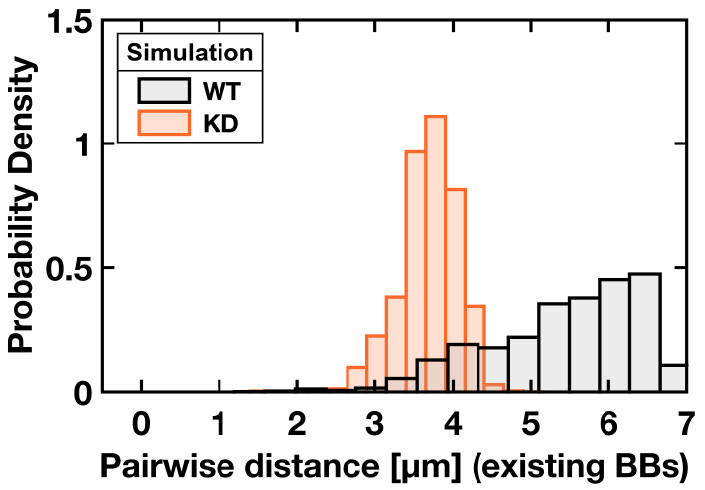}\par
    \refstepcounter{modelfigure}
    \textbf{Figure \themodelfigure.} 
    \justifying (Left) The mean pairwise spacing between the basal bodies at any given instant for WT (black) and KD (orange) based simulations. The error bars represent mean $\pm$ SD of data averaged over 20 simulations, using $5~{\rm \mu m^2}$ area bins. (Right) Histogram for the pairwise spacing. Data were obtained from 20 simulation runs.
    \label{fig:davg}
\end{figure}

\begin{figure}[htb]
    \centering
    \includegraphics[width=0.4\linewidth]{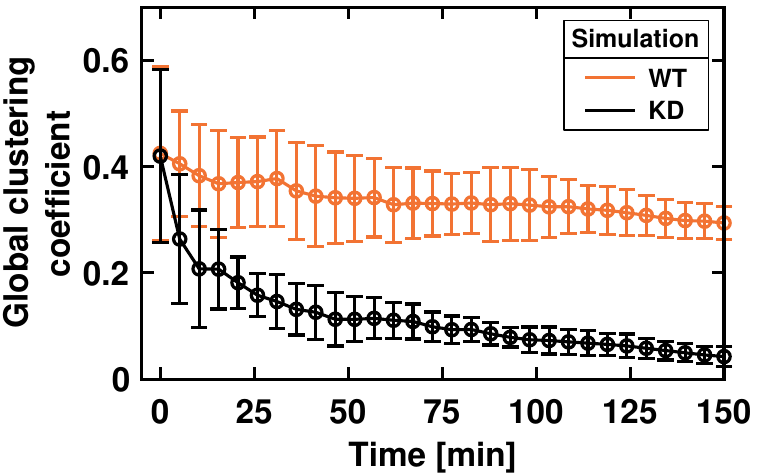}
        \includegraphics[width=0.4\linewidth]{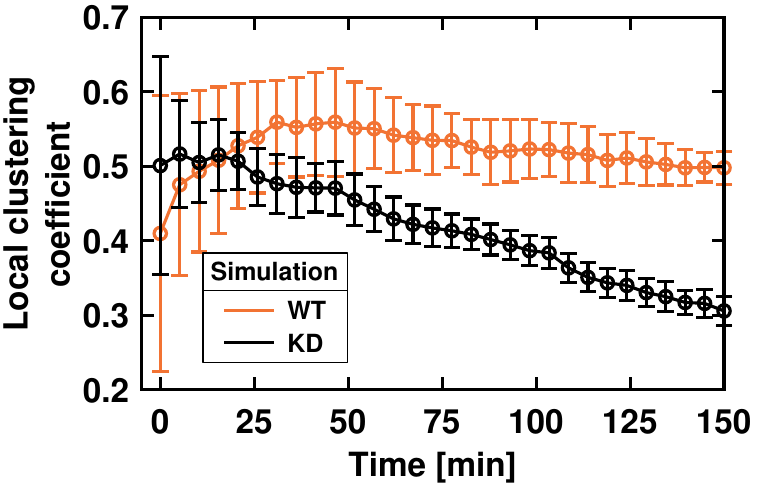}
    \par
    \refstepcounter{modelfigure}
    \textbf{Figure \themodelfigure.} 
    \justifying (Left) Global clustering coefficient for WT and KD based simulations. (Right) Local clustering coefficient for WT and KD based simulations. The error bars represent mean $\pm$ SD of data averaged over 20 simulations, using 5 min bins.
    \label{fig:GC_LC}
\end{figure}

First, we analyze the organization of the basal bodies within the apical surface as a function of time. For example, we would like to quantify if the basal bodies are uniformly organized or more randomly scattered within the apical surface. One of the simplest approaches for that is to quantify the variation in apical area $\tilde{a}$ associated with individual basal bodies -- $\tilde{a}$ is easiest obtained from the Voronoi tessellation of basal body spatial collection and delimited with the apical surface boundary. If $\tilde{a}$ is similar across the basal bodies, it is indicative of their uniform surface distribution. On the other hand, variation in this quantity is indicative of more scattered spatial distribution of the basal bodies. In Figure~\ref{fig:AreaVar}, as done for the experimental data, we plot the normalized statistical variance in $\tilde{a}$ as a function of time for WT and KD. As expected, the variance is higher at earlier times, indicating that the basal bodies are scattered in the apical surface due to their low numbers. However, as the number of basal bodies increase over time, the density assisted mechanical interactions between the basal bodies leads to their getting organized in the apical surface, as reflected in the decreasing normalized variance. Finally, we note that the variance in $\tilde{a}$ for KD case is mostly comparable to that for WT. 
\FloatBarrier
Another way to quantify the organization of basal bodies in the simulations is by calculating $d_{\rm min}$, i.e., the minimum pairwise distance between basal bodies. The average minimum distance between neighboring basal bodies provides insight into the typical spacing that develops within the population. Figure~\ref{fig:dmin} shows how this average minimum distance evolves over time for both conditions and also the full distribution of minimum distances to capture the variability in local spacing patterns. We find that, similar to the experiments, the average value of $d_{\rm min}$ decreases with time and saturates to a value greater than $1.0~{\rm \mu m}$ for the WT case and lesser than $1.0~{\rm \mu m}$ for the KD case. In addition to calculating $d_{\rm min}$, we also quantified the probability distribution for $d_{\rm min}$ across all basal bodies at all times and over multiple ($N = 20$) simulation realisations. As expected, we find that for the KD cases, lower basal body spacings are favored since the actin pockets in KD that resist the pairwise approach of basal bodies towards each other are less developed as compared to the WT. We also note here that the specifics of area variance (Figure~\ref{fig:AreaVar}) are dependent on the kinetics of appearance of basal bodies in the apical surface (Figure~\ref{fig:BB}) and the associated apical surface dynamics (Figure~\ref{fig:Apical}). Similarly, the mechanics of interaction between basal bodies via actin pockets is also instrumental in dictating the spacing between the basal bodies and hence their overall arrangement in the apical surface.
\FloatBarrier
The average pairwise distance $d_{\rm avg}$ between all basal bodies offers a complementary perspective on the global organization of the basal body population. This metric reflects the overall dispersion and clustering tendencies within the system. Figure~\ref{fig:davg} tracks the temporal evolution of average inter-basal body distances and also show the complete distribution of pairwise distances at specific time points. We find that, in general, $d_{\rm avg}$ increases with increasing area. However, the probability distribution for $d_{\rm avg}$ is not uniform, but instead shows a peak at $\approx 6~{\rm \mu m}$ -- the final radius of the apical surface is $\approx 8~{\rm \mu m}$, for reference. The location of this peak is influenced by the basal body appearance probabilities (Figure~\ref{fig:Appearence}). In the case of KD, the peak appears at $\approx 3.5~{\rm \mu m}$ -- the final radius of the apical surface in the case of KD is $\approx 5.6~{\rm \mu m}$ (Figure~\ref{fig:Apical}). In this case, the lower value of the peak, as compared to WT, is indicative of the smaller apical area, preferred clustering in one quadrant, and preferential appearance of basal bodies. 
\FloatBarrier
We do yet another independent analysis to quantify how well-spread globally the basal bodies in the apical surface and if they form any local clusters. Such clustering analysis would provide insight into both local and global organization patterns within the basal body population. We quantify clustering tendencies using appropriate statistical measures that capture the degree to which basal bodies aggregate into distinct groups rather than maintaining uniform spatial distributions. The global clustering coefficient (GC) is calculated as follows
\begin{align}
    GC &= \sqrt{\frac{(\text{Clust}_{\theta})^2 + (1 - \text{Disp}_{r,\text{norm}})^2}{2}},
    \label{eq:GC}
\end{align}
where
\begin{align}
    \text{Clust}_{\theta} &= \left| \frac{1}{N} \sum_{k=1}^{N} e^{i \theta_k} \right| = \frac{1}{N} \sqrt{\left(\sum_{k=1}^{N} \cos \theta_k\right)^2 + \left(\sum_{k=1}^{N} \sin \theta_k\right)^2},\label{eq:GC_theta}\\
    \text{Disp}_{r,\text{norm}} &= \min\left(1, \frac{\sigma_r}{R / \sqrt{18}}\right) \label{eq:GC_r}
\end{align}
The idea is to calculate if the basal bodies are uniformly spread across the apical surface both azimuthally and radially. The azimuthal coefficient ${\rm Clust}_\theta$ (Eq.~\ref{eq:GC_theta}) would be zero if all the basal bodies are uniformly distributed. Similarly, if the basal bodies are radially uniformly distributed in a circle of radius $R$, then the ideal standard deviation of their distribution would evaluate to $R/\sqrt{18}$. Hence, for basal bodies uniformly scattered across the apical surface ${\rm Disp}_{\rm r} \approx 1$.  For such case $GC \approx 0$ indicate the absence of global clustering. In the extreme limit in which all the basal bodies are gathered in a narrow region, then $GC \approx 1$, indicating extreme clustering. On the other hand, even if $GC = 0$, i.e., no global clustering, it could still be possible that there are patches in which the basal bodies form local clusters. Hence, in order to quantify local clustering, we calculate
\begin{align}
    LC &= \max \left( 0, 1 - \frac{d_{\text{obs}}}{d_{\text{hex}}} \right) \label{eq:LC}
\end{align}
where,
\begin{align}
    d_{\text{obs}} &= \frac{1}{N} \sum_{i=1}^{N} \left( \min_{j \neq i} \sqrt{(x_i - x_j)^2 + (y_i - y_j)^2} \right) \\
    d_{\text{hex}} &= R \sqrt{\frac{2 \pi}{N \sqrt{3}}}
\end{align}
Here, the average value of $d_{\rm min}$ is compared with the value expected for hexagonal packing. For regular arrangement, $LC \ll 1$, whereas for clustered patches $d_{\rm obs} \ll d_{\rm hex}$ due to which $LC \approx 1$ (Eq.~\ref{eq:LC}). Figure~\ref{fig:GC_LC} shows the temporal evolution of both global and local clustering coefficients for WT and KD simulations. We find that, as expected, at early times, local and global clustering coefficients are both large for WT as well as KD case. As time-progresses, the global clustering decreases and tends to $0$ for WT as the basal bodies get scattered across the apical surface due to stochastic appearance in the apical surface followed by anomalous diffusion. In the case of KD, since the basal bodies remain confined in one quadrant, the global clustering coefficient remains saturated at a relatively higher value even at later times. 
\FloatBarrier
The local clustering, too is larger at early times for the WT, and decreases slowly to a lower but non-zero value at later times, indicating that the distribution of the basal bodies is getting more ordered but not reached hexagonal distribution, at least within the observed time-window of $150~{\rm min}$. For the KD case, the local clustering too is higher when compared to the WT since the actin pockets are smaller thus facilitating closer approach of basal bodies to each other. 
\begin{figure}[htb]
    \centering
    \includegraphics[width=0.4\linewidth]{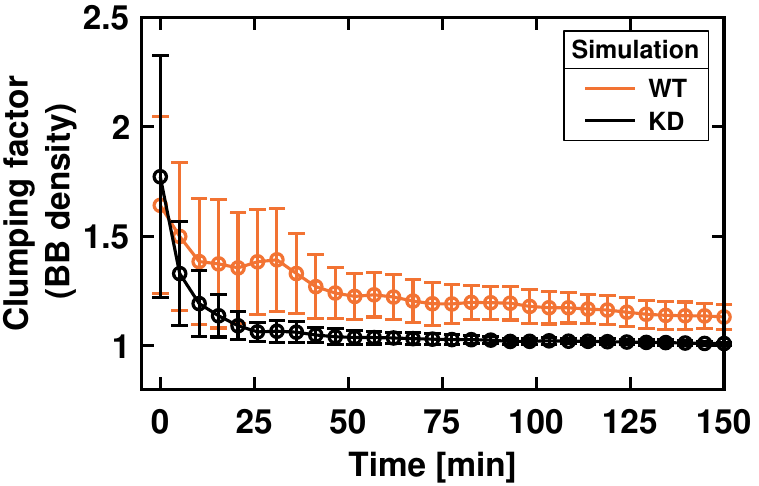}\par
    \refstepcounter{modelfigure}
    \textbf{Figure \themodelfigure.} 
    \justifying Clumping of basal bodies in one quadrant from the model for WT and KD case. The error bars represent mean $\pm$ SD of data averaged over 20 simulations, using 5 min bins.
    \label{fig:clumping}
\end{figure}
\FloatBarrier
The last aspect of our basal body configurational analysis examines clumping behavior to understand how basal body segregation patterns differ between WT and KD simulations. In this analysis, for any time-frame, we obtain $n_1, n_2, n_3, n_4$, the number of basal bodies in quadrants $1, 2, 3, 4$ and then calculate the clumping factor
\begin{align}
    {\rm CF} = 4\frac{n_1^2 + n_2^2 + n_3^2 + n_4^2}{(n_1 + n_2 + n_3 + n_4)^2}
\end{align}
averaged over multiple repeats of simulations $N_{\rm rep} = 20$ and within a time-window $t, t + \Delta t$. Here, if the basal bodies are uniformly distributed across all quadrants, $CF \approx 1$. If all the basal bodies are confined within one quadrant then $CF \approx 4$.  Figure~\ref{fig:clumping} quantifies the degree of clumping over time, showing clear differences in segregation dynamics between the WT and KD simulations. 
\begin{figure}[htb]
    \centering
    \includegraphics[width=0.4\linewidth]{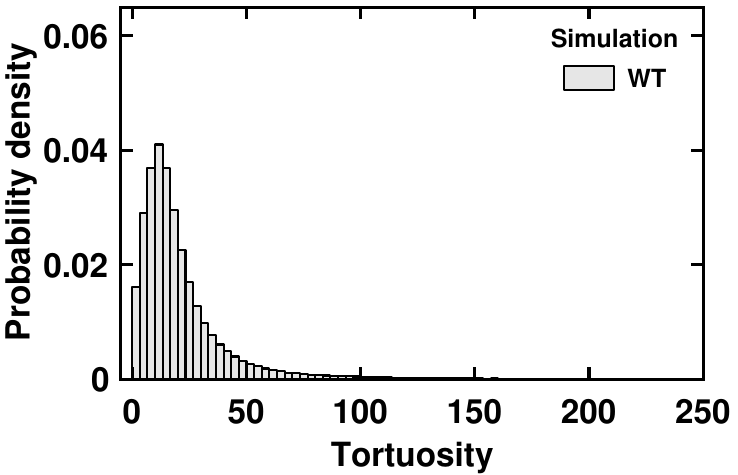}
     \includegraphics[width=0.4\linewidth]{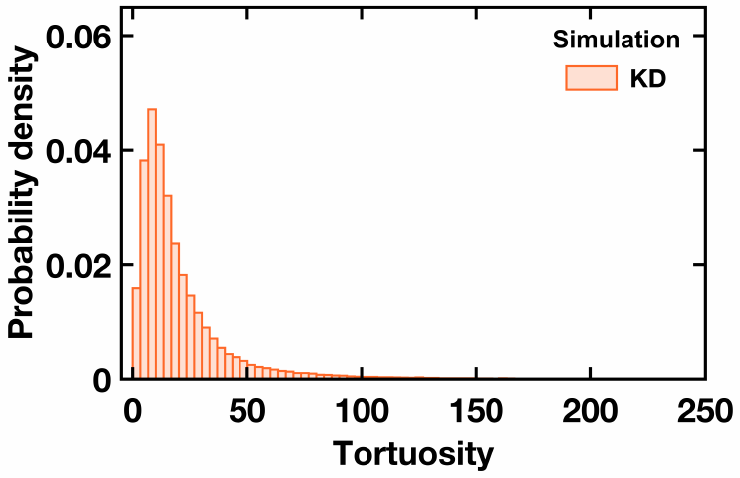}\par
    \refstepcounter{modelfigure}
    \textbf{Figure \themodelfigure.} 
    \justifying Tortuosity of WT (left) and KD (right) probability distribution. Data were obtained from 20 simulation runs.
    \label{fig:tortousity}
\end{figure}
\FloatBarrier
Finally, we also examine the trajectory evolution of individual basal bodies by calculating the tortuosity of their movement paths over time. Tortuosity quantifies the degree to which basal body trajectories deviate from straight-line motion. For a given trajectory 
\begin{align}
    {\rm To}(t) = \frac{\sum_t |\mathbf{r}(t+\Delta t) - \mathbf{r}(t)|}{|\mathbf{r}(t) - \mathbf{r}(0)|}
\end{align}
providing insight into the local constraints and interactions that influence movement patterns. The combined probability distribution of tortuosity values across the population reveals the heterogeneity in movement behavior. Figure~\ref{fig:tortousity} presents the probability distribution of tortuosity values. Interestingly, we find that that the tortuosity distribution for both WT and KD simulations show similar behavior indicating that the differences between WT and KD is likely not governed by basal body movements within the apical surface, but by confinement within domains of apical surface and mechanical interactions between the basal bodies. 

All simulations and postprocessing was performed in MATLAB and the code is publicly available on GitHub~\cite{inamdar2026code}.
\section{Simulation parameters}
\begin{table}[H]
\centering
\caption{Fine-grained model simulation parameters}
\label{tab:params_fg}
\begin{tabular}{|l|l|p{8.5cm}|}
\hline
\textbf{Parameter} & \textbf{Value} & \textbf{Physical meaning} \\
\hline
$D_w$ & $0.3~{\rm \mu m^2/s}$ & Diffusivity of the BB within the harmonic trap; sets the short-time caging MSD plateau $\Delta R_0^2 \approx 4D_w\tau_w$. The Stokes--Einstein value for a sphere (BB) of radius $r_b = 0.25~{\rm \mu m}$ in plain water is $k_BT/6\pi\eta r_b \approx 1~{\rm \mu m^2/s}$ at room temperature; $D_w = 0.3~{\rm \mu m^2/s}$ is $\approx 30\%$ of this, consistent with the BB moving within a viscous actin-rich pocket rather than free solution.\\
\hline
$\tau_w$ & $0.005~{\rm s}$ & Trap relaxation time; sets the short-time plateau via $\Delta R_0^2 = 4D_w\tau_w \sim 10^{-2}~{\rm \mu m^2}$. Chosen so that the plateau is flat well before the earliest plotted lag time ($t = 0.02~{\rm s}$). Note: $\tau_w$ governs short-time caging and is distinct from the elbow time $\tau \sim 10~{\rm s}$ at which the crossover to anomalous diffusion occurs. \\
\hline
$D_{\rm raw}$ & $0.0161~{\rm \mu m^2/min}$ & Mean anomalous diffusion strength of the trap before normalisation; fitted from fine-timed experimental trajectories. \\
\hline
$t_{\rm ref}$ & $10~{\rm s}$ & Reference time for normalising $D$; set to the mean elbow transition time $\tau \sim 10~{\rm s}$ from experiments. The normalisation $D = D_{\rm raw}\cdot t_{\rm ref}^{1-\alpha}$ ensures ${\rm MSD}(t_{\rm ref}) = 4D_{\rm raw}\,t_{\rm ref}$ is identical across realisations with different $\alpha$. \\
\hline
$\alpha$ & $\mu = 0.85$, $\sigma = 0.2$ & Anomalous exponent of the trap (${\rm MSD} \propto \Delta t^\alpha$, $H = \alpha/2$); sampled per realisation from a Gaussian, clipped to $[0.4, 1.4]$. \\
\hline
$T_{\rm max}$ & $25~{\rm min}$ & Total simulation duration, matching fine-grained experimental imaging window. \\
\hline
$dt$ & $0.0005~{\rm s}$ & Integration time step; $dt/\tau_w = 0.0005/0.005 = 0.1 \ll 1$ ensures Euler--Maruyama accuracy for the harmonic trap. \\
\hline
$N_{\rm fun}$ & $30$ & Number of independent realisations over which the MSD is ensemble-averaged. \\
\hline
\end{tabular}
\end{table}
\begin{table}[H]
\centering
\caption{Parameters for coarse-grained simulations}
\label{tab:params}
\begin{tabular}{|l|l|l|p{7.0cm}|}
\hline
\textbf{Parameter} & \textbf{WT} & \textbf{KD} & \textbf{Physical meaning} \\
\hline
$k_{b}$ & $1.0$ & $1.0$ & BB-core repulsion stiffness. Drag coefficient set to unity in the simulation; effective unit upon dimensionalisation is s$^{-1}$. \\
\hline
$k_{s}^{\rm fac}$ & $1000$ & $1000$ & Dimensionless ratio; soft actin-shell stiffness $k_s = k_{b}/k_s^{\rm fac}$, i.e.\ $10^3\times$ softer than the BB core. \\
\hline
$r_o$ & $0.8~{\rm \mu m}$ & $0.4~{\rm \mu m}$ & Outer actin-pocket radius ($r_o = r_b + r_s$); sets the range of soft-shell repulsion. Reduced in KD, reflecting weaker actin pockets after $\alpha$-actinin-1 knockdown. \\
\hline
$r_b$ & $0.25~{\rm \mu m}$ & $0.25~{\rm \mu m}$ & BB hard-core radius; minimum realisable centre-to-centre separation is $2r_b = 0.5~\mu$m. \\
\hline
$D_r$ & $0.0285~{\rm \mu m^2/min}$ & $0.0317~{\rm \mu m^2/min}$ & Anomalous diffusion strength, normalised so ${\rm MSD}(t_{\rm ref}) = 4D_r t_{\rm ref}$ at $t_{\rm ref} = 30~{\rm s}$. Fitted to experimental BB displacement distributions. \\
\hline
$\alpha$ & $\mu = 0.85$, $\sigma = 0.2$ & $\mu = 0.72$, $\sigma = 0.2$ & Anomalous exponent (${\rm MSD} \propto \Delta t^\alpha$); sampled per BB from a Gaussian. Lower mean in KD reflects reduced mobility due to weakened actin cross-linking. \\
\hline
$N_{\rm bb}$ & $120$ & $45$ & Final BB count at saturation, set by experimentally observed values at end of apical expansion. \\
\hline
$T_{\rm sat}^{\rm BB}$ & $140~{\rm min}$ & $130~{\rm min}$ & Time at which BB number reaches $N_{\rm bb}$ from starting value $N_{\rm bs}$; appearance rate $k_{\rm on} = (N_{\rm bb} - N_{\rm bs})/T_{\rm sat}^{\rm BB}$. \\
\hline
$dt$ & $0.0625~{\rm s}$ & $0.0625~{\rm s}$ & Integration time step; chosen so that $k_{\rm bb} \cdot dt \ll 1$, ensuring numerical stability of the stiff BB-core repulsion. \\
\hline
\end{tabular}
\end{table}

%

\clearpage
\phantomsection
\majorheading{Methods}
\addcontentsline{toc}{section}{\hyperref[sec:Methods]{\large\bfseries Methods}}
\label{sec:methods}
The sections below describe the image-processing workflows and data-analysis methods. All image processing and analyses were performed using ImageJ macros and MATLAB, and the corresponding code is publicly available on GitHub~\cite{bb_analysis_RT}.

\section{Image analysis for coarse-timed experiments}
In this section of methods, image analysis performed on coarse-timed datasets acquired by confocal microscopy at 30–60 s intervals over a total duration of 2.5–3 h are described. After the completion of acquisition, only cells meeting the following criteria were selected for analysis: (i) an initial apical surface area of $\leq$ \SI{75}{\micro\meter\squared}, (ii) complete migration of BBs from the basal to the apical surface by the end of the acquisition, and (iii) sustained cellular viability for at least 2.5 h of imaging.

\subsection{Preprocessing of image stacks}
\subsubsection{3D drift correction}
The first preprocessing step consisted of correcting three-dimensional drift across time to ensure spatial alignment of all z-slices. This was performed using the ImageJ plugin Poorman3Dreg (developed by Michael Liebling, University of California Santa Barbara), applying a rigid body transformation with maximum intensity projection based registration.

\subsubsection{Z registration}
Following 3D drift correction, image stacks were cropped in xy to isolate the MCC and remove surrounding tissue. Z registration was then performed to correct for misalignment of z-slices across time points arising from repeated manual adjustments of the z-stack range during acquisition. Such misalignment can result in the apical plane appearing at different z positions over time, leading to inconsistent visualization of BBs during apical expansion. Z registration was achieved using a custom made script that exploits the high actin signal at the apical cortex as a reference. For each time point, the script identifies the z-slice with the highest actin intensity, corresponding to the apical plane, and assigns it to a fixed z index across the entire time series. All other z-slices are repositioned accordingly. This transformation is then applied identically to the BB channel, resulting in actin and BB image stacks that are registered in z and aligned to a common apical reference plane across time. Consequently, both actin and BB channels were required for this registration step.

\subsubsection{Z projection of the apical plane}
After z registration, a z-projection of the apical plane was performed to account for curvature of the apical surface during expansion. In some cells, the apical cortex exhibited slight curvature along the z-axis, with peripheral regions appearing in adjacent z-slices relative to the center of the apical domain. This curvature can introduce variability in BB fluorescence intensity across the apical surface and impair reliable BB detection and tracking. To correct for this effect, a custom made script was used to perform maximum-intensity z-projection over those slices spanning the curved apical plane. The extent of curvature was assessed manually by scrolling carefully through the z-stack, and typically encompassed 3–5 z-slices (corresponding to $\sim$\SIrange{1.2}{2}{\micro\meter}). These slices were selectively projected while leaving all other z-slices unchanged. The resulting projected slice defined a flattened representation of the apical surface. This procedure was applied identically to both actin and BB channels, and the projected apical slice was isolated to generate a 2D time series.

\subsubsection{2D registration of apical surface stacks}
The resulting 2D apical surface time stacks were further registered in xy to remove any residual lateral drift remaining after earlier corrections. Registration was performed jointly on actin and BB channels using the Poorman3Dreg plugin with rigid-body transformation and maximum-intensity projection options.

Unless stated otherwise, all subsequent image processing, quantification, and analysis were performed exclusively on these isolated, maximum-intensity–projected, and 2D-registered apical surface time stacks. Throughout the Methods and main text, references to the “apical surface” or "apical domain" correspond to this processed dataset.

\subsection{Image processing}

\subsubsection{Apical area contour extraction}
The first step of image processing was the identification of the apical boundary, defined as the periphery of the apical surface of MCCs. This boundary was extracted from the actin channel, which clearly delineates the apical surface. To achieve this, contrast across the apical surface was first normalized using the \textit{Enhance Contrast} function in Fiji, applied uniformly across the entire time series. The images were then smoothed using a Gaussian blur with a high sigma value ($\sim{8-12}$) to homogenize intensity across the apical surface. The processed images were subsequently binarized to distinguish the apical surface from the background. These steps were iteratively repeated using different contrast enhancement parameters, Gaussian blur sigma values, and binarization algorithms to obtain an optimally binarized stack with a uniform apical surface. The resulting binary stacks were carefully inspected for defects within the apical domain, which were manually corrected where necessary. ROIs corresponding to the apical domain at each time point were then generated using the \textit{Analyze Particles} function and added to the \textit{ROI Manager} in Fiji. All extracted ROIs were verified against the original actin image stacks used for binarization. Any discrepancies were corrected either by repeating the above processing steps or through manual adjustment until the ROIs were well matched with the actin-defined apical surface.

\subsubsection{Identification of the apical peripheral cortex}
The next step involved extracting ROIs corresponding to the actin rich peripheral cortex within the apical domain. Along the z-axis, the apical side of MCCs is enriched in actin and is referred to as the apical cortex. In contrast, within the plane of the apical surface, the cell periphery exhibits higher actin intensity than the interior of the apical area. This actin enriched region is referred here as the apical peripheral cortex. Throughout this study, intensity measurements within the apical domain, particularly for high-resolution imaging data, were normalized to the actin intensity of the apical peripheral cortex. For simplicity, this normalization is referred to as "normalized by cortex intensity". To generate ROIs corresponding to the apical peripheral cortex, the previously obtained apical domain contour ROIs were radially shifted inward by varying distances to identify the appropriate inner boundary of the peripheral cortex. Once the optimal shift value matching the thickness of the apical peripheral cortex was determined, peripheral cortex ROIs were generated for the entire time series.

The ROIs of the apical domain contour and apical peripheral cortex were applied to the original actin image stacks to extract geometric and intensity-based parameters for each time point, including apical surface area, centroid position, major and minor axis lengths, mean apical actin intensity, as well as the area and actin intensity of the apical peripheral cortex. All analyses, except for contrast enhancement, Gaussian blurring, and binarization, were performed using custom-made scripts.

\subsubsection{BB detection}
BB detection was performed on the apical surface BB channel to enable reliable tracking. A constant intensity offset was first subtracted from the BB channel using the \textit{Subtract} operation under \textit{Process$\to$Math} in Fiji, followed by intensity normalization using \textit{Enhance Contrast}. BBs were then enhanced using either Difference of Gaussians (DoG) or Laplacian of Gaussian (LoG) filtering implemented via the GDSC plugin \cite{etheridge2022gdsc} to improve spot definition. BB positions were identified using the \textit{Find Maxima} function with an empirically determined prominence value and the output set to single-point detections. This resulted in a binarized image stack in which BBs were represented as discrete dots across time. The binarized BB stack was manually verified against the original images, and detection parameters were iteratively refined until maximum correspondence was achieved.

\subsubsection{BB tracking}
BB tracking was performed using a two-step approach consisting of manual tracking followed by automated tracking, with manual tracking used to guide parameter selection for automated analysis.

\paragraph{Manual tracking:} Approximately 5–10 BBs with the longest visible lifetimes were manually tracked in the original BB image stack using the Manual Tracking plugin (by Fabrice Cordeli\`{e}res) in ImageJ. Track overlays and trajectory files containing pixel coordinates were exported. From these data, average step distances were calculated and used to inform automated tracking parameters.

\paragraph{Automated tracking:} Automated tracking was performed on the binarized BB image stack using the Particle Tracker plugin \cite{sbalzarini2005feature} in Fiji. The particle radius (typically 1–3 pixels) was chosen such that each BB was detected as an individual particle in the preview. The link range was set to 3, and the displacement field was defined using the average step distance obtained from manual tracking. Particle dynamics were modeled as Brownian motion. Following tracking, trajectories were visualized using the \textit{Visualize All Trajectories} option. Trajectories were filtered based on minimum duration (number of frames) to retain long-lived trajectories corresponding to persistent BBs. Filter thresholds were empirically chosen such that the number of retained trajectories matched the number of BBs present at the end of apical expansion. Typically, this threshold corresponded to approximately one-third to one-fifth of the total number of frames. Trajectories retained after filtering were verified for close visual agreement with manually tracked BB trajectories. At the end of automated tracking, inflated trajectory numbers were observed and multiple short trajectories corresponding to a single BB were manifesting as individual trajectories. This effect was attributed primarily to fragmentation of long trajectories by the tracking algorithm rather than to spurious particle detection, as BB identification was based on a binarized image stack containing only BB signals. Therefore, to correct this inflation from the inclusion of shorter trajectories, downstream analyses were restricted to those trajectories with sufficient tracking confidence, as determined by their persistence over a minimum number of frames. From the filtered trajectories, BB dynamics were quantified, including directionality, entry and exit behavior, step size distributions, contour length, end-to-end displacement, drift velocity, mean squared displacement (MSD), pairwise distance distributions, and clustering metrics, which are described in the following subsections. 

\subsection{Parameters calculated for the coarse-timed data}
Image-derived quantities were used for downstream quantitative analysis. All downstream analyses were performed using custom-written scripts in MATLAB~\cite{bb_analysis_RT}.

\subsubsection{Morphogenetic and experimental time axes}
For each cell, the apical surface area was measured at every time point using the extracted contour ROIs over the entire apical expansion period ($\sim{2.5}$ h). The experimental acquisition time provided the time axis, whereas the instantaneous apical area was used as a morphogenetic time variable. All parameters were analyzed and plotted as functions of both experimental time and morphogenetic time.

When plotting parameters against apical area, the analysis was restricted to the range between \SI{50}{\micro\meter\squared} and \SI{300}{\micro\meter\squared}. The lower limit of \SI{50}{\micro\meter\squared}  was chosen for two reasons: (1) MCCs that had been inserted but had not yet begun expanding could remain static for extended periods, so starting the image acquisition with cells that had roughly \SI{50}{\micro\meter\squared} ensured higher possibility of apical expansion; (2) in most MCCs, the first BBs reached the apical surface around \SI{50}{\micro\meter\squared} . The upper limit of \SI{300}{\micro\meter\squared} corresponds to the approximate maximum area reached during the plateau of apical surface area expansion in MCCs, although fluctuations were observed during this plateau phase.

To prepare the parameter data for plotting against apical area, the following procedure was applied:
\begin{enumerate}
    \setlength{\parskip}{0pt}
    \setlength{\itemsep}{0pt}
    \item Parameter data corresponding to apical areas $\geq$\SI{50}{\micro\meter\squared} were isolated.
    \item The time corresponding to \SI{50}{\micro\meter\squared} was normalized to zero by subtracting it from all subsequent time points.
    \item Parameter data were clipped at the maximum apical area of \SI{300}{\micro\meter\squared}.
    \item Data were further clipped at the maximum experimental time of 2.5 h.
\end{enumerate}
This procedure ensured that all parameter data corresponding to apical areas between \SI{50}{\micro\meter\squared} and \SI{300}{\micro\meter\squared}, and within the total experimental duration of 2.5 h, were included. It also accounted for fluctuations in apical area during the plateau phase around \SI{300}{\micro\meter\squared}. The resulting adjusted area and time axes were then used for plotting all parameters as functions of apical area and time.

\subsubsection{BB density}
At each time point, the number of BBs was obtained by counting the binarised dots of BBs in the apical surface image. BB density was defined as the ratio of the number of BBs to the corresponding apical area measured from the contour ROI.

\subsubsection{Voronoi tessellation analysis}
Voronoi tessellations were generated from the binarized BB positions using the Voronoi function in Fiji (\textit{Process$\to$Binary$\to$Voronoi}). Tessellation edges extending beyond the apical domain were removed by applying the apical area contour ROI and clearing regions outside the contour. The areas of individual Voronoi cells were then measured using the \textit{Analyze Particles} function in Fiji. To characterize temporal changes in BB spatial organization within a single cell, Voronoi cell areas were grouped into bins corresponding to 10\% increments of apical area expansion and plotted as cumulative distribution functions (CDFs). To quantify changes in BB distribution uniformity over time, the variance of Voronoi cell areas was computed for each frame, yielding one variance value per time point. The temporal evolution of this variance was then averaged across cells. For visualization of temporal trends, Voronoi area variance was log-transformed to reduce skewness and enhance the decreasing trend over time. The log-transformed variance was subsequently normalized using min–max normalization to generate averaged plots with standard deviation. For statistical comparisons between experimental conditions, raw (non-transformed) variance values were used to preserve interpretability of absolute variability. However, variance was normalized by dividing by the square of the mean Voronoi area, equivalent to the squared coefficient of variation. This normalization yields a dimensionless measure, as variance has units of \SI{}{\micro\meter\tothe{4}}, which are cancelled by the squared mean Voronoi area.

\subsubsection{Pairwise and Minimum distance calculations}
To characterize BB spatial organization, three categories of distances were analyzed:
\begin{enumerate}
\setlength{\parskip}{0pt}
\setlength{\itemsep}{0pt}
    \item Distances between BBs within the same frame (\sfigref{sfigoned}); 
    \item Distances between newly appearing BBs in the current frame and newly appearing BBs in the preceding frame (\sfigref{sfigtwoc});
    \item Distances between newly appearing BBs and pre-existing BBs within the same frame (\sfigref{sfigthreea}).
\end{enumerate}
For each category, two distance metrics were computed:
\begin{itemize}
\setlength{\parskip}{0pt}
\setlength{\itemsep}{0pt}
    \item Minimum distance, defined as the shortest Euclidean distance from each BB in one group to its nearest neighbor in the other group;
    \item Pairwise distance, defined as the set of all Euclidean distances between BBs in one group and all BBs in the other group.
\end{itemize}

Distances were computed frame-by-frame. Here, a frame corresponds to a time point, and a group refers to the set of BBs between which distances were calculated. For example, in case (i), both groups consist of BBs within the same frame; in case (ii), the two groups correspond to newly appearing BBs in subsequent frames; and in case (iii), the two groups correspond to newly appearing and pre-existing BBs within the same frame.

\subsubsection{BB entry rate}
BB trajectories were obtained from tracking data, where initiation of a trajectory corresponds to the appearance of a newly docked BB at the apical surface. For each trajectory, the frame of first appearance and spatial position within the apical contour were recorded. The number of newly appearing BBs in each frame was counted, and BB entry rates were quantified by binning these counts either by apical area increments of \SI{10}{\micro\meter\squared} or by time intervals of 5 min, allowing visualization of the temporal evolution of BB entry.

\subsubsection{Spatial bias in BB appearance}
To assess spatial biases in BB appearance during apical expansion, all frames containing newly appearing BBs were identified, along with the corresponding BB coordinates. For each such frame, two quantities were computed (\sfigref{sfigtwob}):
\begin{enumerate}
\setlength{\parskip}{0pt}
\setlength{\itemsep}{0pt}
\item \textit{Area fraction}, defined as the ratio of the apical contour area at a given earlier frame to the apical contour area of the current frame. This ratio corresponds to a concentric annular (“donut”) region representing newly added apical area during expansion.
\item \textit{BB count within area fraction}, defined as the number of newly appearing BBs located within each area fraction.
\end{enumerate}
Because the apical surface expands over time, the contour from each previous frame defines a nested region within the current frame. Thus, for a given frame $n$, all contours from frames $1$ to $n$ are retained. These nested contours partition the current apical surface into concentric regions (annular zones), each corresponding to the incremental area added between successive frames.

To illustrate the procedure, consider a frame in which four new BBs appear (e.g., Frame 4), as illustrated in \sfigref{sfigtwob}, and suppose these new BBs are appearing as follows:
\begin{itemize}
    \setlength{\parskip}{0pt}
    \setlength{\itemsep}{0pt}
    \item $1$ BB within the innermost region defined by contour $A_1$,
    \item $0$ BBs in the region between contours $A_1$ and $A_2$,
    \item $1$ BB in the region between contours $A_2$ and $A_3$,
    \item $2$ BBs in the outermost region between contours $A_3$ and $A_4$.
\end{itemize}
In this frame, the cumulative apical areas are $A_1$, $A_2$, $A_3$, and $A_4$, where $A_4$ represents the total apical area at that time point. The corresponding area fractions are therefore defined as $A_1/A_4$, $A_2/A_4$, $A_3/A_4$, $A_4/A_4$, with associated BB counts: $1$, $0$, $1$, $2$, respectively. Notably, the final area fraction ($A_4/A_4 = 1$) always corresponds to the newest region added during expansion at that frame.
This procedure is repeated for every frame containing newly appearing BBs. As the frame number increases, the number of nested contours also increases, since all prior contours are retained. For example, in a 300 frame acquisition, when analyzing BB appearance in frame 300, the contours from frames 1 to 299 define the concentric regions within which newly appearing BBs in frame 300 are assigned. Consequently, the number of evaluated area fractions progressively increases with frame number.
After processing all frames of a given cell, all pairs of $A_i/A_n$, and $BB~count~in~corresponding~annular~region$ are pooled and binned according to area fraction. The binned area fractions are plotted on the x-axis and the corresponding BB counts on the y-axis (\figref{figtwoc}). The resulting distribution is then interpreted as preferential appearance towards peripheral regions if the area fraction is closer to 1 and towards apical center if the area fraction is closer to 0. For spatial interpretability, the area fraction was additionally transformed into \textit{distance fraction} given by $\sqrt{A_i/A_n}$. Because apical area scales with the square of radial distance, this transformation converts the area fraction into a normalized radial coordinate, where 0 corresponds to the apical center and 1 corresponds to the periphery. Plots of BB count versus distance fraction therefore directly quantify radial bias in BB appearance across the expanding apical domain.

\subsubsection{BB exit rate calculation}
BB exit events were identified as the termination points of BB trajectories. The number of BB exits per frame was counted and binned either by apical area increments of 10 µm² or by 5 min time intervals. Averaged BB exit rates were then plotted to visualize the temporal evolution of BB removal from the apical surface.

\subsubsection{Directionality analysis using dot product}
To quantify the directional bias of BB trajectories, the dot product between the displacement vector $\vec{u}$ and the positional vector $\vec{r}$ was computed for each trajectory:
\[
\vec{u} \cdot \vec{r} = \lvert\vec{u}\rvert \, \lvert \vec{r} \rvert \ \cos\theta
\]
Here, the displacement vector is defined as $\vec{u} = (u_x, u_y) = (x_{\mathrm{end}} - x_{\mathrm{start}},\, y_{\mathrm{end}} - y_{\mathrm{start}})$, and the positional vector is defined as $\vec{r} = (r_x,r_y) = (x_{\mathrm{start}} - x_{\mathrm{centroid}},\, y_{\mathrm{start}} - y_{\mathrm{centroid}})$. The resulting $\cos\theta$ values range from -1 to 1, where negative values indicate trajectories oriented away from the apical periphery and positive values indicate trajectories oriented toward the periphery. The magnitude of $\lvert \cos\theta \rvert$ reflects the strength of directional alignment. Cumulative counts of trajectories with positive and negative $\cos\theta$ values were used to quantify directional biases over time.

\subsubsection{Drift estimation}
To quantify the advective (drift) component of BB motion, the mean velocity vector was computed for each trajectory. The angle of this vector relative to the image x-axis was used to define the trajectory’s principal direction of motion. Each trajectory was then rotated such that its mean direction aligned with the x-axis. This procedure was applied independently to each trajectory, followed by ensemble averaging across all trajectories. The mean displacement along the aligned x-axis captures the coherent drift component, whereas displacement along the orthogonal y-axis reflects stochastic fluctuations around the drift direction.

\subsubsection{Mean squared displacement analysis}
The mean squared displacement (MSD) was computed for each trajectory at multiple lag times $\tau$ as:
\[
\mathrm{MSD}(\tau) = \frac{1}{N_\tau} \sum_{i} \left[ x(t_i + \tau) - x(t_i) \right]^2
\]
where $x(t)$ denotes the BB position, and $N_\tau$ is the number of displacement pairs contributing to lag time $\tau$. For each trajectory, log–log plots of MSD versus lag time were fitted using linear regression over lag times up to 700 s to avoid noise from low-statistics at longer times. The MSD was assumed to scale as $\mathrm{MSD}(\tau) \sim \tau^\alpha$, and the scaling exponent $\alpha$ and intercept c were extracted from $\log(\mathrm{MSD}) = \alpha \log(\tau) + c$. For visualization, MSDs from all trajectories were plotted together with their ensemble average, and the same fitting procedure was applied to the ensemble-averaged MSD.

\subsubsection{Actin and BB area estimation}
BB tracking data were used to generate circular ROIs centered at BB positions for all time points. ROIs were drawn with a radius of $0.25~{\rm \mu m}$, determined from manual measurements of BB diameters across multiple experiments. These ROIs defined BB occupied regions within the apical surface. Using ROI exclusion, the remaining apical area, predominantly containing actin signal, was identified. BB area and actin-dominated area were computed separately at each time point to quantify relative spatial occupancy during apical expansion.

\subsubsection{Clustering coefficients}
Clustering coefficients were used to quantify BB spatial organization at local and global scales. Both coefficients are dimensionless and range from 0 (uniform distribution) to 1 (maximal clustering).

\paragraph{Local clustering (LC):} Local clustering quantifies how closely BBs are positioned relative to their immediate neighbors within a single frame. For each BB, the shortest distance to all neighboring BBs which is referred to as the nearest neighbor distance, is calculated. The local clustering coefficients for a frame is then defined as:
\[
\mathrm{LC} = 1 - \frac{\langle d_{\mathrm{nn}} \rangle}{1/\sqrt{\rho_{BB}}}
\]
where $\langle d_{\mathrm{nn}} \rangle$ is the mean nearest-neighbor distance and $\rho_{BB}$ is the BB density. Here, BB density is defined as the number of BBs divided by the apical area. By normalizing to the inter-BB distance ($1/\sqrt{BB~density}$), LC accounts for differences in cell size and BB number. Values of LC close to 0 indicate a uniform, well-spaced local distribution of BBs, whereas values approaching 1 indicate tight local clustering of BBs.

\paragraph{Global clustering (GC):} Global clustering quantifies the overall spatial organization of BBs within the apical domain, specifically the large scale asymmetry in BB distribution, integrating both angular and radial distributions. For a given frame, GC is calculated as a normalized Euclidean distance combining angular clustering ($Clust_\theta$) and radial dispersion ($Disp_{r,norm}$):
\[
\mathrm{GC} = \sqrt{\frac{{Clust_\theta}^2 + (1-Disp_{r,norm})^2}{2}}
\]
Angular clustering ($Clust_\theta$) is computed from the polar coordinates of the BB positions relative to the apical centroid. For each BB, the angle $\theta$ is calculated, and the angular clustering is defined as:
\[
\mathrm{Clust_\theta} = \left| \frac{1}{N} \sum_{k=1}^{N} e^{i \theta_k} \right|
\]
where values range from 0 (uniform angular distribution) to 1 (all BBs aligned in the same angular direction).

Radial dispersion normalization ($Disp_{r,norm}$) accounts for differences in apical domain size and shape. For each frame, the periphery of the apical area contour is used to generate a mesh grid. Simulated particle positions are randomly distributed within this contour, and their radial distances from the centroid are computed. The standard deviation of these simulated radial distances is used to normalize the actual BB radial distances, producing a frame-specific $Disp_{r,norm}$ value. This ensures that radial dispersion is comparable across frames and cells, regardless of changes in apical area. Thus, GC integrates the normalized angular and radial components, giving equal weight to angular asymmetry and radial clustering. Higher GC values indicate stronger global clustering and asymmetry, whereas lower values indicate a more uniform distribution of BBs across the apical domain.

\subsubsection{Clumping factor}
To compare BB spatial organization with actin intensity distributions, a quadrant-based clumping factor (CF) was computed. The clumping factor was defined as:
\[
\mathrm{CF} = \frac{\langle \rho^2 \rangle}{\langle \rho \rangle^2}
\]
which is computed across four angular quadrants of the apical domain. This metric, commonly used to quantify spatial inhomogeneity, was applied both to BB density and to actin intensity, enabling direct comparison of heterogeneity across matched spatial regions.

\section{Image analysis for fine-timed experiments}
Fine-timed datasets consisted of 30,000 frames acquired over 10.5 min at 1 frame every 21 ms. Tracking of BBs across this large dataset was performed in two stages:
\begin{enumerate}
    \item \textbf{Averaged stack generation}: Maximum intensity projections were performed for every 30 frames of the original stack, producing an averaged stack of 1000 frames.
    \item \textbf{Tracking strategy}: BBs were first tracked in the averaged stack to obtain robust initial tracks. These tracks were then used as references to construct BB trajectories in the original stack. 
\end{enumerate}
All subsequent processing involves both the averaged and original stacks until tracking step is complete.

\subsection{Preprocessing of image stacks}
\subsubsection{Drift correction}
Lateral (xy) drift was corrected to ensure that measured BB movements reflect true dynamics and not stage or sample drift. This is especially important at small time intervals, where even minor drift can introduce apparent BB motion. Drift correction was performed using the StackReg plugin \cite{thevenaz1998pyramid} in Fiji with rigid body transformation. The drift-corrected stack was then maximum intensity projected every 30 frames to generate the averaged stack of 1,000 frames. Both the averaged and original stacks were saved as separate files.

\subsubsection{BB detection}
BBs in both the averaged and original stacks were detected using the ThunderSTORM plugin \cite{ovesny2014thunderstorm} in Fiji. Camera parameters were set as follows: pixel size = 106.667 nm, photoelectrons per A/D count = 5, base level A/D counts = 5, EM gain = 100. The resulting detected BB positions were saved separately for both stacks.

\subsection{Image processing}
\subsubsection{BB tracking}
Tracking was performed using a nearest-neighbor algorithm for both averaged and original stacks.
\begin{enumerate}
    \item \textbf{Averaged stack}: Tracks were constructed by searching for the subsequent BB position in consecutive frames within a radius of 300 nm. If no particle was found, the track was advanced using the previous BB position. This process generated trajectories for all detected BBs.
    \item \textbf{Original stack}: BB tracks from the averaged stack were used as references. For each BB in a track from the averaged stack, a corresponding BB was searched in the original stack within a 300 nm radius. If found, this position was used to extend the track; otherwise, the averaged stack position was used. Particles not found for >5 consecutive frames were excluded. Completed tracks for the original stack were saved for downstream analysis.
\end{enumerate}

\subsection{Parameters calculated for the fine-timed data}
\subsubsection{MSD calculation}
Mean squared displacement (MSD) was calculated as described for coarse-timed data, using only trajectories with $\geq20,000$ steps.

\subsubsection{Alpha value extraction}
MSDs were ensemble averaged across all BBs within a single apical area over 10.5 min. Assuming MSD behaves as $MSD(\tau) \sim{} \tau^\alpha$, linear fits of $log(MSD)$ versus $log(\tau)$ were performed to extract the slope ($\alpha$) and intercept (c). fine-timed MSDs exhibited an initial constrained diffusion phase followed by a generalized diffusion phase (sub-, normal-, or superdiffusive). To capture this, two $\alpha$ values were calculated:
\begin{enumerate}
    \item \textbf{Short-time $\bm{\alpha}$}: Linear fitting was iteratively extended from the smallest lag time until R² > 0.9 or RMSE < 0.02.
    \item \textbf{Long-time $\bm{\alpha}$}: Fitting began immediately after the short-time regime, with maximum lag time set to $\sim{}trajectory~ length/4$ to avoid noisy tails, iteratively extended until R² > 0.99 or RMSE < 0.02.
\end{enumerate}

Both short- and long-time $\alpha$ values were extracted from ensemble averaged MSDs for each cell and plotted across the apical area range.

\subsubsection{Transition time extraction}
The lag time at which BB motion transitions from short time constrained diffusion to a longer time generalized diffusive regime was defined as the transition time. This timescale captures the point at which BB trajectories escape local confinement and begin to exhibit sustained transport or diffusion. To extract this transition time, we used the two power law regimes identified in the MSD analysis. Linear fits to the ensemble averaged MSD in log–log space were performed separately for the short time regime and the long time regime, yielding slopes ($\alpha_s, \alpha_l$) and intercepts ($y_s, y_l$), respectively. The transition time ($\tau_{transition}$) was then defined as the lag time at which these two fitted lines intersect (\sfigref{sfigfourc}). The intersection point along the lag time axis was calculated as:
\[
\tau_{transition} = \frac{y_l - y_s}{\alpha_s - \alpha_l},
\]
where $y_s$ and $\alpha_s$ correspond to the short-time fit and $y_l$ and $\alpha_l$ to the long-time fit. This intersection provides an objective estimate of the timescale at which BB dynamics change regime.

\subsubsection{Diffusion strength calculation}
The long-time MSD behavior of BB trajectories exhibited a broad range of dynamical regimes in an area dependent manner, including subdiffusive, diffusive, and superdiffusive motion, with $\alpha$ values spanning approximately 0.2–1.6 in both fine- and coarse-timed data (see figure below). Under such conditions, the conventional diffusion coefficient D, obtained from the generalized MSD relation
\[
MSD(\tau) = 4D\tau^{\alpha},
\]
is not directly comparable across trajectories. Specifically, when $\alpha \neq 1$, the units of D become \si{\micro\meter\squared\second\tothe{-\alpha}}, rather than the standard \SI{}{\micro\meter\squared\per\second}. As a result, diffusion coefficients extracted from trajectories with different $\alpha$ values cannot be pooled or compared directly without violating dimensional consistency. To enable meaningful comparison across trajectories and across apical areas, we therefore defined a normalized diffusion strength by rescaling the diffusion coefficient using the transition time as a characteristic timescale. For each trajectory, the diffusion strength was calculated as:
\[
D_{strength} = D \cdot \tau_{transition}^{\alpha - 1}.
\]
This normalization restores consistent units (\SI{}{\micro\meter\squared\per\second}) while preserving information about the underlying dynamical regime. The diffusion strength thus provides a physically interpretable measure of BB mobility that can be compared across trajectories with different $\alpha$ values.

Using this approach, the diffusion strength measured from the fine-timed data peaked at approximately \SI{0.02}{\micro\meter\squared\per\minute}. A similar normalization strategy was applied to the coarse-timed data, where the frame interval was used as the characteristic timescale, yielding a comparable diffusion strength of approximately \SI{0.03}{\micro\meter\squared\per\minute} (see figure below).

\begin{figure}[H]
    \centering
    \includegraphics[width=0.9\linewidth]{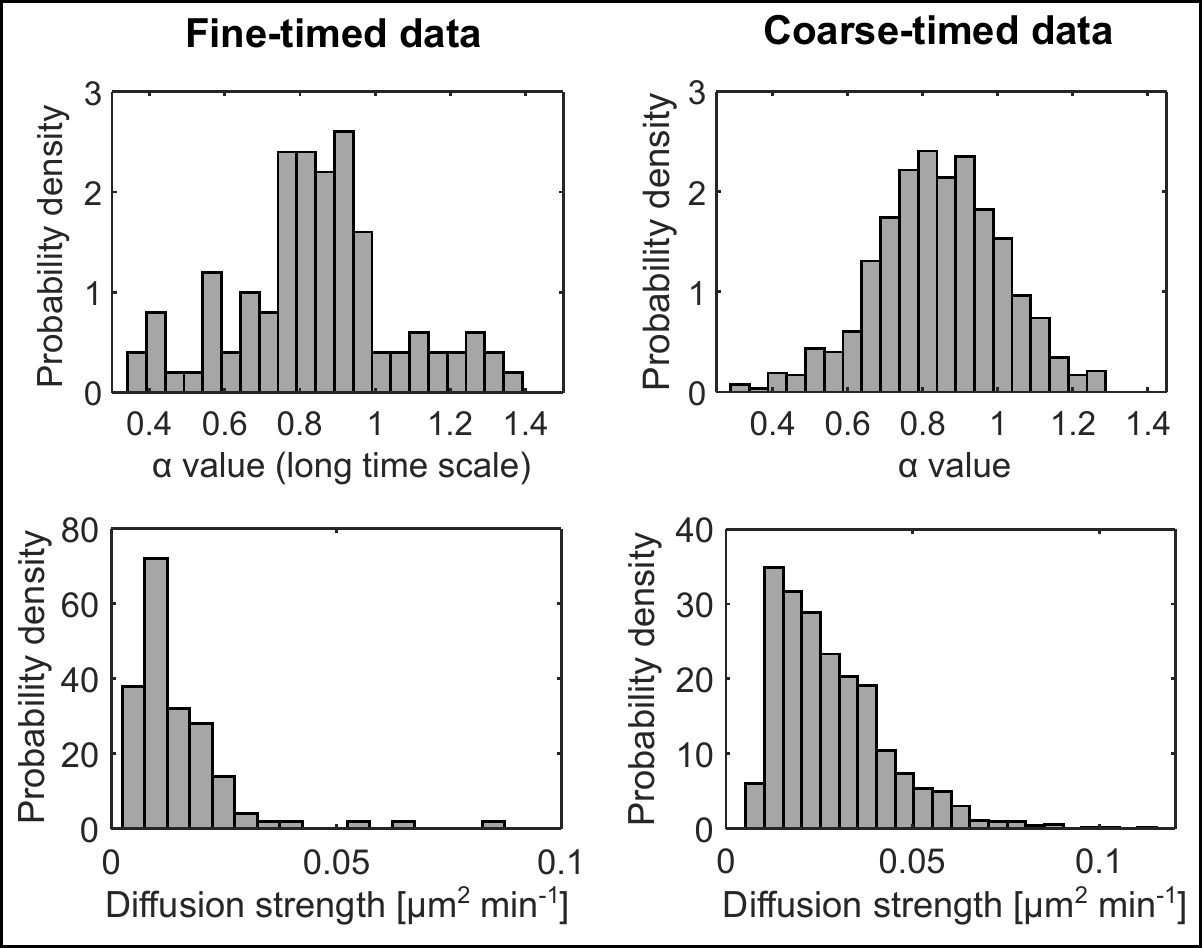}\par
    \justifying \textbf{Figure:} All data correspond to WT experiments. Plots in the first column correspond to fine-timed data, and those in the second column to coarse-timed data. The first and second rows show the probability densities of the $\alpha$ values and diffusion strengths, respectively.
    \label{fig:alphavalue_diffstrength_methods_fig}
\end{figure}

\subsubsection{Breakpoint estimation}
To identify potential changes in the relationship between apical area and dynamical parameters such as $\alpha$ values and transition times, we performed a piecewise linear regression analysis with a variable breakpoint. This analysis tests whether the data are better described by two linear regimes separated by a characteristic apical area. Breakpoint values were allowed to vary between approximately 70 and \SI{250}{\micro\meter\squared} in steps of \SI{1}{\micro\meter\squared}. For each candidate breakpoint, the dataset was split into two subsets (below and above the breakpoint), and independent linear regressions were fit to each subset. The goodness of fit was quantified using the sum of squared errors (SSE) across both segments. The breakpoint yielding the minimum SSE was selected as the best-fit breakpoint. The resulting breakpoint locations were consistently found within the range of $\sim{150}$–\SI{200}{\micro\meter\squared} across parameters. To assess whether the identified breakpoint corresponded to a genuine change in behavior, values before and after the breakpoint were compared using appropriate statistical tests.

\subsubsection{Assessment of localization noise}
To assess whether the features observed in the short-timescale MSDs of BBs could arise from localization noise or instrumental drift, we performed MSD analysis on immobilized fluorescent beads imaged under identical conditions. Fluorescent beads of approximately \SI{0.5}{\micro\meter} diameter, comparable to the size of BBs, were attached to the coverslip and imaged using the same frame rate and acquisition duration as the fine-timed BB experiments. MSDs were calculated for individual bead trajectories and ensemble averaged using the same analysis pipeline applied to BBs. The ensemble-averaged bead MSD at the smallest lag times provides an upper bound on localization noise and residual mechanical drift in the system. We found that the ensemble-averaged MSD of the beads was approximately an order of magnitude lower than the corresponding MSD values of BBs (\sfigref{sfigfourj}).

To further evaluate the legitimacy of BB displacements at short timescales, we compared the root mean squared displacement (RMSD) at the first lag time for ensemble-averaged BB and bead trajectories (\sfigref{sfigfourj}). The RMSDs of BBs and beads were 85 nm and 25 nm, respectively. Since the BB displacement at the first lag time was more than threefold larger than that of the beads, this strongly supports that the observed BB motion reflects genuine biological displacement rather than localization noise alone. 

However, immobilized beads are expected to exhibit higher fluorescence intensity than BBs, enabling more accurate localization. In contrast, the lower fluorescence signal of BBs introduces additional localization uncertainty, which contributes an additive offset to the MSD according to Martin et al.\cite{martin2002apparent} $MSD = 4Dt + 2\sigma^2$, where $\sigma$ represents the localization uncertainty. Since the beads were immobilized on the surface, we used the bead RMSD of 25 nm as an empirical estimate of the experimental noise floor. To estimate the contribution of this noise floor to the measured BB displacement, we subtracted the squared bead RMSD from the squared BB RMSD and subsequently calculated the square root of the resulting value. This yielded a noise-corrected BB displacement of 77 nm, which remains approximately threefold larger than the bead displacement (25 nm). These results indicate that localization noise alone cannot account for the measured BB displacement at the first lag time.

Notably, the measured BB displacement (85 nm) is smaller than the physical pixel size (106.667 nm), necessitating verification that the localization algorithm achieves reliable sub-pixel precision and that the theoretical localization uncertainty agrees with the empirical estimate obtained from the bead measurements. BB localization in the fine-timed experiments was performed using the ThunderSTORM plugin \cite{ovesny2014thunderstorm} with parameters optimized for sub-pixel localization (\textit{see Methods section 8.1.2}). The plugin reports individual $\sigma$ values for each localization event. These values were independently validated by manually calculating $\sigma$ using the expression described by Quan et al.\cite{quan2010localization}, yielding consistent results. Across all BB detections in all cells, the mean and mode of the localization uncertainty distribution were approximately 20–25 nm, consistent with the experimental noise floor estimated from the immobilized beads. Together, these analyses confirm that the localization algorithm achieves reliable sub-pixel precision, that the bead displacement provides a robust empirical estimate of the localization noise floor, and that the noise-corrected BB displacement of 77 nm at the first lag time represents a legitimate biological displacement.

Taken together, these findings indicate that localization noise does not dominate the observed BB dynamics, and that the short-timescale constrained diffusion and subsequent dynamical regime transitions reflect genuine BB motion rather than measurement artifacts.

\section{Common parameters for coarse- and fine-timed analysis}
\subsubsection{Step distance}
The BB step distance was defined as the Euclidean distance between two consecutive positions of a BB along its trajectory. For the coarse-timed dataset (frame interval: 30–60 s; total duration: $\sim{2.5}$ h), step distances from all trajectories across all cells were pooled, and their probability density distributions were computed.

\subsubsection{Contour length}
The contour length of a BB trajectory was calculated as the cumulative sum of all step distances along that trajectory. For the coarse-timed data, contour lengths from all BB trajectories across all cells were pooled and plotted as probability density distributions. For the fine-timed data, contour lengths were computed for trajectories spanning the entire apical area range. These values were binned in \SI{50}{\micro\meter\squared} apical area intervals and plotted as mean $\pm$ SD to visualize how contour length evolves with increasing apical area.

\subsubsection{End-to-end length}
The end-to-end length was defined as the direct Euclidean distance between the starting and ending coordinates of a BB trajectory. For the coarse-timed data, end-to-end lengths from all trajectories across all cells were pooled and plotted as probability density distributions. For the fine-timed data, end-to-end lengths were calculated for trajectories across the full apical area range, binned in \SI{50}{\micro\meter\squared} intervals, and plotted as mean $\pm$ SD to examine their evolution during apical expansion.

\subsubsection{Tortuosity}
Tortuosity of BB trajectories was quantified as the ratio of contour length ($d_{l}$) to end-to-end length ($d_{r}$), i.e., Tortuosity = $d_{l} / d_{r}$, and was calculated for both coarse-timed and fine-timed datasets. For the coarse-timed data, tortuosity was computed for each individual trajectory, and the resulting values were pooled and plotted as probability density distributions. For the fine-timed data, trajectories are substantially shorter in duration but sampled at much higher temporal resolution. As a result, tortuosity values were computed for trajectories grouped by apical area, binned in \SI{50}{\micro\meter\squared} intervals, and plotted to assess how trajectory complexity evolves across the apical expansion. Notably, tortuosity values obtained from the fine-timed data were substantially larger than those from the coarse-timed data. This arises because the high temporal resolution captures thousands of very small BB displacements, leading to large cumulative contour lengths, while the corresponding end-to-end displacements remain small due to the short observation window. Consequently, the ratio $d_{l} / d_{r}$ becomes large. To facilitate visualization and comparison across apical area bins, tortuosity values from the fine-timed dataset were therefore natural log-transformed prior to plotting.

\section{Image analysis for high-resolution experiments}
\subsection{Airyscan processing}
The first step in the high resolution imaging analysis pipeline was the processing of raw Airyscan data. Airyscan raw images were processed using ZEN 3 software, where the raw data were deconvolved using the Super Resolution filter, where the value was automatically estimated by the algorithm to generate the corresponding high resolution image. The default filter setting was chosen after testing a range of filter values across multiple images, during which the default value was found to consistently provide optimal deconvolution performance without introducing artifacts.

\subsection{Image processing}
Subsequent analysis steps like the apical plane isolation, apical area contour identification, BB detection, and BB position extraction, closely follow the analysis pipeline described for the coarse-timed experiments. Therefore, only those steps that were specifically adapted or tuned for high resolution imaging are described in detail below. For a complete description of shared steps, refer to the corresponding sections in the coarse-timed analysis.

\subsubsection{Apical plane isolation}
To isolate the apical plane containing the BBs and the associated actin meshwork, the acquired z-stack spanning the entire apical cortex was carefully examined. The apical cortex was manually identified and isolated. In cases where curvature of the apical surface was observed, the z-slices encompassing the curved apical plane were projected along the z-axis. The resulting projected image was used as a representative apical surface for subsequent analysis of the actin meshwork surrounding BBs.

\subsubsection{Apical area contour extraction and BB detection}
The apical area contour was manually annotated using the actin channel. From this contour, the apical peripheral cortex was defined by inwardly shifting the apical boundary by a specified distance using a custom written script. This inward shifted contour defined the ROI for the peripheral cortex, which was used to quantify actin intensity at the apical periphery. Actin intensities measured within this peripheral cortex ROI were used for normalization.

BBs were detected in the BB channel by transforming the signal into binarized, discrete puncta using either a Difference of Gaussians (DoG) or Laplacian of Gaussian (LoG) filter implemented via the GDSC plugin\cite{etheridge2022gdsc}, followed by identification of central maxima using the \textit{Find Maxima} function in Fiji.

\subsubsection{BB position identification}
BB positions were extracted from the binarized images using the Particle Tracker plugin\cite{sbalzarini2005feature}. The same workflow described in the Automated tracking section (7.2.4.2) of the coarse-timed analysis was followed, with the exception that all parameters were left at their default values. As these datasets consist of single frame images rather than time stacks, the plugin returns BB positions only for the first frame.

\subsubsection{Definition of BB and surrounding regions}
Using the BB positions obtained above, circular ROIs were drawn around each BB with a radius of \SI{0.25}{\micro\meter}. A second concentric circular ROI with a radius of \SI{0.5}{\micro\meter} was then generated. The inner ROI (\SI{0.25}{\micro\meter} radius) was used to quantify actin intensity at the BB position, while the annular region between the \SI{0.25}{\micro\meter} and \SI{0.5}{\micro\meter} radii was used to quantify actin intensity in the surrounding region. The choice of \SI{0.25}{\micro\meter} and \SI{0.5}{\micro\meter} radii was guided by repeated manual measurements of BB diameters and the spatial extent of high intensity actin enrichment surrounding BBs. These ROI definitions enabled automated and consistent extraction of actin intensities at BB positions and in their immediate surroundings for all BBs within the apical area. All ROI generation and intensity extraction steps were implemented using a custom written script.

\subsection{Parameters calculated for high-resolution imaging analysis}
\subsubsection{Actin intensity ratios at different CDF percentiles}
For each apical domain, actin intensities measured at BB positions and in the surrounding regions were plotted as cumulative distribution functions (CDFs). These CDFs were used to visualize systematic differences in actin enrichment between the two regions as a function of apical area. From each CDF, the corresponding actin intensity at $25^{\text{th}}$, $50^{\text{th}}$, and $75^{\text{th}}$ percentiles were extracted. For each percentile, the ratio of actin intensity in the surrounding region to that at the BB position was calculated. These ratios were then plotted across the entire apical area range to assess how the relative actin enrichment evolves during apical expansion, across all cells.

\subsubsection{Ratio of ensemble-averaged actin intensity}
While the CDF based analysis rigorously incorporates actin intensity measurements from individual BBs, it may introduce pseudoreplication by treating multiple BBs within the same apical domain as independent data points. To address this, a complementary ensemble averaged analysis was performed. For each apical domain, actin intensities at BB positions and in surrounding regions were averaged across all BBs within that apical domain, yielding one mean value per domain for each region. The ratio of the ensemble averaged actin intensity in the surrounding region to that at the BB position was then computed, resulting in a single ratio per apical domain. These ratios were plotted across the full apical area range and used as a complementary measure to the CDF based ratios.

\subsubsection{Breakpoint estimation}
Changes in the relationship between apical area and (i) the CDF based actin intensity ratios and (ii) the ratio of ensemble averaged actin intensities were assessed using breakpoint estimation, following the same procedure described for the fine-timed experiments. Following breakpoint identification, data points before and after the breakpoint were statistically compared to assess the nature and significance of changes in the relationship between apical area and relative actin enrichment around BBs.

\section{Image analysis for figures and movies}
\textbf{\figref{figoneb, figtwoa, figfivea, sfigonea, sfigsixa}}: For visualization purposes, the contrast of the BB channel in these figures was enhanced by subtracting a constant intensity offset from the images using the \textit{Subtract} operation under the Math functions in Fiji.

\textbf{\figref{figonec, figfiveb}}: Orthogonal views shown in these figures were generated using the ClearVolume plugin\cite{royer2015clearvolume} in Fiji.

\textbf{\figref{figtwod} and Movie 6}: BB entry and exit events shown in this figure and movie were obtained as follows. A BB of interest was first identified in the xy plane. The \textit{Orthogonal Views} function in Fiji was then used to locate the same BB in the corresponding xz and yz planes. Images corresponding to the xy, xz, and yz planes were saved separately. This procedure was repeated for each successive step of BB movement observed in the xy plane. This approach allowed capture of BB entry and exit events at the apical plane in the xy view, while simultaneously tracking BB motion below the apical surface in the xz and yz views. The process was continued until the BB ceased its entry–exit cycles and remained stably associated with the apical domain. Following this, BB movement along the z-axis during the entry–exit cycles was quantified by manually tracking the BB in the xz plane using the Manual Tracking plugin (by Fabrice Cordeli\`{e}res) in Fiji, yielding a measure of BB displacement along the apico–basal axis.

\section{Statistical analysis}
Data are presented as the mean of binned values (corresponding to the intersection of the vertical and horizontal error bars), with error bars representing the standard deviation (SD), for the time- and area-evolution plots. For violin plots, individual data points are shown as dotted markers within the distribution when sample sizes are moderate. For plots with very large sample sizes, individual points are not displayed to avoid visual overcrowding. In all violin plots, the central line indicates the median, and the upper and lower lines represent the $75^\text{th}$ and $25^\text{th}$ percentiles, respectively.

For plots showing probability distributions, the peak of unimodal histogram distributions was estimated by fitting a normalized Gaussian function to the binned data. The fitted mean was used as an estimate of the peak. This procedure was used solely to obtain a smooth and consistent peak estimate and was not intended to infer the underlying distribution shape. Across all unimodal distributions analyzed, this approach yielded peak estimates that were visually consistent with the data.

For statistical comparisons, normality of the data was assessed using the Shapiro–Wilk, D’Agostino–Pearson, Kolmogorov–Smirnov, and Anderson–Darling tests implemented in GraphPad Prism. The tests generally yielded consistent conclusions regarding normality. When minor discrepancies occurred, the overall assessment was based on the majority of tests.

For unpaired data with two groups, normally distributed data with large sample sizes were assessed using an unpaired t-test, whereas non-normally distributed data or data with small sample sizes were analyzed using the Mann–Whitney test. For unpaired data involving more than two groups, a Kruskal–Wallis test was performed, followed by Dunn’s post-hoc test for pairwise comparisons.

A significance threshold of $p < 0.05$ was applied. Exact $p$ values are reported in the figure legends, and significance is denoted as follows: $p < 0.05 \text{(*)},~p < 0.01 \text{(**)},~p < 0.001 \text{(***)},~p < 0.0001 \text{(****)}$. All statistical analyses were performed using GraphPad Prism 11. Each experiment was repeated at least three times, and the number of cells, trajectories, and independent experiments is reported in the figure legends.


 \begin{table}[H]  
  \centering
  \small 
  \majorheading{List of materials}
  \addcontentsline{toc}{section}{\hyperref[sec:List of materials]{\large\bfseries List of materials}}
  \begin{tabular}{p{8cm} p{3.5cm} p{3cm}}
  \addlinespace[1em]
    \hline
    \addlinespace[0.5em]
    \multicolumn{3}{c}{\textbf{Chemicals and staining reagents}} \\
    \addlinespace[0.5em]
    \hline
    \addlinespace[0.2em]
    \textbf{Product} & \textbf{Source} & \textbf{Identifier} \\
    \addlinespace[0.2em]
    \hline
    \addlinespace[0.5em]
    Chorulon & MSD Animal Health & 422741 \\
    \ce{Nacl} & Fisher Scientific & BP358 \\
    \ce{KCL} & Fisher Scientific & BP366 \\
    \ce{CaCl2} & Scharlab & CA01941000 \\
    \ce{MgCl2} & Sigma-Aldrich & M2670 \\
    Cysteine & Sigma-Aldrich & C7880 \\
    Ficoll & Sigma-Aldrich & F4375 \\
    3-(N-Morpholino)propanesulfonic acid (MOPS) & Sigma-Aldrich & M1254 \\
    Ethylene Glycol-bis(beta-aminoethyl ether) - N,N,N',N' - tetraacetic acid tetrasodium salt (EGTA) & Sigma-Aldrich & E8145 \\
    \ce{MgSO4} & Fisher Scientific & BP213 \\
    Formaldehyde & Sigma-Aldrich & 47608 \\
    Tris-HCl (Trizma\textsuperscript{\textregistered} Hydrochloride) & Sigma-Aldrich & T5941 \\  Triton X-100 & Sigma-Aldrich & T8787 \\
    Fetal bovine serum (FBS) & Sigma-Aldrich & F2442 \\
    Dimethyl Sulfoxide (DMSO) & Sigma-Aldrich & D2650 \\
    Phalloidin 555 & Invitrogen & A34055 \\
    Phalloidin Atto-594 & Sigma-Aldrich & 51927 \\
    Phalloidin 647 & Invitrogen & A22287 \\
    Ultra low melting point agarose & Sigma-Aldrich & A2576 \\
    Low melting point agarose & Promega & V2111 \\
    Cover slips (Thickness 1.5 H; 25 mm) & Marienfeld & 0117650 \\
    Imaging chambers (Attofluor Cell chamber\textsuperscript{\texttrademark} for microscopy) & Invitrogen & A7816 \\
    High vacuum Silicone grease & Sigma-Aldrich & Z273554 \\
    Fluorescent beads (\SI{0.5}{\micro\meter}; Yellow) & Spherotech & FP-0552-2\\
    \end{tabular}
    \begin{tabular}{p{8cm} p{6cm} p{0.1cm}}
    \addlinespace[1em]
    \hline
    \addlinespace[0.5em]
    \multicolumn{3}{c}{\textbf{Plasmids}} \\
    \addlinespace[0.5em]
    \hline
    \addlinespace[0.2em]
    \textbf{Plasmids} & \textbf{Source} & \textbf{} \\
    \addlinespace[0.2em]
    \hline
    \addlinespace[0.2em]
    alpha-Tubulin-LifeAct-GFP \newline alpha-Tubulin-LifeAct-RFP & John Wallingford group\\
    alpha-Tubulin-Chibby-GFP \newline alpha-Tubulin-Chibby-RFP  &  Brian Mitchell group \\
    \addlinespace[0.2em]
    \hline
  \end{tabular}
\end{table}

\begin{table}[H]  
  \centering
  \small 
  \begin{tabular}{p{8cm} p{3.5cm} p{3cm}}
  \hline
    \addlinespace[0.5em]
    \multicolumn{3}{c}{\textbf{Oligomorpholinos}}\\
    \addlinespace[0.5em]
    \hline
    \addlinespace[0.2em]
    \textbf{Morpholino} & \textbf{Company} & \textbf{Sequence (5'-3')} \\
    \addlinespace[0.2em]
    \hline
    \addlinespace[0.5em]
    $\alpha$-Actinin-1 (ACTN1.L) Translation blocking & Gene Tools USA & \texttt{\seqsplit{CATAATGATCCATCCTGAGCTGCTG}}\\
    $\alpha$-Actinin-1 (ACTN1.L) Splice blocking & Gene Tools USA & \texttt{\seqsplit{CCTGTGGCTCAAAATCTTACCTTCA}}\\  
    \end{tabular}
    \begin{tabular}{p{3.5cm} p{4.5cm} p{7cm}}
    \addlinespace[1em]
    \hline  
    \addlinespace[0.5em]
    \multicolumn{3}{c}{\textbf{Softwares}} \\
    \addlinespace[0.5em]
    \hline
    \addlinespace[0.2em]
    \textbf{Product} & \textbf{Company} & \textbf{Link} \\
    \addlinespace[0.2em]
    \hline
    \addlinespace[0.5em]
    ImageJ & - & \url{https://imagej.net/ij/}\\
    Fiji & - & \url{https://fiji.sc}\\
    MATLAB R2023b & Mathworks & \url{https://se.mathworks.com/products/matlab.html}\\
    GraphPad Prism 11 & GraphPad Software, LLC & \url{www.graphpad.com}\\
    Inkscape 1.4 & - & \url{https://inkscape.org}\\
    Adobe Illustrator 27 & Adobe & \url{https://www.adobe.com/products/illustrator.html}\\
    \hline
  \end{tabular}
\end{table}

\bibliography{literature_refs}
\addcontentsline{toc}{section}{\hyperref[sec:References]{\large\bfseries References}}